\documentclass[modern]{aastex631}

\newcommand{\kms}{${\rm km~s}^{-1}$}

\newcommand{\dm}{$\Delta{\rm m}_{15}(B)$}
\newcommand{\sbv}{$s_{BV}$}

\submitjournal{ApJS}

\shorttitle{1991T-like Supernovae}
\shortauthors{Phillips et al.}
\begin{document}

\title{1991T-like Supernovae\footnote{This paper includes data gathered with the 6.5 meter Magellan telescopes at Las Campanas Observatory, Chile.}}

\author[0000-0003-2734-0796]{M.~M.~Phillips}
\affiliation{Carnegie Observatories, Las Campanas Observatory, Casilla 601, La Serena, Chile}

\author[0000-0002-5221-7557]{C.~Ashall}
\affiliation{Department of Physics, Virginia Polytechnic Institute and State University, 850 West Campus Drive, Blacksburg, VA 24061, USA}

\author[0000-0001-6272-5507]{Peter J. Brown}
\affiliation{George P. and Cynthia Woods Mitchell Institute for Fundamental Physics and Astronomy, Texas A\&M University,
Department of Physics and Astronomy,  College Station, TX 77843, USA}

\author[0000-0002-1296-6887]{L.~Galbany}
\affiliation{Institute of Space Sciences (ICE, CSIC), Campus UAB, Carrer de Can Magrans, s/n, E-08193 Barcelona, Spain.}
\affiliation{Institut d'Estudis Espacials de Catalunya (IEEC), E-08034 Barcelona, Spain.}

\author[0000-0002-2471-8442]{M. A. Tucker}
\altaffiliation{CCAPP Fellow}
\affiliation{Center for Cosmology and Astroparticle Physics, The Ohio State University, 191 West Woodruff Ave, Columbus, OH, USA}
\affiliation{Department of Astronomy, The Ohio State University, 140 West 18th Avenue, Columbus, OH, USA}
\affiliation{Department of Physics, The Ohio State University, 191 West Woodruff Ave, Columbus, OH, USA}

\author[0000-0003-4625-6629]{Christopher R.~Burns}
\affiliation{Observatories of the Carnegie Institution for Science, 813 Santa Barbara St., Pasadena, CA 91101, USA}

\author[0000-0001-6293-9062]{Carlos~Contreras}
\affiliation{Carnegie Observatories, Las Campanas Observatory, Casilla 601, La Serena, Chile}

\author[0000-0002-4338-6586]{P.~Hoeflich}
\affiliation{Department of Physics, Florida State University, 77 Chieftan Way, Tallahassee, FL  32306, USA}

\author[0000-0003-1039-2928]{E.~Y.~Hsiao}
\affiliation{Department of Physics, Florida State University, 77 Chieftan Way, Tallahassee, FL  32306, USA}

\author[0000-0001-8367-7591]{S. Kumar}
\affiliation{Department of Physics, Florida State University, 77 Chieftan Way, Tallahassee, FL  32306, USA}
\affiliation{Department of Astronomy, University of Virginia, 530 McCormick Rd, Charlottesville, VA 22904, USA}

\author[0000-0003-2535-3091]{Nidia~Morrell}
\affiliation{Carnegie Observatories, Las Campanas Observatory, Casilla 601, La Serena, Chile}

\author[0000-0002-9413-4186]{Syed~A.~Uddin}
\affiliation{George P. and Cynthia Woods Mitchell Institute for Fundamental Physics and Astronomy, Texas A\&M University,
Department of Physics and Astronomy,  College Station, TX 77843, USA}

\author[0000-0001-5393-1608]{E.~Baron}
\affiliation{Planetary Science Institute, 1700 East Fort Lowell Road, Suite 106, Tucson, AZ 85719-2395, USA}
\affiliation{Hamburger Sternwarte, Gojenbergsweg 112, D-21029 Hamburg, Germany}
\affiliation{Homer L. Dodge Department of Physics and Astronomy, University of Oklahoma, 440 W. Brooks, Rm 100, Norman, OK 73019-2061, USA}

\author[0000-0003-3431-9135]{Wendy~L.~Freedman}
\affiliation{Department of Astronomy and Astrophysics, University of Chicago, 5640 S. Ellis Ave, Chicago, IL 60637, USA}

\author[0000-0002-6650-694X]{Kevin~Krisciunas}
\affiliation{George P. and Cynthia Woods Mitchell Institute for Fundamental Physics and Astronomy, Texas A\&M University,
Department of Physics and Astronomy,  College Station, TX 77843, USA}

\author[0000-0003-0554-7083]{S.~E.~Persson}
\affiliation{Observatories of the Carnegie Institution for Science, 813 Santa Barbara St., Pasadena, CA 91101, USA}

\author[0000-0001-6806-0673]{Anthony L. Piro}
\affiliation{Observatories of the Carnegie Institution for Science, 813 Santa Barbara St., Pasadena, CA 91101, USA}

\author[0000-0003-4631-1149]{B. J. Shappee}
\affiliation{Institute for Astronomy,
University of Hawai`i at M\={a}noa,
2680 Woodlawn Dr., Honolulu, HI, USA}

\author[0000-0002-5571-1833]{Maximilian Stritzinger}
\affiliation{Department of Physics and Astronomy, Aarhus University, Ny Munkegade 120, DK-8000 Aarhus C, Denmark}

\author[0000-0002-8102-181X]{Nicholas~B.~Suntzeff}
\affiliation{George P. and Cynthia Woods Mitchell Institute for Fundamental Physics and Astronomy, Texas A\&M University,
Department of Physics and Astronomy,  College Station, TX 77843, USA}

\author[0000-0003-2183-148X]{Sudeshna Chakraborty}
\affiliation{Department of Physics, Florida State University, 77 Chieftan Way, Tallahassee, FL  32306, USA}

\author[0000-0002-1966-3942]{R.~P.~Kirshner}
\affiliation{Gordon and Betty Moore Foundation, 1661 Page Mill Road, Palo Alto, CA 94304, USA}
\affiliation{Harvard-Smithsonian Center for Astrophysics, 60 Garden Street, Cambridge, MA 02138, USA}

\author[0000-0002-3900-1452]{J. Lu}
\affiliation{Department of Physics, Florida State University, 77 Chieftan Way, Tallahassee, FL  32306, USA}

\author[0000-0002-2966-3508]{G.~H.~Marion}
\affiliation{University of Texas at Austin, 1 University Station C1400, Austin, TX 78712-0259, USA}

\author[0000-0002-1633-6495]{Abigail Polin}
\affiliation{Observatories of the Carnegie Institution for Science, 813 Santa Barbara St., Pasadena, CA 91101, USA}
\affiliation{TAPIR, Walter Burke Institute for Theoretical Physics, Caltech, 1200 East California Boulevard, Pasadena, CA 91125, USA}

\author[0000-0002-9301-5302]{M. Shahbandeh}
\affiliation{Department of Physics, Florida State University, 77 Chieftan Way, Tallahassee, FL  32306, USA}




\begin{abstract}

Understanding the nature of the luminous 1991T-like
supernovae is of great importance to supernova cosmology as they are likely to have
been more common in the early universe.
In this paper we explore the observational properties of 1991T-like supernovae to 
study their relationship to other luminous, slow-declining Type~Ia supernovae (SNe~Ia).
From the spectroscopic and photometric criteria defined in \citet{phillips22},  we identify 17 1991T-like supernovae from the literature.
Combining these objects with ten 1991T-like supernovae from the Carnegie Supernova Project-II, the spectra, light curves, and colors of these events, along with their host galaxy 
properties, are examined in detail.  We conclude
that 1991T-like supernovae are closely related in essentially all of their UV, optical, and near-infrared
properties --- as well as their host galaxy parameters --- to the slow-declining subset of Branch core-normal supernovae
and to the intermediate 1999aa-like events, forming a continuum of luminous 
SNe~Ia.  The
overriding difference between these three subgroups appears to be the extent to which $^{56}$Ni mixes into 
the ejecta, producing the pre-maximum spectra dominated by \ion{Fe}{3} absorption, the broader UV light curves, and the higher luminosities that characterize the 1991T-like events.
Nevertheless, the association of 1991T-like SNe with the rare Type~Ia~CSM supernovae 
would seem to run counter to this hypothesis, in which case 
1991T-like events may form a separate subclass of SNe~Ia, possibly
arising from single-degenerate progenitor systems. 

\end{abstract}



\section{Introduction}

\label{sec:intro}

The degree to which Type~Ia supernovae (SNe~Ia) represented a homogenous class of explosions was a matter of considerable
debate in the 1980s \citep[e.g.][]{branch81,pskovskii84,cadonau85}.  The first indisputable evidence for significant differences 
came with the observation of high-velocity intermediate mass elements (IMEs) in SN~1984A   \citep{branch87} and 
the discovery of three 
peculiar events: SN~1986G \citep{phillips87}, SN~1991T \citep{filippenko92b,phillips92,ruiz-lapuente92}, and 
SN~1991bg \citep{filippenko92a,leibundgut93}.  SN~1991T was extremely well observed
\citep{filippenko92b,phillips92,ruiz-lapuente92,jeffery92,mazzali95,gomez96,lira98,blondin12,silverman12} and displayed 
reasonably normal optical light curves, albeit with a relatively slow decline rate after maximum of
\dm~$= 0.95 \pm 0.05$ mag\footnote{\dm~is defined as the amount in magnitudes that
the supernova fades in the first 15 days since the time of $B$-band maximum \citep{phillips93}.} \citep{lira98}.  Nevertheless,
the optical spectrum at pre-maximum phases was dominated by strong absorption features of \ion{Fe}{3} instead of the 
\ion{Si}{2}, \ion{Ca}{2}, and \ion{S}{2} lines that are typically observed in the spectra of other SNe~Ia
 at these phases.  By two weeks after maximum light, however, the spectrum of SN~1991T had evolved to one closely resembling a normal 
Type~Ia event\footnote{In this paper, a ``normal Type~Ia 
supernova'' refers to the ``core-normal'' definition of \citet{branch06}.}.

Attempts to interpret the spectra of SN~1991T focussed on the strong \ion{Fe}{3} absorption and the absence of  
IMEs in the earliest spectra.  \citet{filippenko92b} suggested that a ``double detonation'' initiated at
the boundary between the C-O core and He envelope of a white dwarf \citep{nomoto82} could explain the prominent Fe-group elements 
in the outer layers of the SN, although they could not rule out a ``delayed detonation'' (DDT) model of a Chandrasekhar-mass C-O white 
dwarf \citep{khokhlov91a}.
More recently, \citet{townsley19} proposed that the detonation of a thin layer of helium on the surface of a $1~M_\sun$ C-O white dwarf can 
reproduce the spectra and light curves of ``normal'' SNe~Ia, but the question of whether 1991T-like (henceforth ``91T-like'') SNe can also be successfully modeled as 
a double detonation was not addressed.  However, \citet{polin21} found that the [\ion{Ca}{2}]/[\ion{Fe}{3}] ratio in the optical nebular spectra of 
luminous ($M_B < -19$~mag) SNe Ia, including SN 1991T, cannot be reproduced by the double detonation mechanism.
\citet{ruiz-lapuente92} favored a DDT that did not produce significant IMEs in the outermost layers, but pointed 
out \citep[as did][]{filippenko92b} that models of DDT explosions provide a better match to normal SNe~Ia than they do to SN~1991T.  
\citet{mazzali95} carried out extensive modeling of both pre-maximum and post-maximum spectra and concluded that SN~1991T was 
likely produced by a late transition from deflagration to detonation \citep[e.g.,][]{yamaoka92} that led to the production of $\sim0.6~M_\sun$ 
of $^{56}$Ni in the outermost $1~M_\sun$ of the ejecta.  \citet{fisher99} speculated that SN~1991T was a ``super-Chandraskhar'' (SC) 
explosion resulting from the merger of two white dwarfs, but this was motivated by an incorrectly assumed distance that led to an 
overestimate of the peak luminosity.  Mergers will also
produce asymmetric envelopes that are inconsistent with 
polarization measurements (see \S\ref{sec:discussion}).
An abundance tomography study by \citet{sasdelli14} concluded that DDT models provided the 
best match to SN~1991T, but that the peculiar distribution of elements with the IMEs dominant only in a narrow shell wedged between 
$^{56}$Ni-rich and O-rich zones required an {\em early} (high-density) transition from deflagration to detonation.   \citet{fisher15} have suggested 
that the gravitationally-confined detonation \citep{plewa04} of an accreting Chandrasekhar-mass white dwarf could explain 91T-like 
supernovae, where the outer Fe-group elements are produced by a strongly off-center deflagration.  However, detailed calculations
of this model by \citet{seitenzahl16} were not found to be in good agreement with the observed properties of SN~1991T.
A pulsational-driven detonation \citep[PDD;][]{khokhlov91b},
in which an initial deflagration stalls and the inner region collapses until a detonation occurs at the border between processed and unprocessed elements,
was suggested for SN~1991T by \citet{hoeflich94}.
\citet{hoeflich96}, \citet{quimby07}, and, more recently,
\citet{aouad22} have emphasized that a PDD
could lead to the IMEs being confined to a relatively narrow velocity range.  Nevertheless, 
such models predict 
significant amounts of unburned carbon in the outermost layers of the ejecta \citep{baron08},
which is not observed in 91T-like SNe.

A few years after the discovery of SN~1991T, \citet{nugent95} argued that the differences observed in the maximum light
optical spectra of SNe~Ia --- from the fast-declining, sub-luminous SN~1991bg, to the slow-declining, luminous SN~1991T --- could be 
ascribed to differences in temperature corresponding to the amount of $^{56}$Ni produced in the explosion,
with 91T-like events representing the highest temperatures (largest $^{56}$Ni masses).
\citet{hoeflich_etal96} independently showed that the luminosity--decline
rate relation \citep{phillips93} could be modelled by the
explosions of Chandrasekhar mass white dwarfs that produce
varying amounts of $^{56}$Ni, affecting both the peak
luminosity as well as the diffusion time for photons escaping
from the outer ejecta.
Observations of SN~1999aa 
\citep{filippenko99}, which showed a pre-maximum spectral evolution intermediate between that of SN~1991T and normal SNe~Ia
\citep{li01b,garavini04}, supported the idea that 91T-like events are simply extreme examples of normal SNe~Ia
\citep{branch01,garavini04}.
However, a recent abundance tomography study of SN~1999aa by \citet{aouad22} found abundance peculiarities in the
outer ejecta like those deduced for SN~1991T by \citet{sasdelli14}, concluding that a temperature increase alone cannot 
account for the differences between these SNe and normal events.
\citet{OBrien23} reached a similar conclusion that a combination of both
abundance and ionization differences is required to explain the transition from
normal to 91T-like SNe.
Moreover, the discovery of the Type~Ia-CSM supernovae (SNe~Ia-CSM) events
\citep{hamuy03b,aldering06,prieto07,dilday12,taddia12,silverman13b} that show 91T-like spectra at early epochs before the commencement of a 
strong interaction with circumstellar material (CSM) suggests that 91T-like SNe might be fundamentally different than normal
SNe~Ia  \citep{dilday12}.

\citet{phillips22} (henceforth, Paper~I) described a new method for classifying 91T-like events through the combination of one or more optical spectra
obtained at phases $\leq +10$~days and an $i/I$-band light curve to identify ten 91T-like SNe in the 
second phase of the Carnegie Supernova Project \citep[CSP-II;][]{phillips19,hsiao19}.
This sample was used to demonstrate that 91T-like SNe are
$0.1$--$0.6$~mag more luminous than normal SNe~Ia with similar post-maximum decline rates \citep[see also][]{boone21,yang22}. 
In this second paper, we explore the optical, near-infrared (NIR), and utraviolet (UV) characteristics
of 91T-like SNe, and compare them with the properties of 1999aa-like (henceforth ``99aa-like'') and slow-declining Branch 
``core-normal'' (CN) SNe.
The new classification method for identifying 91T-like SNe is reviewed in \S\ref{sec:literature} and applied to produce a sample of 17 91T-like events culled
from the literature.
Next, in \S\ref{sec:opt_spectra} and \S\ref{sec:nir_spectra}, we discuss different aspects of the optical and NIR spectra of
91T-like events, comparing them with 99aa-like and other luminous SNe~Ia.
In \S\ref{sec:lcurves}, general characteristics of the optical and ultraviolet light curves, including color evolution and 
pseudo-bolometric light curves are considered. 
In \S\ref{sec:hosts}, host galaxy properties are discussed, while in \S\ref{sec:Ia-csm}, observations of the best-observed SNe~Ia-CSM at 
early phases when the ejecta--CSM interaction was weakest are reviewed.
Finally, in \S\ref{sec:discussion}, we summarize our findings and briefly speculate on the general question of the progenitors and explosion 
mechanisms of luminous SNe~Ia.

\section{Literature Sample of 91T-like Supernovae}
\label{sec:literature}

The defining feature of 91T-like SNe is the unusual weakness in  pre-maximum and maximum-light optical spectra of features due to IMEs,
specifically \ion{Si}{2} $\lambda6355$, \ion{Ca}{2}~H~\&~K, and \ion{S}{2} $\lambda\lambda$5449,5622.  In Paper-I, we demonstrated that
91T-like events could be identified from one or more measurements of the pseudo-equivalent width (pEW) of the \ion{Si}{2} $\lambda6355$ absorption
obtained at phases $\leq +10$~days\footnote{In this paper, all phases are given with respect to
the epoch of $B$ maximum, unless otherwise specified.}.  An $i$/$I$-band light curve is also 
important to differentiate 91T-like SNe from 02cx-like (a.k.a. ``Type~Iax'') 
and 03fg-like (formerly called ``Super-Chandrasekhar'') events that can also show weak \ion{Si}{2} $\lambda6355$ 
and blue continua at similar epochs, as the latter SNe
do not show a distinct secondary maximum and their $i$/$I$-band light curves peak after $B$~maximum \citep{gonzalez14,ashall20}.

To summarize, we employ both spectroscopic and photometric requirements for classifying a SN as a 91T-like event, which we define
as follows:

\begin{itemize}
 
  \item Spectroscopic:  At least one pEW(\ion{Si}{2} $\lambda6355$) measurement before $+10$~days that is consistent with the trajectory of SN~1991T
   in a plot of pEW(\ion{Si}{2} $\lambda6355$) versus light curve phase.  In practical terms, this translates to pEW(\ion{Si}{2} $\lambda6355$) $\lesssim20$~\AA\ at $-10$~days,
   $\lesssim40$~\AA\ at maximum, and $\lesssim$50~\AA\ at $+10$~days.  As discussed in Paper~I, precise phase information is critical to 
   distinguishing between 99aa-like and 91T-like events (e.g., the spectrum of a 99aa-like SN obtained a week before maximum is virtually identical to one of
   a 91T-like event observed at maximum).
   
   \item Photometric:  An $i$/$I$-band or NIR light curve that reaches maximum before the epoch of $B$ maximum and displays a clear secondary maximum.
   
\end{itemize}

\begin{figure*}[t]
\epsscale{1.05}
\plotone{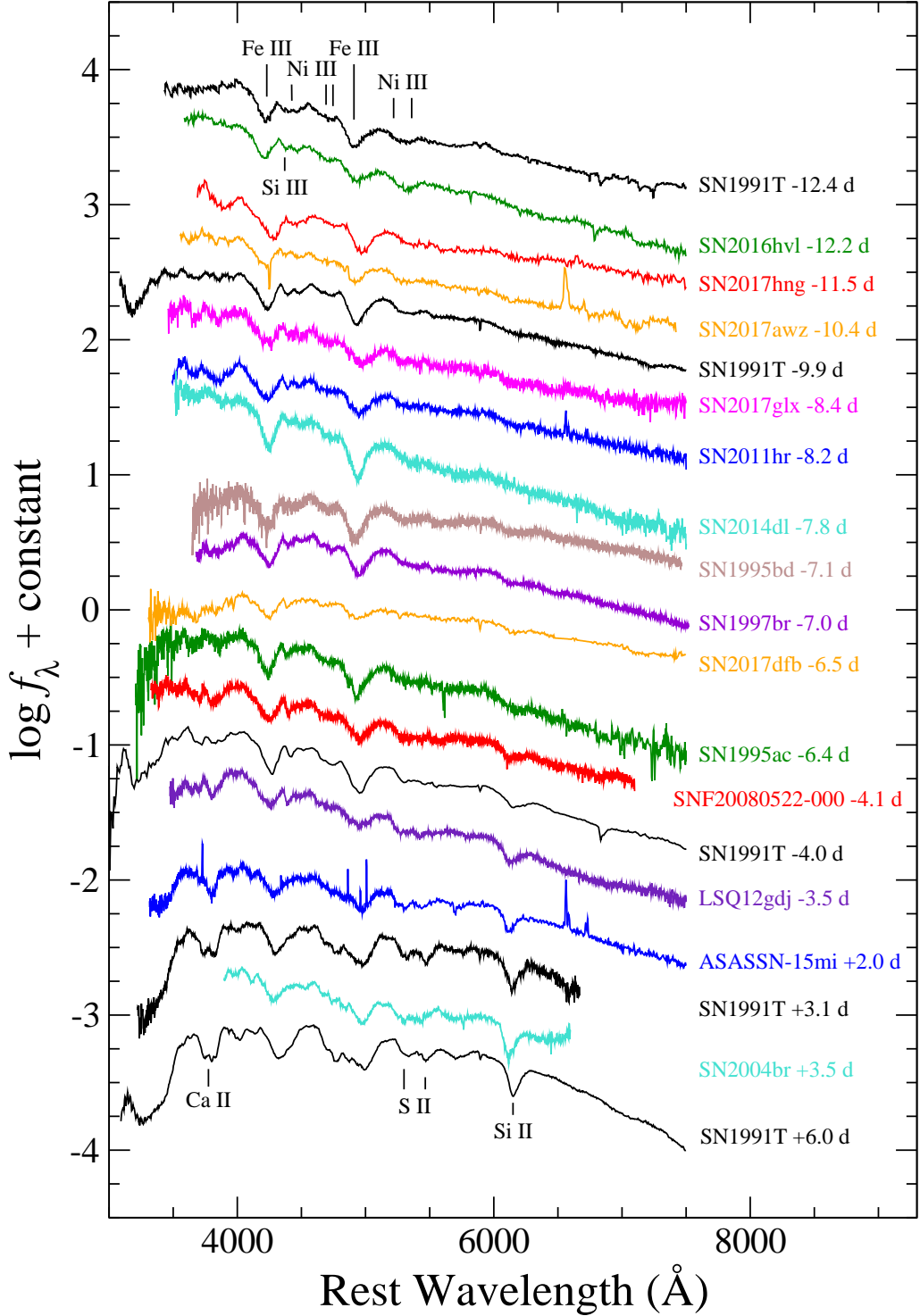}
\caption{Optical spectra of 15 of the literature sample of 91T-like SNe obtained from $-12$ to $+4$~days.
Missing from this figure are SNF20070803-005 and SNF20080723-012 whose spectra are not publicly available but are reproduced in \citet{scalzo12}. 
Spectra of SN~1991T at five phases covering this range are plotted in black for comparison.  All of the spectra 
have been corrected for Galactic reddening.  Identifications of the strongest features at pre- and post-maximum phases are included.}
\label{fig:historical_91T_spec}
\end{figure*}

Using the above criteria in Paper~I, a total of ten 91T-like events were discovered in the CSP-II sample of 214 SNe~Ia.
Applying the same criteria to spectra and light curves available in the published literature
through 2019,
we have identified an additional 17 91T-like SNe, along with  
eight suspected members of this subclass.  Details regarding the identification of these objects are given in Appendix \ref{sec:historical}, with
Table~\ref{tab:historical_91T} providing the SN and host galaxy names, heliocentric redshifts, decline rate and color stretch\footnote{The color stretch 
parameter, \sbv, is a dimensionless parameter defined as the time difference between $B$-band 
maximum and the reddest point in the $(B-V)$ color curve divided by 30 days, where typical SNe~Ia have \sbv $\sim$1.  As demonstrated by \citet{burns14}, \sbv\ does a better job than \dm\ of sorting in luminosity
SNe~Ia with \dm~$>1.7$~mag.}
measurements, along with spectroscopic and 
photometric references.  A sampling of their optical spectra 
 covering phases from $-12$ to $+4$~days is illustrated in Figure~\ref{fig:historical_91T_spec}.
All spectra have been corrected for Galactic reddening using the \citet{schlafly11} recalibration of the \citet{schlegel98} infrared-based dust map and 
assuming $R_V = 3.1$.  However, no corrections for host galaxy reddening have been
applied, which likely accounts for the differences in the continuum slopes of the
spectra of objects such as SNe~1995bd and 2017dfb.

In the next two sections, we discuss the optical and NIR spectroscopic and photometric characteristics of 91T-like SNe based on observations
of the CSP-II and literature samples.

\section{Optical Spectra}
\label{sec:opt_spectra}

For a detailed description of the early evolution of the optical spectra of 91T-like SNe, the reader is referred to \S3.3 of Paper~I.  As illustrated in 
Figure~\ref{fig:historical_91T_spec}, at the earliest epochs, the only strong features visible are the two absorption lines at observed wavelengths of approximately
4230~\AA\ and 4910~\AA, which correspond to \ion{Fe}{3} $\lambda\lambda$4404,5129.  Much weaker absorption ascribed to blends of \ion{Ni}{3} lines and 
to the \ion{Si}{3} $\lambda\lambda\lambda$4553,4568,4575
multiplet 2 (hereafter referred to as \ion{Si}{3} $\lambda$4560)
has also been identified at these early epochs \citep[e.g., see][]{mazzali95}.  The spectra remain essentially unchanged until $-7$~days
when a weak \ion{Si}{2} $\lambda6355$ absorption feature becomes visible.  A few days before maximum, features corresponding to
\ion{Ca}{2}~H~\&~K and \ion{S}{2} $\lambda\lambda$5449,5622 can also be identified.  Note that the two \ion{Fe}{3} features remain strong throughout these phases.
However,  by $+14$~days, the optical spectra of 91T-like SNe 
have evolved to closely resemble those of normal SNe~Ia, making it more difficult to distinguish them
(see the bottom panel of Figure~1 of Paper~I).

\begin{figure*}[t]
\epsscale{1.}
\plottwo{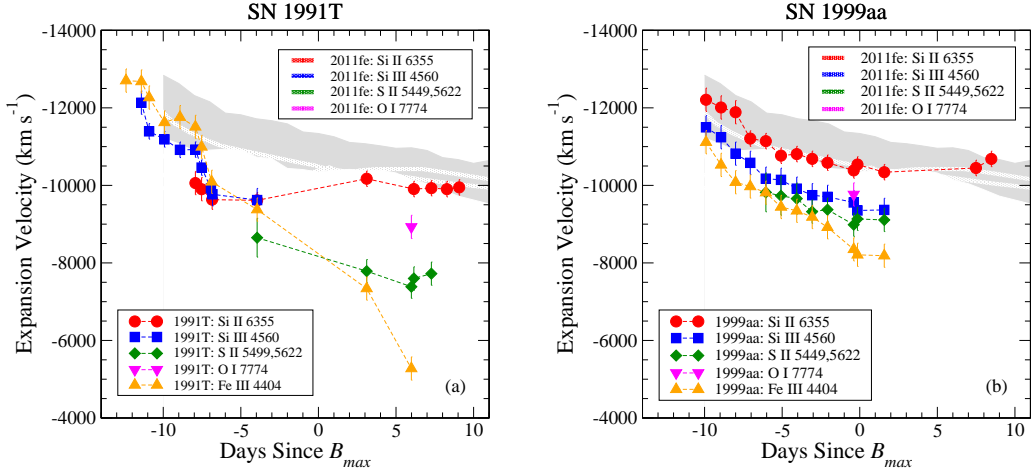}{Fig2b.eps}
\caption{(left) Comparison of velocity measurements of the \ion{Si}{2}, \ion{Si}{3}, \ion{S}{2}, \ion{Fe}{3}, and \ion{O}{1} ions for SNe~1991T and 2011fe. 
The measurements correspond to the absorption minima of each line.  The gray shaded band displays the RMS dispersion of CN SNe Ia
from \citet{folatelli13}. (right) Same plot comparing velocity measurements for SNe~1999aa and 2011fe.}
\label{fig:SiII_SiIII_S_II_FeIII_vels}
\end{figure*}

\begin{figure*}[t]
\epsscale{0.7}
\plotone{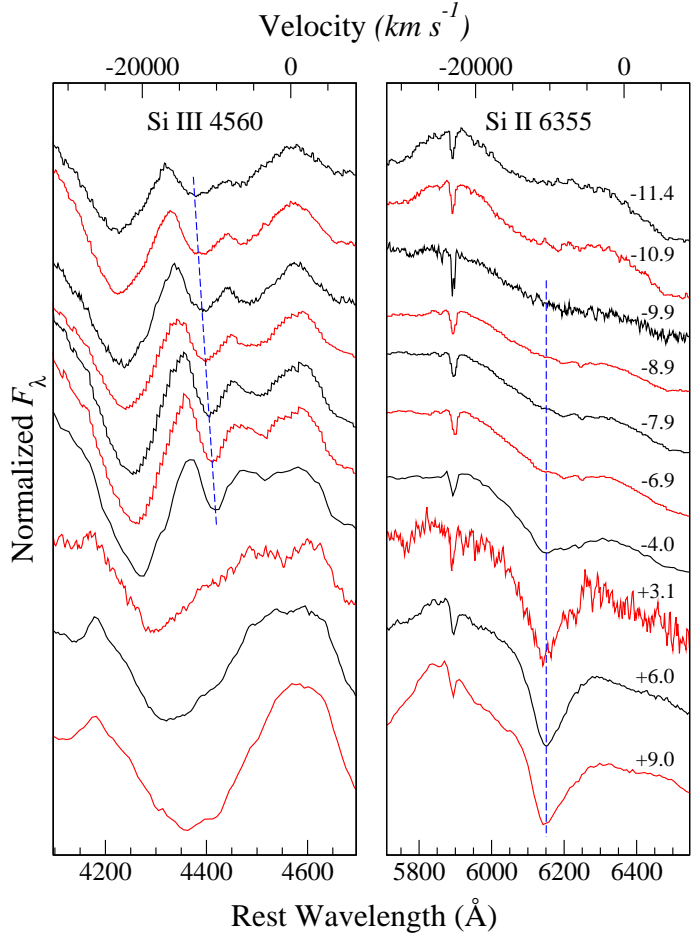}
\caption{Series of optical spectra of SN~1991T showing the time evolution of the \ion{Si}{3} $\lambda$4560 (left) and
\ion{Si}{2} $\lambda$6355 (right) absorption features.  The blue dashed lines illustrate the approximate velocity evolution of both
lines.}
\label{fig:SiII_SiIII_evolution}
\end{figure*}

The left panel of Figure~\ref{fig:SiII_SiIII_S_II_FeIII_vels} shows the evolution of the expansion velocities\footnote{Note that all expansion velocities quoted or plotted in this
paper were calculated using the relativistic Doppler formula \citep[see equation 6 
of][]{blondin06}.} of SN~1991T, measured for the minima of the absorption features of
\ion{Si}{2} $\lambda$6355, \ion{Si}{3} $\lambda$4560, \ion{S}{2} $\lambda\lambda$5449,5622, and \ion{Fe}{3} $\lambda$4404 compared with similar measurements for 
the Branch CN SN~2011fe.  Examining this plot, the first thing to note is that the evolution of the \ion{Si}{3} $\lambda$4560 velocities closely 
mimics that observed for SN~2011fe.  The \ion{Si}{2} $\lambda$6355 velocities, on the other hand, show a strikingly different behavior, with those of SN~2011fe falling monotonically with time, 
hugging the lower 1$\sigma$ edge of the expansion velocities for CN SNe, whereas those of SN~1991T remain essentially flat at a value of $\sim -10,000$~\kms,
which is approximately the same value that the \ion{Si}{2} velocities of SN~2011fe had leveled to by $+10$~days.
This peculiar behavior of SN~1991T is illustrated in Figure~\ref{fig:SiII_SiIII_evolution} where the time evolution of the spectra of SN~1991T in the wavelength
regions of the \ion{Si}{3} $\lambda$4560 and \ion{Si}{2} $\lambda$6355 lines is plotted.  Absorption due to \ion{Si}{3} $\lambda$4560
is clearly present in the $-11.4$~days spectrum, and increases in visibility while also decreasing in velocity through $-4.0$~days.  This feature may still be faintly
present at $+3$~days, but by $+6$~days cannot be clearly discerned.  In contrast, \ion{Si}{2} $\lambda$6355 absorption
does not become obvious until $-6.9$~days, although it may be faintly present as early as $-9.9$ days.  In contrast
to \ion{Si}{3} $\lambda$4560, the velocity of the minimum of the \ion{Si}{2} $\lambda$6355 absorption remains
nearly constant during the entire period that it can be identified in the spectrum.

\ion{S}{2} $\lambda\lambda$5449,5622 
absorption is  visible in the spectra of SN~1991T for a $\sim$10-day period centered on maximum light and at slowly decreasing velocities which closely
resemble the measurements for SN~2011fe.  Note, however, that the \ion{S}{2} lines in SN~2011fe were clearly visible for a much longer period, extending from the 
first spectrum obtained at $-15$~days until nearly $+10$~days.  This difference likely reflects the higher ionization of the outer ejecta of
SN~1991T, as evidenced by the strength of the \ion{Fe}{3} absorption compared to \ion{Fe}{2} at these same epochs.    Taken together, these
measurements demonstrate that the IMEs in both SNe were present at a similar range of expansion velocities in spite of the odd behavior of the \ion{Si}{2} $\lambda$6355 
velocities.

As also shown in Figure~\ref{fig:SiII_SiIII_S_II_FeIII_vels}, expansion velocities measured for the minimum of the
strong \ion{Fe}{3} $\lambda$4404 absorption in SN~1991T are observed to decrease rapidly from near $-12,500$~\kms\ at $-12$~days to $-5,000$~\kms\ at $+6$~days.
Although \ion{Fe}{3} is likely present in the earliest spectra of SN~2011fe, this feature is dominated by absorption due to \ion{Mg}{2} $\lambda$4481 \citep{parrent12}
and, thus, a comparison cannot be made.

The right panel of Figure~\ref{fig:SiII_SiIII_S_II_FeIII_vels} displays a plot of the expansion velocities of the same absorption features in the spectra of SN~1999aa.
In contrast to what was observed for SN~1991T, the pre-maximum expansion velocities of \ion{Si}{2} $\lambda$6355 steadily declined, following
the evolution of the measurements for SN~2011fe until approximately maximum light, and then plateaued at a value of $-10,500$~\kms.  The evolution of the 
\ion{Si}{3} $\lambda$4560 velocities is similar to that observed for SN~2011fe, although at somewhat higher values and perhaps approaching an asymptote of
$\sim -9,500$~\kms at maximum.  The \ion{S}{2} $\lambda\lambda$5449,5622 lines show a slow decline in velocity similar to that observed for SN~2011fe at the same
epoch, but displaced by $\sim500$~\kms\ to higher values.  In the case of \ion{Fe}{3} $\lambda$4404, a monotonic decline is observed for SN~1999aa, very similar to that found for
SN~1991T. Again, the conclusion is that the IMEs in SN~1999aa occupy a similar range of expansion velocities as they do in SN~1991T and SN~2011fe.

The similarity of the velocity evolution of the \ion{Si}{3} $\lambda$4560 line in all three SNe is particularly 
interesting, as this is a high excitation multiplet (19~eV versus 8~eV for \ion{Si}{2} $\lambda$6355)
that is strongly subject to non-LTE effects.
Particularly where ionization/excitation effects play a strong role, the behavior of this feature might be expected to vary significantly between SNe~Ia subclasses.
The observed similarity of the behavior of the \ion{Si}{3} $\lambda$4560 line 
is thus a strong argument that these three SNe are physically very closely related to each other.

\begin{figure}[t]
\epsscale{0.85}
\plotone{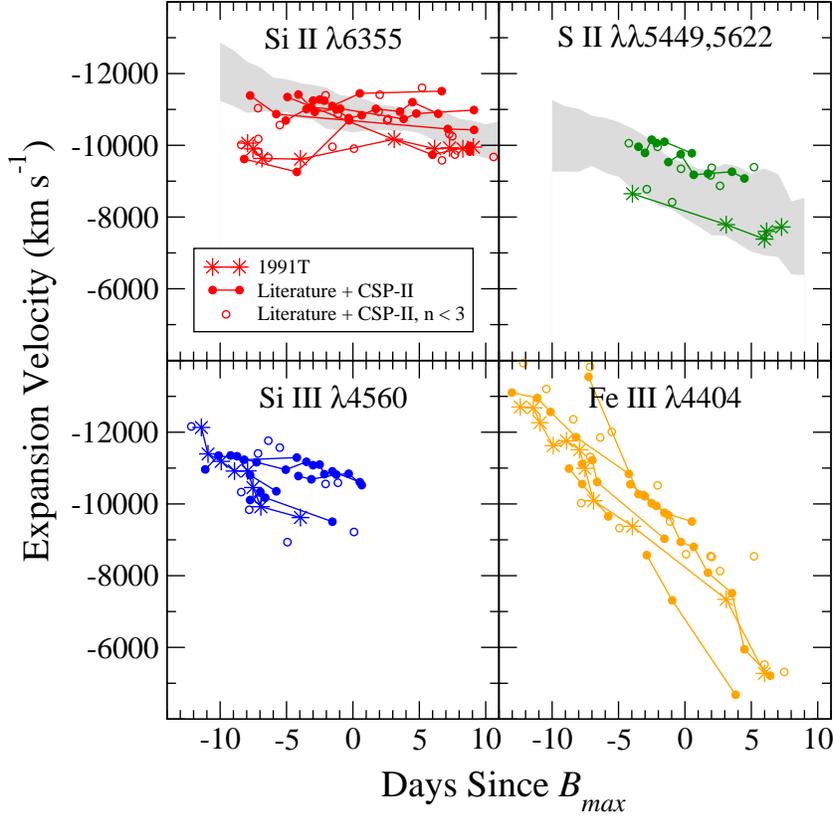}
\caption{Expansion velocity measurements of the \ion{Si}{2} $\lambda$6355, \ion{Si}{3} $\lambda$4560, \ion{S}{2} $\lambda\lambda$5449,5622, and \ion{Fe}{3} $\lambda$4404
absorption features for all of the 91T-like SNe in the literature and CSP-II samples.  
Where $\geq 3$ measurements for the same SN could be made, these are joined by solid lines.
Velocities for the same features in the spectra of SN~1991T are plotted for 
comparision.  Error bars are not shown in order to keep the plot uncluttered, but typically are $\pm500$~\kms\ or less.  The gray shaded bands display the RMS
dispersion of CN SNe Ia from \citet{folatelli13}.}
\label{fig:vels_91T-like}
\end{figure}

Unfortunately, none of the 91T-like SNe in the literature and CSP-II samples have as many spectral observations as SN~1991T.  However, we can still get an idea of how these
events compare by plotting expansion velocities for all of the available spectra in the same figure.  Such plots are shown for the \ion{Si}{2} $\lambda$6355, 
\ion{Si}{3} $\lambda$4560, \ion{S}{2} $\lambda\lambda$5449,5622, and \ion{Fe}{3} $\lambda$4404 features in Figure~\ref{fig:vels_91T-like}.  Within the distribution 
of the velocities of all four of these features, those for SN~1991T fall in the lower part of the envelope of measurements.   The behavior of \ion{Si}{2} $\lambda$6355 
is particularly interesting.  The measurements for nearly all of the SNe fall uniformly between $-9,000$ and $-12,000$~\kms, with no hint of a decline in the average 
velocity between $-10$ and $+10$~days.  91T-like SNe are thus the extreme members of the ``low-velocity gradient''
group of SNe~Ia as defined by \citet{benetti05}.  The apparent ubiquity of a \ion{Si}{2} ``velocity plateau'' in 91T-like SNe was pointed out previously by
\citet{scalzo12,scalzo14b} based on fewer objects, and has been interpreted as evidence of the IMEs being confined to a narrow region in velocity space 
\citep{scalzo14b,taubenberger17}, perhaps due to a density enhancement in the outer layers of the ejecta \citep{quimby07}.
Nevertheless, Figure~\ref{fig:vels_91T-like} shows that absorption due to \ion{Si}{3} $\lambda$4560 is observed to range from $-8,000$ to $-12,500$~\kms\ for the 91T-like SNe in our samples.
Likewise, the \ion{S}{2} $\lambda\lambda$5449,5622 absorption ranges between $-7,500$ and 
$-10,000$~\kms.   

\begin{figure}[t]
\epsscale{0.7}
\plotone{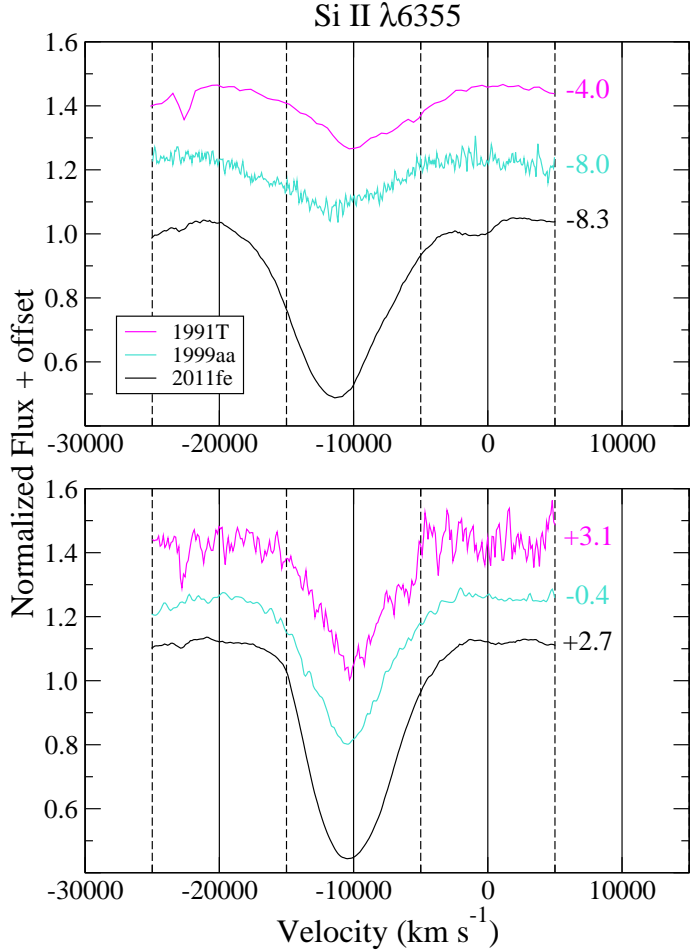}
\caption{Profiles of the \ion{Si}{2} $\lambda$6355 absorption in SN~1991T, SN~1999aa, and SN~2011fe at pre-maximum epochs (top), as well as near maximum (bottom). 
Pseudo-continuum points to the blue and red were used to define a continuum that has been subtracted from the spectra.}
\label{fig:SiII_profiles}
\end{figure}

It is instructive to compare the \ion{Si}{2} $\lambda$6355 absorption line profiles of SN~1991T, SN~1999aa, and SN~2011fe.  This is shown in Figure~\ref{fig:SiII_profiles} 
where the profiles are plotted at pre-maximum epochs and also within a few days of maximum.  The epochs for SN~1991T and SN~1999aa were chosen
so that the strength of the absorption was similar.  Early-on, the blue wings of all three SNe reach velocities of approximately $-20,000$~\kms, once again 
demonstrating that the IMEs cover a similar range in velocity space.  At maximum, the  profiles are somewhat sharper, with those of SN~1991T and 
SN~1999aa taking on a nearly triangular shape.

\begin{figure}[t]
\epsscale{.75}
\plotone{Fig6.eps}
\caption{Expansion velocity measurements of \ion{Si}{2} $\lambda$6355 and \ion{O}{1} $\lambda$7774 plotted versus each other for a sample of
SNe~Ia observed by the CSP-I.  Velocities for \ion{Si}{2} $\lambda$6355 correspond to maximum light and are taken from \citet{folatelli13} and
\cite{morrell24}.  The velocities for \ion{O}{1} $\lambda$7774 were measured from spectra taken within $\pm$4~days of maximum.  
Measurements for two 99aa-like (2013hh and ASASSN-15as) and three 91T-like SNe (LSQ12gdj, 2013U, and 2014eg) from the CSP-II sample are 
included in the plot, as are points for SN~1991T and SN~1999aa.  The dotted line is for comparison, and indicates a one-to-one trend.}
\label{fig:vels_SiII_vs_OI}
\end{figure}

\ion{O}{1} $\lambda$7774 was observed in a spectrum of SN~1991T obtained $\sim$1~week after maximum light \citep{filippenko92b}.
In DDT models of SNe~Ia, primordial C and O are found in a thin outer shell.  Oxygen is also a product of explosive C burning in SNe~Ia, and its velocity is expected 
to be constant once the photosphere recedes below the inner edge of the C-burning zone which, in both normal and sub-luminous events, has occurred by maximum 
light \citep[e.g., see Figure~1 of][]{hoeflich17}.  Indeed,
constant velocity in the \ion{O}{1} line is observed in the 
prototypical Branch CN SN~2011fe (see Figure~\ref{fig:SiII_SiIII_S_II_FeIII_vels}).  

As illustrated in Figure~\ref{fig:vels_SiII_vs_OI}, the expansion velocities of \ion{O}{1} $\lambda$7774 and \ion{Si}{2} $\lambda$6355 at maximum light are strongly correlated
for SNe~Ia in the CSP-I sample.  SN~1991T lies clearly in this trend, as do the 99aa-like and  91T-like SNe in the CSP-II sample for which we were able to 
measure \ion{O}{1} $\lambda$7774 velocities. 
 
\begin{figure}[ht]
\epsscale{.75}
\plotone{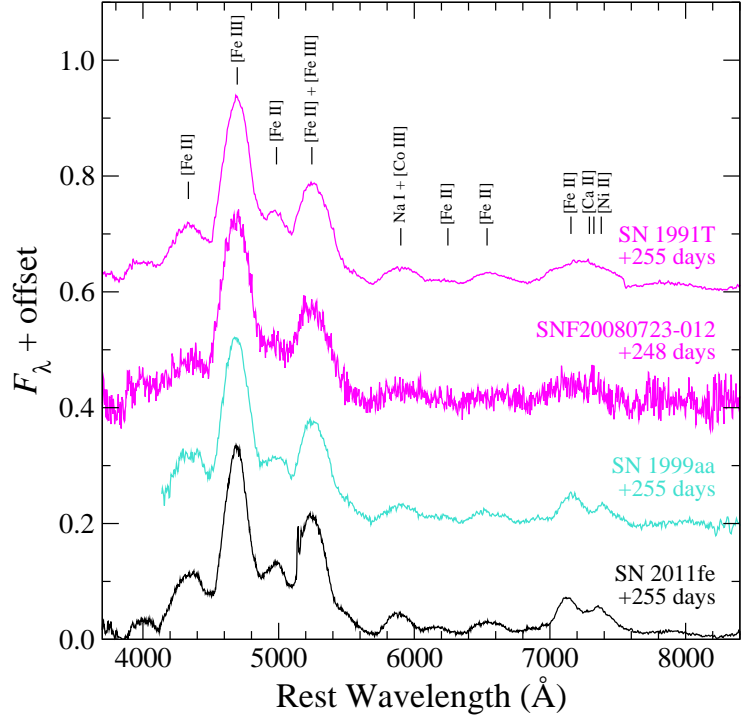}
\caption{Nebular spectra of SN~1991T, the 91T-like SNF20080723-012, SN~1999aa, and 
the Branch CN SN~2011fe.  Phases with respect to the time of maximum light in the 
$B$~band are indicated, as are identifications of the strongest emission features.  References for the spectra are: SN~1991T \citep{gomez98}, 
SNF20080723-012 \citep{taubenberger13}, SN~1999aa \citep{silverman12}, and SN~2011fe \citep{zhang16b}.}
\label{fig:nebular}
\end{figure}

Nebular-phase optical spectra of 91T-like SNe are extremely scarce --- none were 
obtained of the ten 91T-like SNe observed by the CSP-II, and spectra of only two
have been published in the literature: SN~1991T 
\citep{gomez98,mazzali98,silverman12} and SNF20080723-012 \citep{taubenberger13}.
Figure~\ref{fig:nebular} displays spectra obtained $\sim$250~days after
$t(B_{max})$ of SN~1991T and SNF20080723-012.  Shown for comparison are 
similar-epoch spectra of SN~1999aa and the Branch CN SN~2011fe. All four spectra are
remarkably similar, with the strongest emission features due to blends of 
forbidden lines of \ion{Fe}{2} and \ion{Fe}{3}.  \citet{mazzali98} found that the width 
of the [\ion{Fe}{3}]~4700 feature was inversely correlated with \dm, 
implying a relationship between the amount of $^{56}$Ni produced in
the explosion and the extent to which it has been mixed outward in the ejecta.
Later studies \citep{blondin12,silverman13a,graham22} have concluded that
this correlation is weak (or non-existent) for SNe~Ia with \dm~$ \lesssim 1.5$~mag,
although for the most-luminous events this may reflect the breakdown of \dm\ as
a luminosity indicator \citep{phillips22}.
For the four spectra shown in Figure~\ref{fig:nebular}, the two 91T-like SNe
display the broadest emission features compared to the less-luminous SN~1999aa
and SN~2011fe.  The only other clear difference between the spectra of the 91T-like
events and the other two SNe is the profile of the blend of the 
[\ion{Fe}{2}] and [\ion{Ni}{2}] features between 7000 and 7600~\AA.  For the
two 91T-like SNe, it appears that emission due to [\ion{Ca}{2}] $\lambda$7291,7324
is also contributing to the blend.  In contrast, for normal bright SNe such as SN~2011fe,
[\ion{Ca}{2}] emission does not appear strongly in the nebular spectrum
until $\sim+400$~days \citep[e.g., see][]{tucker22a,kumar23}.

In DDT models, Ca is produced at the interface between Si/S and the lower density 
portion of the ejecta where nuclear statistical  equilibrium (NSE) conditions
were reached \citep[see Figure 3 of][]{hoeflich02}. As emphasized by \citet{wilk20},
[\ion{Ca}{2}] emission is strongly coupled to the radiation field --- and, by 
extension, to the $^{56}$Ni. Thus, the earlier appearance of the [\ion{Ca}{2}]
lines in 91T-like SNe may reflect the greater mixing of $^{56}$Ni into
the IMEs.

\section{Near-Infrared Spectra}
\label{sec:nir_spectra}

\begin{figure*}[t]
\epsscale{1.}
\plottwo{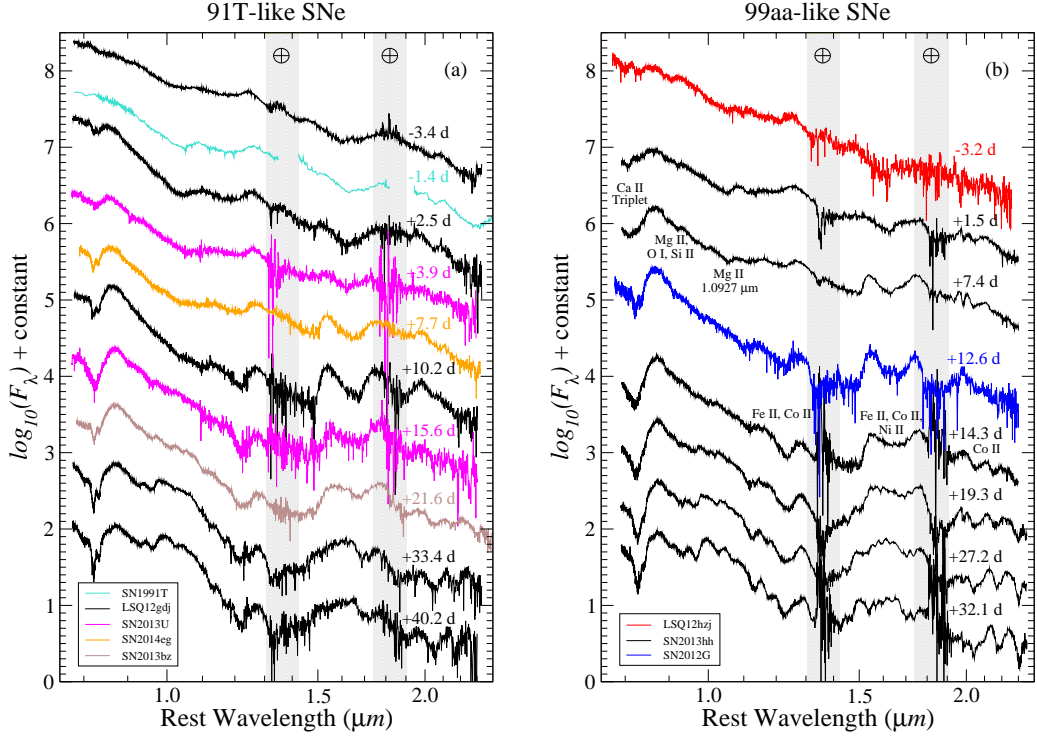}{Fig8b.eps}
\caption{(left) NIR spectra of four 91T-like SNe observed in the course of the CSP-II. Included for comparison is the spectrum of SN~1991T published by
\citet{meikle96}.  Phases with respect to $t(B_{max})$ are indicated for each spectrum.
The spectra have been corrected for telluric absorption; wavelength regions of the strongest telluric features are shaded in gray and indicated by the earth symbol.  
(right) Similar plot of the NIR spectra of three 99aa-like SNe observed by the CSP-II.  Some of the more notable absorption and emission features are labelled by their major contributors.}
\label{fig:91T_99aa_nir_spectra}
\end{figure*}

A total of nine NIR spectra of four 91T-like SNe were obtained during the course of the CSP-II\footnote{These spectra were published by \citet{lu23} and
 are available on the CSP website
(\url{https://csp.obs.carnegiescience.edu/data}).}.  Apart from a spectrum of SN~1991T at maximum light
published by \citet{meikle96}, as far as we are aware, these are the first NIR spectra to be obtained in the photospheric phases
of 91T-like events.  These data, which are plotted in the left panel of Figure~\ref{fig:91T_99aa_nir_spectra}, were acquired with the Folded-port IR Echellette 
(FIRE) on the 6.5~m Magellan Baade telescope at Las Campanas Observatory.  FIRE spectra of several 99aa-like SNe were also obtained during the 
CSP-II.  Spectra of three such objects covering a similar range of phases as the 91T-like SNe are plotted in the right panel of Figure~\ref{fig:91T_99aa_nir_spectra} for 
comparison.  Details of the acquisition and reduction of the FIRE spectra are given by  \citet{hsiao19}.  A log of these FIRE observations is listed in Table~\ref{tab:NIR_spectra}.  

\begin{deluxetable}{ccccc}
\tabletypesize{\scriptsize}
\tablecolumns{4}
\tablewidth{0pt}
\tablecaption{Log of FIRE NIR Spectroscopic Observations\label{tab:NIR_spectra}}
\tablehead{
 \colhead{UT Date} &
\colhead{MJD} &
\colhead{$t(B_{max})$\tablenotemark{a}} &
\colhead{Exposure\tablenotemark{b}}
}
\startdata
\sidehead{{\bf SN2012G}}
\hline
2012-02-04.28 & 55961.28 & $+12.1$ &  8.5 \\
\hline
\sidehead{{\bf LSQ12gdj}}
\hline
2012-11-19.14  &  56250.14  &  $-3.4$  &  20.0 \\
2012-11-25.17  &  56256.17  &  $+2.5$  &  31.7 \\
2012-12-03.10  &  56264.10  &  $+10.2$  &  37.0 \\
2013-12-27.06  &  56288.06  &  $+33.4$  &  44.9 \\
2013-01-03.06  &  56295.06  &  $+40.2$  &  47.6 \\
\hline
\sidehead{{\bf LSQ12hzj}}
\hline
2013-01-06.24  &  56298.24  &  $-3.2$   &  26.4 \\
\hline
\sidehead{{\bf SN2013U}}
\hline
2013-02-16.08  &  56339.08  &  $+3.9$   &  21.1 \\
2013-02-28.20  &  56351.20  &  $+15.6$  &  31.7 \\
\hline
\sidehead{{\bf SN2013bz}}
\hline
2013-05-19.19  &  56431.69  &  $+21.6$  &  18.5 \\
\hline
\sidehead{{\bf SN2013hh}}
\hline
2013-12-14.35  &  56640.35  &  $+2.0$  &  12.7 \\
2013-12-20.34  &  56646.34  &  $+7.9$  &  16.9 \\
2014-01-01.34  &  56658.34  &  $+19.8$  &  8.5 \\
2014-01-09.35  &  56666.35  &  $+27.9$  &  8.5 \\
2014-01-14.32  &  56671.32  &  $+27.9$  &  8.5 \\
\hline
\sidehead{{\bf SN2014eg}}
\hline
2014-12-08.05  &  56999.05  &  $+7.7$  &  8.5 \\
\enddata
\tablenotetext{a}{Phase in days with respect to the epoch of $B$ maximum, $t(B_{max})$, corrected for time dilation.}
\tablenotetext{b}{Total exposure time in minutes.}
\end{deluxetable}

\begin{figure}[ht]
\epsscale{0.75}
\plotone{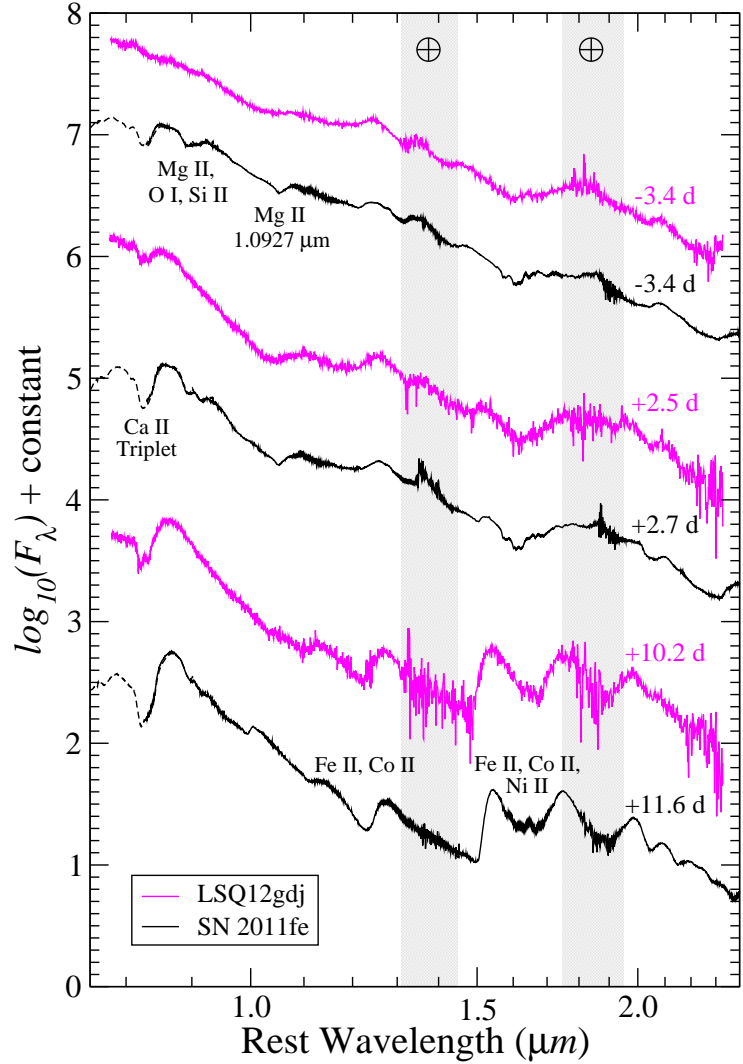}
\caption{Comparison of the NIR spectra of the 91T-like event LSQ12gdj and the Branch CN SN~2011fe at three different phases.  Identifications of the strongest
features are given.  The spectra of SN~2011fe have been combined with the optical spectrophotometry published by \citet{pereira13} in order to include the 
\ion{Ca}{2} $\lambda\lambda$8498,8542,8662 triplet feature.  These extensions are denoted by the dashed black lines.  Wavelength regions of the strongest telluric 
features are shaded in gray and indicated by the earth symbol.}
\label{fig:LSQ12gdj_11fe_nir_spectra}
\end{figure}

In general, the evolution of the 91T-like and 99aa-like SNe in the NIR is quite similar.  The principal exception is the \ion{Ca}{2} 
$\lambda\lambda$8498,8542,8662 triplet absorption which, as expected from the
optical spectra, is stronger in the 99aa-like SNe at comparable epochs.  However, by one month after maximum, the strength is nearly the same in both subtypes.
In Figure~\ref{fig:LSQ12gdj_11fe_nir_spectra}, the three earliest spectra of the 91T-like event LSQ12gdj are compared
with NIR spectra of the Branch CN SN~2011fe at similar phases \citep{hsiao13}.  Published spectrophotometry by \citet{pereira13} has been used to 
extend the NIR spectra of SN~2011fe to cover the \ion{Ca}{2} $\lambda\lambda$8498,8542,8662 triplet.  The spectra of LSQ12gdj are seen to closely resemble
those of SN~2011fe, with the major exception again being the weakness of the \ion{Ca}{2} triplet absorption in LSQ12gdj, especially in the first two 
epochs ($-3.4$ and $+2.5$~days).  At these same epochs, 
\ion{Mg}{2} $\lambda$1.0927~$\mu$m absorption is also clearly visible in the spectra of SN~2011fe, but is not readily identified in LSQ12gdj.

As observed at optical wavelengths, at phases beyond $+7$~days, the NIR spectra of 91T-like (and 99aa-like) SNe are not easily distinguished 
from the spectra of Branch CN SNe.  At these epochs, the NIR spectra become dominated by the \ion{Ca}{2} triplet and features due to Fe-group elements  
(see Figures~\ref{fig:91T_99aa_nir_spectra} and \ref{fig:LSQ12gdj_11fe_nir_spectra}). 
At a phase of $+10 \pm 5$~days, a prominent double-peaked emission feature is observed in the $H$-band 
window between the two telluric bands.  This is the so-called ``$H$-band break'' \citep{wheeler98,hoeflich02,hsiao13,ashall19} which is composed of a blend of \ion{Fe}{2},
\ion{Co}{2}, and \ion{Ni}{2} emission lines.  The iron and cobalt at these phases are the product of the radioactive decay of $^{56}$Ni produced in the explosion.

\begin{figure}[t]
\epsscale{.8}
\plotone{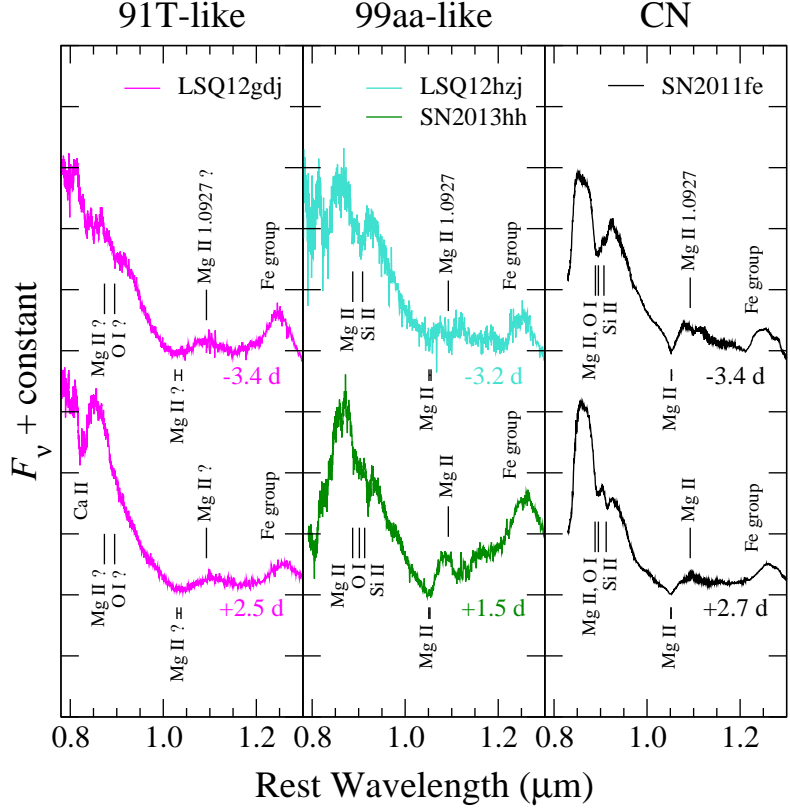}
\caption{Comparison of the 0.8--1.3~$\mu$m spectra of the 91T-like SN LSQ12gdj, the 99aa-like events LSQ12hzj and SN~2013hh, and the Branch CN SN~2011fe 
a few days before and after maximum light.  Identifications of the strongest features are given.  See text for further details.}
\label{fig:Y+J_nir_spectra}
\end{figure}

One of the few lines in the NIR for which it is possible to measure an expansion velocity is \ion{Mg}{2} $\lambda$1.0927~$\mu$m.
Figure~\ref{fig:Y+J_nir_spectra} shows a blow-up of the 0.8--1.3~$\mu$m spectral region for the 91T-like LSQ12gdj, the 99aa-like
events LSQ12hzj and SN~2013hh, and the Branch CN SN~2011fe.  Absorption due to \ion{Mg}{2} $\lambda$1.0927~$\mu$m was clearly visible in the 
earliest spectra obtained of SN~2011fe, and remained detectable until a few days after maximum \citep{hsiao13}.  This line is identified
in the two spectra plotted in the right-hand panel of Figure~\ref{fig:Y+J_nir_spectra}.  Also observed is an absorption
feature at $\sim$1~$\mu$m which is due to a blend of \ion{Mg}{2} $\lambda\lambda$0.9218,0.9244~$\mu$m, \ion{O}{1} $\lambda$0.9265~$\mu$m, 
and \ion{Si}{2} $\lambda$0.9413~$\mu$m.  These same features appear to be present in the spectra of the 99aa-like SNe LSQ12hzj and
SN~2013hh.  However, in the 91T-like LSQ12gdj,
the situation is not so clear.  While there are broad dips in the spectra of all three SNe near where \ion{Mg}{2} $\lambda$1.0927~$\mu$m might be 
expected, the evidence for detectable absorption near 1~$\mu$m is weak at best.  Likewise, there are no obvious absorption features in the spectra of 
LSQ12gdj at the expected positions of the \ion{Mg}{2} $\lambda\lambda$0.9218,0.9244~$\mu$m, \ion{O}{1} $\lambda$0.9265~$\mu$m, 
and \ion{Si}{2} $\lambda$0.9413~$\mu$m lines.
The absence of these \ion{Mg}{2}, \ion{Si}{2}, and \ion{O}{1} features in the pre-maximum- and maximum-light NIR spectra of LSQ12gdj
is consistent with the high level of ionization displayed by the optical spectra of 91T-like SNe at these same epochs.

\begin{figure}[t]
\epsscale{0.7}
\plotone{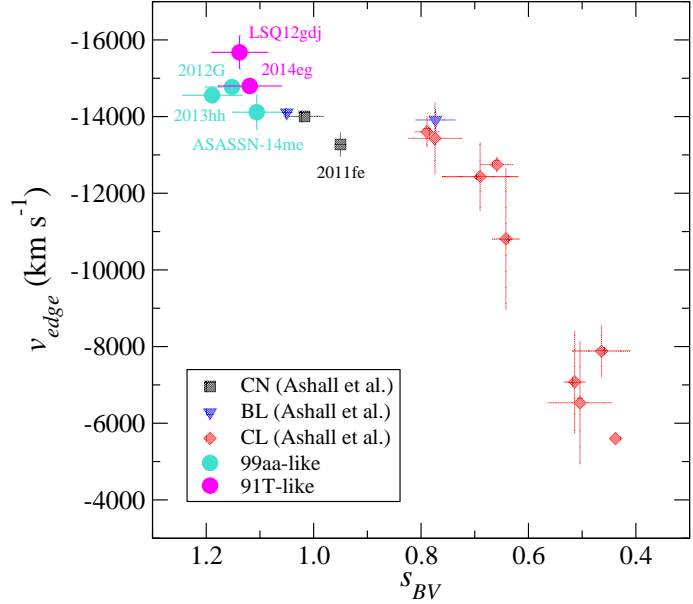}
\caption{Velocity measurements of the edge of the $H$-band break plotted versus the color stretch parameter, \sbv.  The values for the Branch CN, BL (``broad-line''), and CL (``cool'')
SNe are reproduced from Figure~4 of \citet{ashall19}.  New measurements for three 99aa-like events (SN~2012G, SN~2013hh, and ASASSN-14me) and two 91T-like SNe
(LSQ12gdj and SN~2014eg) from the CSP-II have been added.}
\label{fig:H_break}
\end{figure}

\begin{figure}[b]
\epsscale{0.7}
\plotone{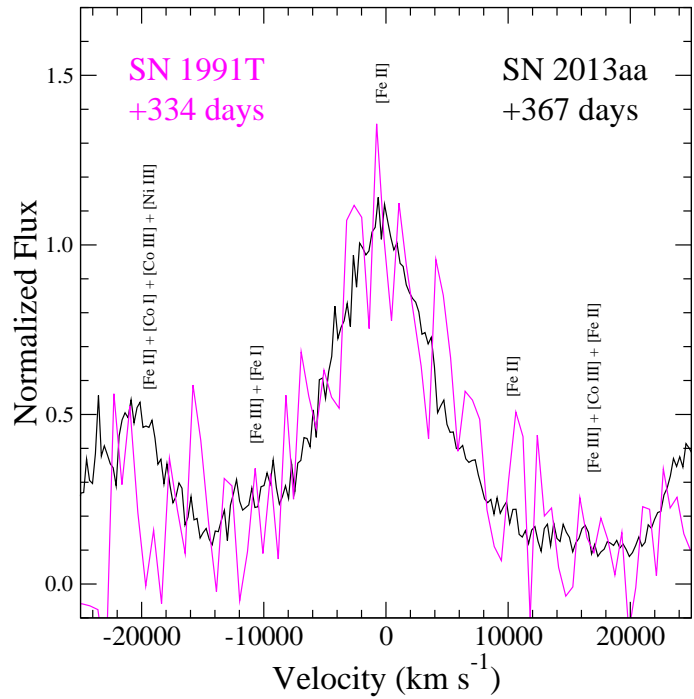}
\caption{Spectra covering the wavelength region of the [\ion{Fe}{2}] 1.644~$\mu$m
lines of SN~1991T and the slow-declining Branch CN SN~2013aa, both obtained
approximately one year past $B$~maximum. References for the spectra are
\citet{spyromilio92} for SN~1991T and \citet{kumar23} for SN~2013aa.}
\label{fig:nebular_IR}
\end{figure}

\citet{ashall19} have demonstrated that the maximum expansion 
velocity of the $^{56}$Ni distribution in the SN ejecta may be directly determined by measuring the wavelength of the blue edge of the $H$-band break, $v_{edge}$.
Thus the $H$-band break provides a measure of the extent in velocity space of the $^{56}$Ni in the SN ejecta.
\citet{ashall19} found a strong correlation between the maximum velocity of the $^{56}$Ni and the color stretch parameter, \sbv.  
This result is consistent with the observation that the 
Fe cores of luminous SNe~Ia extend to higher velocities than those of low-luminosity events  \citep{mazzali98,blondin12,silverman13a,graham22}.
Figure~\ref{fig:H_break} shows the expansion velocity of $^{56}$Ni as measured from the $H$-band break, $v_{edge}$, plotted versus \sbv\ for the \citet{ashall19}
sample.  Added to the plot are the $v_{edge}$ measurements derived from our FIRE spectra of three 99aa-like events (SN~2012G, SN~2013hh, 
and ASASSN-14me) and two 91T-like SNe (LSQ12gdj and SN~2014eg).
In keeping with their extreme character and high-ionization spectra at early epochs, the $^{56}$Ni in 99aa- and 91T-like SNe 
is observed to extend to maximum velocities ranging from $-$14,000 to nearly $-$16,000~\kms.  Again, for comparison, the maximum velocity reached by
the $^{56}$Ni in the Branch CN SN~2011fe was $\sim -13,300$~\kms.

To our knowledge, the only NIR nebular spectrum of a 91T-like SN to be published
is of SN~1991T itself, obtained by \citet{spyromilio92} at +338~days with respect 
to $B$~maximum.  Based on this spectrum, these authors
argued that the explosion that produced SN~1991T synthesized a large $^{56}$Ni mass,
which was the same conclusion reached by \citet{bowers97} in an independent analysis.
Figure~\ref{fig:nebular_IR} displays a portion
of this spectrum centered on the [\ion{Fe}{2}] 1.644~$\mu$m emission line.
Unlike the blended nebular emission features observed at optical wavelengths,
this line is relatively isolated (although, as Figure~\ref{fig:nebular_IR}
shows, a few weak lines are likely present in the wings), and thus provides
information on the distribution of Fe in the inner ejecta of the SN.  Although 
the spectrum of SN~1991T is
noisy, it is seen that the wings of the line extend to $\sim$15,000~\kms at
this epoch.  Over-plotted for comparison in the same Figure is the 
[\ion{Fe}{2}] 1.644~$\mu$m line observed for the slow-declining Branch CN SN~2013aa
\citep{kumar23}.  The profiles are generally similar to within the signal-to-noise
ratio of the 1991T spectrum.  In principle, the profile of the [\ion{Fe}{2}] 
1.644~$\mu$m line can be used along with models to discern information on the 
initial white dwarf central density and magnetic field \citep{penney14,diamond15,diamond18,hristov21},
but higher-quality NIR nebular spectra of 91T-like SNe will be needed for such
a study.

\section{Optical and Ultraviolet Light Curves}
\label{sec:lcurves}

\subsection{Optical Light Curve Morphology} 

The 91T-like SNe identified in the literature and CSP-II samples are uniformly characterized by slow-declining (high-color stretch) light curves.
The weighted mean decline rate for the confirmed members of the literature sample is \dm~$=0.99$~mag with an rms
dispersion of 0.17~mag, while for the CSP-II sample it is 0.85~mag with a similar dispersion of 0.18~mag.  The weighted average of the  
color stretch values for the literature sample is \sbv~$=1.17$ with an rms dispersion of 0.08, while for the CSP-II sample the numbers are
nearly identical (\sbv~$=1.18$ with a dispersion of 0.05).

Normalized light curves in the $BVri$ bands for 11 of the literature sample of 91T-like SNe are displayed in Figure~\ref{fig:BVRI_91T}a.\footnote{The
photometry for most of the literature SNe was actually obtained in the Johnson $BV$ and Cousins $RI$ systems, but has been K-corrected to the 
CSP-II $BVri$ bandpasses.}  The $B$ and $V$ light curves of the SNe show a remarkably small amount of dispersion.
In contrast, the $r$- and $i$-band light curves plotted in the lower panels
of this figure show a significant diversity in the morphology of the secondary maximum.  The divergence is greatest at the epoch of the minimum between the
primary and secondary maxima of the $i$ band, which occurs at $\sim+15$~days.  In general, the strength and timing of the secondary
maximum is a strong function of decline rate \citep{hamuy96b,riess96,krisciunas01,burns14,dhawan15}, although large differences in the strength
for SNe with similar decline rates are occasionally observed \citep{krisciunas01}.

\begin{figure*}[t]
\gridline{\fig{Fig13a.eps}{0.8\textwidth}{(a)}}
\gridline{\fig{Fig13b.eps}{0.8\textwidth}{(b)}}
\caption{(a) $BVri$ light curves of 12 of the literature sample of 91T-like SNe.  K-corrections have been applied using the \citet{hsiao07} spectral template.
The light curves in each filter were then shifted vertically to the same magnitude at maximum light after correction for time dilation.  Decline rates, \dm, for each 
SN are given in parentheses. (b) Same as (a), but for $BVri$ light curves of the ten 91T-like SNe observed by the CSP-II.}
\label{fig:BVRI_91T}
\end{figure*}

\begin{figure*}[t]
\gridline{\fig{Fig14a.eps}{0.8\textwidth}{(a)}}
\gridline{\fig{Fig14b.eps}{0.8\textwidth}{(b)}}
\caption{(a) $BVri$ light curves of 12 99aa-like SNe observed by the CSP-I.  K-corrections have been applied using the \citet{hsiao07} spectral template.
The light curves in each filter were then shifted vertically to the same magnitude at maximum light after correction for time dilation.  Decline rates, \dm, for each 
SN are given in parentheses. (b) Same as (a), but for $BVri$ light curves of 
11 slow-declining Branch CN SNe observed by the CSP-II.}
\label{fig:BVRI}
\end{figure*}

Figure~\ref{fig:BVRI_91T}b displays the normalized $BVri$ light curves for the CSP-II sample of ten 91T-like SNe.  The exquisite precision of the CSP-II photometry
reveals just how similar the $B$ and $V$  light curves are over the first month following maximum light.  Again, however, the $i$-band light curves show 
considerable diversity in the depth of the minimum between the primary and secondary maxima.  This diversity is also reflected in the $r$ light curves, 
although to a lesser extent.  

Figures~\ref{fig:BVRI}a and b show equivalent plots of the normalized $BVri$ light curves of 99aa-like and Branch CN SNe observed by the CSP-II.
The Branch CN SNe were selected to have decline-rate values similar to those of the 91T-like events (\dm~$< 1.1$~mag), and will be referred to henceforth
in this paper as ``slow-declining'' CN SNe. 
The weighted average decline rates of these two samples --- \dm~$=0.93$~mag with a dispersion of 0.11~mag for the 99aa-like events and \dm~$=0.94$~mag with a dispersion
of 0.09~mag for the slow-declining CN SNe --- are insignificantly different from the mean value
for the 91T-like SNe, consistent with the conclusion of \citet{phillips22} that decline rate alone cannot be used to distinguish between the most luminous SNe Ia. 
The $B$ and $V$ light curves are quite uniform, whereas the $i$-band light curves, like those of the 91T-like SNe, display a 
dispersion in the depth of the minimum between the primary and secondary maxima.  However, neither sample includes SNe with as small
depths as 91T-like events such as SN~1997br and OGLE-2014-SN-107.

\begin{figure}[t]
\epsscale{1.1}
\plottwo{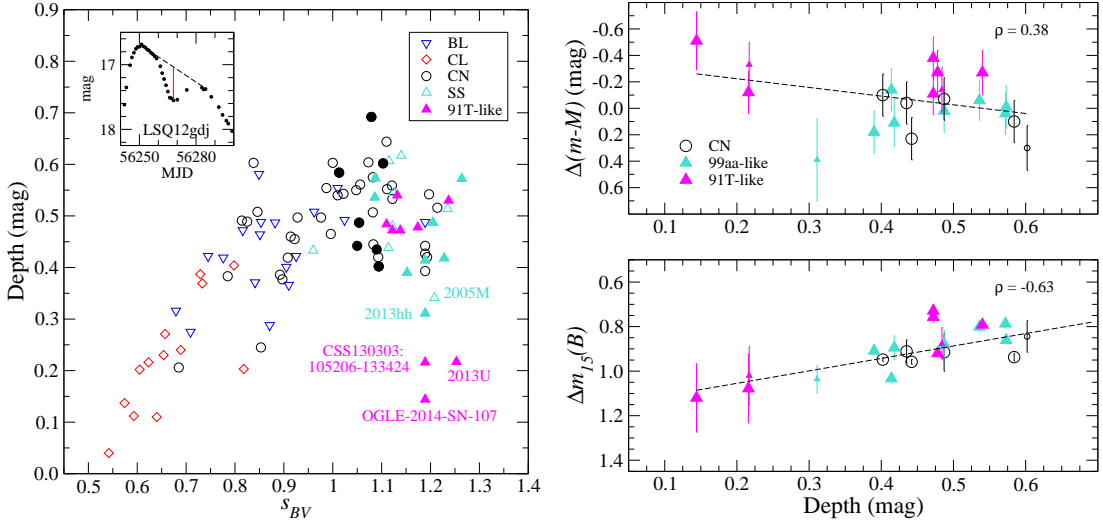}{Fig15b.eps}
\caption{(left) Relationship between the depth of the minimum between the primary and secondary maxima of the $i$-band light curves and the
color stretch parameter, \sbv, for a combined sample of SNe~Ia from the CSP-I and CSP-II with Branch classifications.  Solid symbols 
correspond to objects whose light curves are plotted in Figure~\ref{fig:BVRI_91T}b and Figures~\ref{fig:BVRI}a and b.  The inset plot illustrates an example of how the depth (the red line) is
measured for each SN. (right) $B$-band Hubble residuals in magnitudes (upper panel) and the decline rate parameter, \dm, (lower panel) plotted versus the depth 
of the minimum between the primary and secondary maxima of the $i$-band light curves for 91T-like, 99aa-like, and slow-declining CN SNe 
observed by the CSP-II.  The larger symbols in the lower diagram correspond to SNe whose values of \dm\ were measured directly; the smaller 
symbols correspond to those where \texttt{SNooPy} template fits were used to measure the decline rate. Linear regression fits are shown in both
panels as dashed lines.  Values of the Pearson correlation coefficient, $\rho$, are also indicated.}
\label{fig:depths}
\end{figure}

We attempted to quantify the depth of the minimum between the primary and secondary maxima of the  $i$-band light curves
by fitting a line to both maxima and measuring the amount in magnitudes from this line to the deepest part of the minimum.  This procedure is illustrated for LSQ12gdj 
in the inset to left half of Figure~\ref{fig:depths}. The resulting depth measurements for those SNe~Ia in the CSP-I and CSP-II samples with sufficient sampling of both maxima
are plotted versus color stretch in the same figure.  Although there is significant dispersion in the measurements, a clear trend is observed with the
depths increasing from the Branch CL SNe, through the BL events, and into the CN SNe.  For \sbv~$\ga1.0$ (\dm~$\la 1.1$~mag), the depths of most of the
$i$-band light curves are fairly constant, except for a few outliers with particularly small depth measurements.  The three largest outliers ---
OGLE-2014-SN-107, SN~2013U, and CSS130303:105206-133424 --- are all 91T-like SNe.
Unfortunately, the depth of $i$-band minimum for SN~1991T itself is difficult to measure because of the poor phase coverage of the secondary maximum of the \citet{lira98} photometry, but we estimate that it was between 0.2-0.3~mag (see Figure~\ref{fig:BVRI_91T}a). 

In the upper right half of Figure~\ref{fig:depths}, Hubble diagram residuals are plotted versus the depth of the minimum between the primary 
and secondary maxima of the $i$-band light curves.  The data again correspond to the CSP-II samples of 91T-like, 99aa-like, and slow-declining CN SNe.
A weak correlation is observed in the sense that slow-declining SNe~Ia with smaller depth measurements tend to be more luminous than events 
with deeper minima.  In the lower right panel, \dm\ for the same SNe is plotted
versus the depth of the minimum between the primary and secondary maxima of the $i$-band light curves.  A 
moderately strong negative correlation is also observed between these two parameters.  It should be noted that both correlations are
driven in large part by the three 91T-like SNe with the smallest depth measurements.  However, in the plot of \dm\ versus depth, similar trends are 
observed separately in all  three subclasses of SNe.  

The secondary maximum is a consequence of the recombination of 
\ion{Fe}{3} to \ion{Fe}{2}
\citep{kasen06,jack15}, and its strength and timing can be affected by several factors including the mass and mixing of $^{56}$Ni, the abundance of 
stable iron-group elements synthesized in the core, the progenitor metallicity, and in particular for the $i$-band, the abundance of calcium in the outer
layers of the ejecta \citep{kasen06}.
The large luminosities of 91T-like SNe \citep{boone21,phillips22,yang22} and
the $H$-band break measurements presented in \S\ref{sec:nir_spectra} provide
direct evidence that these objects are characterized by large $^{56}$Ni masses, with
significant mixing to high velocities in the ejecta.
According to the models presented in \citet{kasen06}, the effect of mixing $^{56}$Ni outward 
into the region of the IMEs is to advance the secondary maximum by a few days and to decrease its contrast with respect to the primary maximum.  
This is exactly what is observed for the most-luminous 91T SNe 
observed by the CSP-II, OGLE-2014-SN-107 and SN~2013U (see 
Figure~\ref{fig:BVRI_91T}).  Less clear is the
physics behind the correlation between \dm\ and the $i$-band depth measurements.

\subsection{Optical Color Evolution}
\label{sec:opt_colors}

\begin{figure*}[t]
\epsscale{1.15}
\plotone{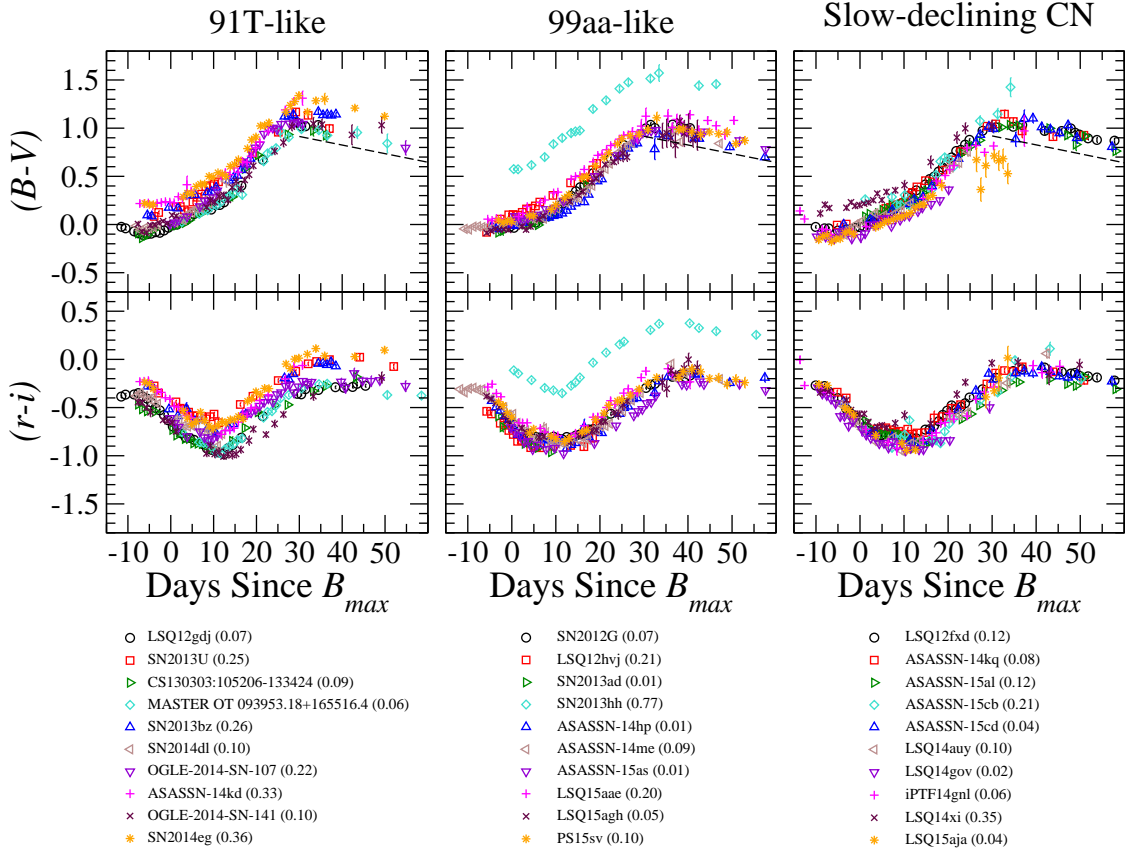} 
\caption{Evolution of the apparent $(B-V)$ and $(r-i)$ colors of 91T-like (left), 99aa-like (middle), and slow-declining CN  SNe (right)
observed by the CSP-II.  These are the same objects
whose $BVri$ light curves are plotted in Figure~\ref{fig:BVRI_91T}b and Figures~\ref{fig:BVRI}a and b.  The color measurements have been K-corrected and also
corrected for Milky Way reddening.  The numbers in parentheses after the names of the SNe in the 
legends are the values of $E(B-V)_{host}$ estimated using \texttt{SNooPy}.  The dashed lines in the $(B-V)$ plots are the Lira relations for 
unreddened SNe as given in equation 6 of \citet{burns14}.}
\label{fig:91T_SS_CN_colors}
\end{figure*}

Figure~\ref{fig:91T_SS_CN_colors} displays the observed $(B-V)$ and $(r-i)$ color evolution of the 91T-like, 99aa-like, 
and the slow-declining CN SNe
in the CSP-II sample whose light curves are plotted in Figure~\ref{fig:BVRI_91T}b and Figures~\ref{fig:BVRI}a and b.  The numbers in the legends in parentheses after the name of 
each SN are the $E(B-V)_{host}$ values derived from \texttt{SNooPy} fits to the combined $uBgVriYJH$ photometry.  Considering
the strong similarity and small dispersion in the shapes of the $B$ and $V$ light curves displayed in Figure~\ref{fig:BVRI_91T}b and Figure~\ref{fig:BVRI}a and b, it is
not surprising that the color evolution in $(B-V)$ of these three groups is also very similar.  The Lira relations \citep{burns14} included in 
Figure~\ref{fig:91T_SS_CN_colors} indicate that the majority of the SNe in each group have suffered host galaxy reddenings of 
$E(B-V)_{host} \la 0.25$~mag, consistent with $E(B-V)_{host}$ values derived from \texttt{SNooPy} fits to the full photometry as well as
equivalent width measurements of the host \ion{Na}{1}~D absorption (see \S\ref{sec:naid})\footnote{While it is not a
given that the 91T-like SNe should follow the \citet{burns14} Lira relation which was constructed from the CSP-I sample that did not include any 91T-like events, the close similarity of the spectra of 91T-like SNe to 99aa-like and slow-declining CN SNe at epochs greater than +30~days (e.g., see Paper~I) suggests that it may apply to all slow-declining SNe~Ia.}.  The
color evolution in $(r-i)$ of the 91T-like SNe is also quite similar to that of the 99aa-like and slow-declining CN SNe except for the larger dispersion observed at phases between 
10--25 days past maximum in the colors of the 91T-like events.  The latter reflects the dispersion in the depth of the minimum between the
primary and secondary maxima of the $i$-filter light curves observed for the 91T-like SNe.

\begin{figure}[t]
\epsscale{1.}
\plotone{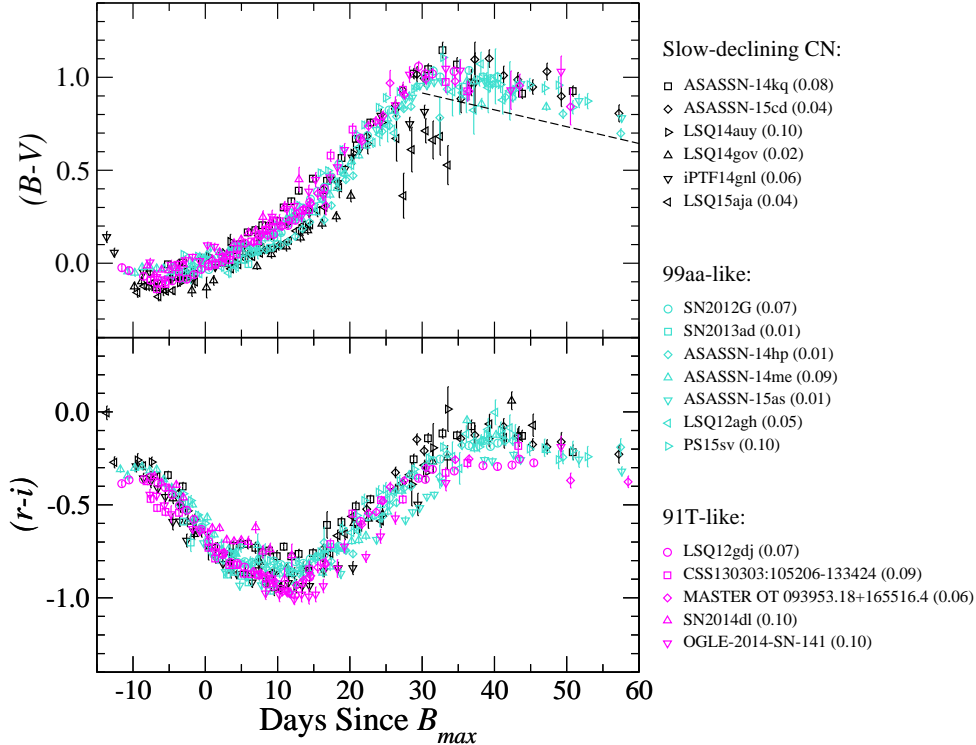} 
\caption{Color evolution in $(B-V)$ and $(r-i)$ of the 91T-like (magenta symbols), 99aa-like (cyan symbols), and slow-declining CN SNe 
(black symbols) with low host reddenings $E(B-V) \leq 0.10$~mag  
observed by the CSP-II.  The dashed line in the $(B-V)$ plot is the Lira relation for 
unreddened SNe as given in equation 6 of \citet{burns14}.}
\label{fig:colors_low_reddening}
\end{figure}

In Figure~\ref{fig:colors_low_reddening}, the $(B-V)$ and $(r-i)$ color evolution of the 91T-like, 99aa-like, and slow-declining CN SNe with
\texttt{SNooPy}-inferred host galaxy reddenings of $E(B-V)_{host} \leq 0.10$~mag is compared.  The agreement is generally  good between the three 
groups.  Some of the observed dispersion is likely due to host reddening, but there is almost certainly an intrinsic component as well,
especially in the $(r-i)$ colors.  The largest dispersion is observed for the slow-declining CN SNe, with the $(B-V)$ colors of LSQ14gov and LSQ15aja at 
phase $\geq 15$~days appearing noticeably bluer than the other slow-declining CN events.

We conclude that the optical colors of the 91T-like SNe are generally similar to those of 99aa-like and slow-declining Branch CN SNe.  
This is especially true within a week of maximum light.  The largest differences in colors are observed in $(r-i)$ at 10--25 days past maximum,
and are ascribed to the larger dispersion in the depth of the minimum between the primary and secondary maxima of the $i$-filter light curves, 
which is peculiar to the 91T-like SNe.  It should be noted, however, that \citet{stritzinger18} published evidence for a strong dichotomy in the ``slopes'' of the
$(B-V)$ colors of Branch SS (``shallow-silicon'') and CN SNe during the first 5~days following explosion.  These authors found that, with
the exception of SN~2012fr (discussed in \S\ref{sec:uv_colors}),
the Branch SS events (which are the combination of 99aa- and 91T-like SNe) were observed to have blue, nearly constant colors over this period, whereas the Branch CN SNe exhibited red colors that evolved rapidly.
Unfortunately, this difference is apparent only in the first few days following explosion, and none of the 91T-like or 99aa-like SNe plotted in Figures~\ref{fig:91T_SS_CN_colors} or \ref{fig:colors_low_reddening} were caught that early.  On the basis of observations of SNe~Ia discovered
 within five days of explosion, \citet{bulla20} have disputed the existence of a dichotomy of slopes in the early evolution 
of the $(g-r)$ colors of Branch SS and CN events, but rather argue that there is a continuous distribution of these.  However, the \citeauthor{bulla20} measurements
have large associated errors, and the range of slopes observed in the evolution of the $(g-r)$ colors over the first five days is approximately half that observed in $(B-V)$.
Clearly, more high precision photometry of SNe caught very soon after explosion will be needed to resolve this point.

\subsection{Ultraviolet Light Curve Morphology}
\label{sec:uv_morph}

\begin{figure}[t]
\epsscale{1.1}
\plotone{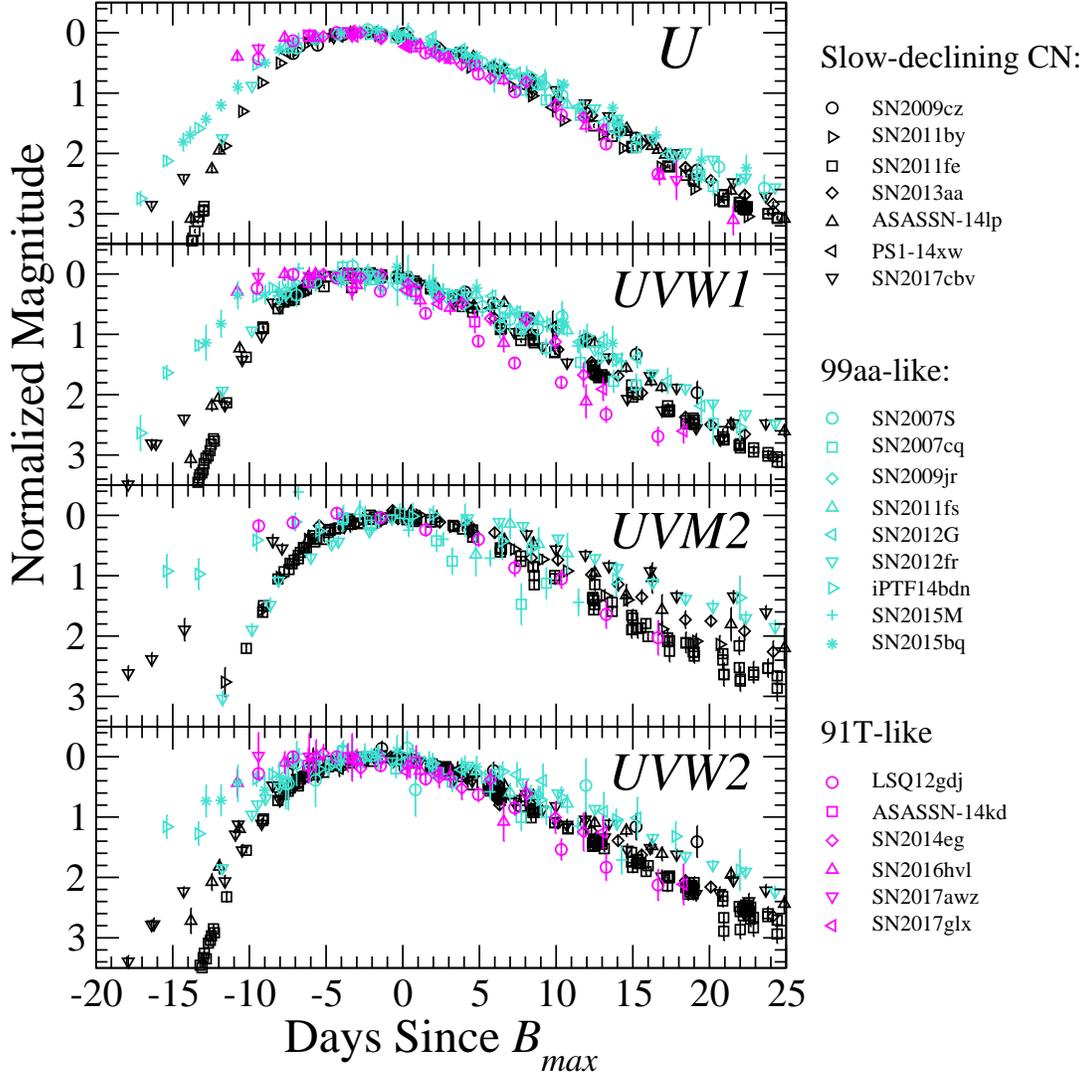} 
\caption{Swift $U$, $UVW1$, $UVM2$, and $UVW2$ light curves of 91T-like (magenta), 99aa-like (cyan), and slow-declining Branch CN (black) SNe.  The observations are 
corrected for time dilation and normalized to the same magnitude at maximum light in each filter.  The time axis is the time (in days) with respect to the phase of $B_{max}$.}
\label{fig:uv_lcurves}
\end{figure}

Since 2005, the {\it Neil Gehrels Swift Observatory} \citep{gehrels04} has provided a unique opportunity to study 
the UV properties of SNe~Ia \citep[see the review by][]{brown15}.
Over the years, Swift Ultra-Violet/Optical Telescope
\citep[UVOT;][]{roming05}
photometry has been obtained for a number of 91T-like, 99aa-like, and slow-declining Branch CN SNe.
In Figure~\ref{fig:uv_lcurves}, the normalized Swift/UVOT light curves in the $U$, $UVW1$, $UVM2$, and $UVW2$ filters are plotted for some of the best-observed 
of these events, which are also listed in 
Table~\ref{tab:host_reddenings}.  Clearly the strong similarity of the light curves of the 91T-like, 99aa-like, and slow-declining Branch CN SNe observed
at optical wavelengths does not extend to the UV.  In particular, the rising light curves to maximum of the 91T-like and 99aa-like events are significantly
broader in all three filters than those of the slow-declining Branch CN SNe.  Moreover, it appears to be the case that the 91T-like SNe reach maximum at UV wavelengths several days earlier
than the 99aa-like and slow-declining Branch CN events.  At post-maximum epochs, the light curves show less dispersion.  The broader pre-maximum UV light curves of
the 99aa-like and 91T-like SNe were previously noted and described by \citet{jiang18} who referred to these objects as luminous early-excess SNe, or ``EExSNe Ia''.
These authors concluded that {\em all} 99aa-like and 91T-like SNe showed such enhanced UV emission when discovered sufficiently early, which they ascribed to a 
``$^{56}$Ni-abundant outer layer''.

\begin{deluxetable}{lllc}
\tabletypesize{\scriptsize}
\tablecolumns{4}
\tablewidth{0pt}
\tablecaption{Host Galaxy Reddenings of SNe with UV Photometry\label{tab:host_reddenings}}
\tablehead{
\colhead{SN Name} &
\colhead{$E(B-V)$} &
\colhead{$R_V$} &
\colhead{Reference\tablenotemark{a}}
}
\startdata
\sidehead {{\bf 91T-like}}
\hline
LSQ12gdj  &  0.03 (0.01) & 3.46 (0.95) & 1 \\
SN2016hvl  &  0.25 (0.02) & 2.99 (0.61) & 2 \\
SN2017awz  &  0.09 (0.02)\tablenotemark{b} & \nodata & 3 \\
SN2017glx  & 0.06 (0.01) & 2.50 (0.27) & 2 \\
ASASSN-14kd  &  0.29 (0.02) & 3.15 (0.38) & 1 \\
SN2014eg  &  0.32 (0.01) & 2.20 (0.28) & 1 \\
\hline
\sidehead {{\bf 99aa-like}}
\hline
SN2007S &  0.47 (0.02) & 2.40 (0.17) & 1 \\
SN2007cq  & 0.12 (0.01) & 2.51 (0.22) & 4 \\
SN2009jr  &  0.61 (0.02) & 1.66 (0.34) & 5 \\
SN2011fs  &  0.08 (0.01) & 3.01 (0.39) & 2 \\
SN2012G  &  0.04 (0.01) & 3.70 (1.02) & 1 \\
SN2012fr  &  0.07 (0.01) & 2.33 (0.67) & 1 \\
iPTF14bdn  & -0.03 (0.03) & 3.00 (0.37) & 6 \\
SN2015M  &  0.09 (0.02) & 2.92 (0.80) & 1 \\
\hline
\sidehead {{\bf Slow-declining Branch CN}}
\hline
SN2009cz  &  0.11 (0.01) & 3.34 (0.72) & 1 \\
ASASSN-14lp  &  0.35 (0.01) & 2.22 (0.20) & 1 \\
PS1-14xw  &  0.12 (0.02) & 3.31 (0.75) & 1 \\
SN2011fe  & 0.06 (0.01) & 2.99 (0.16) & 7 \\
SN2013aa  &  0.04 (0.02) & 3.62 (0.99) & 1 \\
\enddata
\tablecomments{Unless otherwise specified, the $E(B-V)$ and $R_V$ values were derived using the ``color\_model'' model in SNooPy. 
Errors are given in parentheses.}
\tablenotetext{a}{References to photometric data used to derive host extinction parameters with SNooPy.}
\tablenotetext{b}{$E(B-V)$ derived using the ``EBV\_model2'' model in SNooPy''.}
\tablerefs{
(1) CSP-II;
(2) \citet{stahl19};
(3) \citet{foley18};
(4) \citet{ganeshalingam10};
(5) \citet{hicken12};
(6) \citet{brown14};
(7) Photometry in the CSP-I filters synthesized from the spectrophotometric data of \citet{pereira13}
}
\end{deluxetable}

Note that the slow-declining Branch CN SN~2017cbv stands out as an exception in Figure~\ref{fig:uv_lcurves}, displaying a strong UV excess during the
first five days of observations, but then reverting to a pre-maximum rise similar to the other slow-declining Branch CN SNe.
 \citet{hosseinzadeh17}, who discovered and described this phenomenon, suggested that it might be due to an interaction with a 
companion star or circumstellar material, or the presence of $^{56}$Ni in the outermost ejecta. 
In their study of 42 SNe~Ia with early-time optical and UV light curves, \citet{hoogendam23}
classified SN~2017cbv as a ``Double'' event, showing a two-component power-law rise.  These 
authors found that such objects are more luminous than SNe~Ia with single-component power-law rises ten days before peak and earlier.

In interpreting these plots, care must be taken
since no K-corrections have been made.  However, these are likely to be small at these relatively small redshifts \citep[e.g., see Appendix B of][]{milne13}.
It should also be noted that although the peak transmission of the $UVW1$, and $UVW2$ filters is in the UV, they both have 
red tails that extend to optical wavelengths \citep[e.g., see Figure~1 of][]{brown10}.  Thus, the effective wavelengths of these filters can shift significantly 
with light curve phase as the color of the SN evolves, complicating the interpretation of the observed differences.
Nevertheless, the differences at early epochs must come from the UV due to the uniformity of the optical light curves (see Figures~\ref{fig:BVRI_91T}, \ref{fig:BVRI}, and \ref{fig:colors_low_reddening}).

\subsection{Ultraviolet--Optical Color Evolution}
\label{sec:uv_colors}

The UV spectra of SNe~Ia are known to display a greater diversity than is observed at optical wavelengths \citep[e.g., see][]{ellis08,cooke11,foley16}.
Early-on in the explosion, the principal source of opacity at optical and NIR wavelengths is electron scattering, causing the SN ejecta to become rapidly more
transparent \citep[e.g., see Figure 1 of ][]{hoeflich17}.  However, the situation at UV wavelengths is different since the opacity at early epochs is dominated 
by a very large number of overlapping lines of iron-peak elements that allow only the outermost layers to be penetrated
\citep[e.g., see][]{hoeflich98,lentz00,sauer08,derkacy20}.  The brightness of SNe~Ia in the UV compared to the optical has therefore been looked upon as a possible
probe of the progenitor metallicity.  \citet{pan20}  claimed to observe just such an effect from UV spectroscopy, but this assertion has been challenged 
by \citet{brown20} based on an analysis of UV$-$optical colors.  Importantly, 
as argued by \citet{derkacy20}, higher temperatures in the outermost ejecta (e.g., produced by differing density profiles or $^{56}$Ni distributions)
that alter the \ion{Fe}{3}/\ion{Fe}{2} ratio can significantly complicate efforts to determine progenitor metallicity from UV spectra.

\begin{figure*}[ht]
\epsscale{1.1}
\plottwo{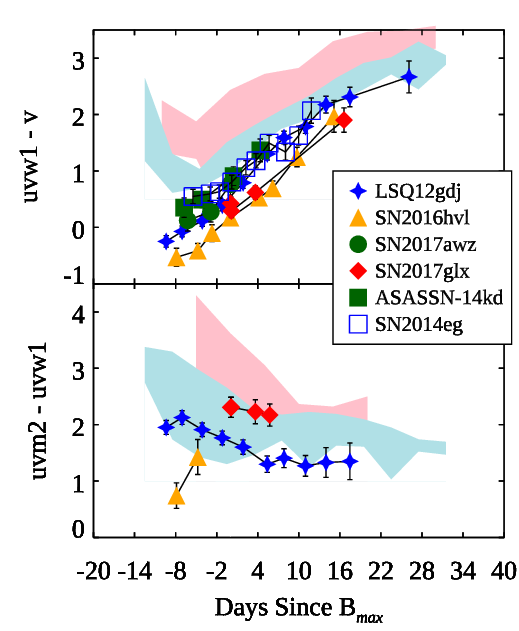}{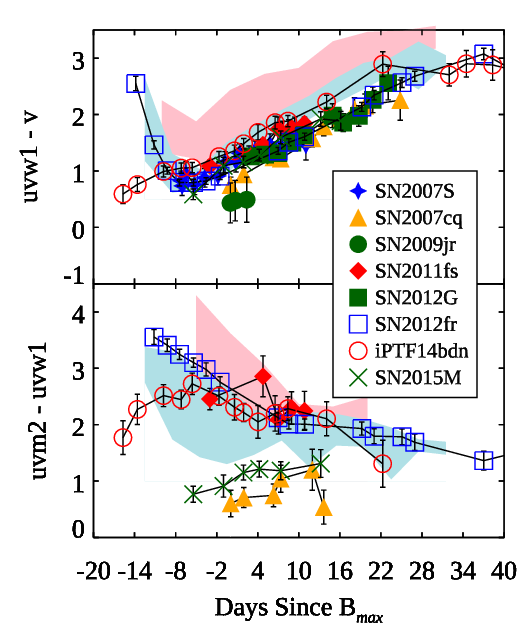}
\epsscale{.55}
\plotone{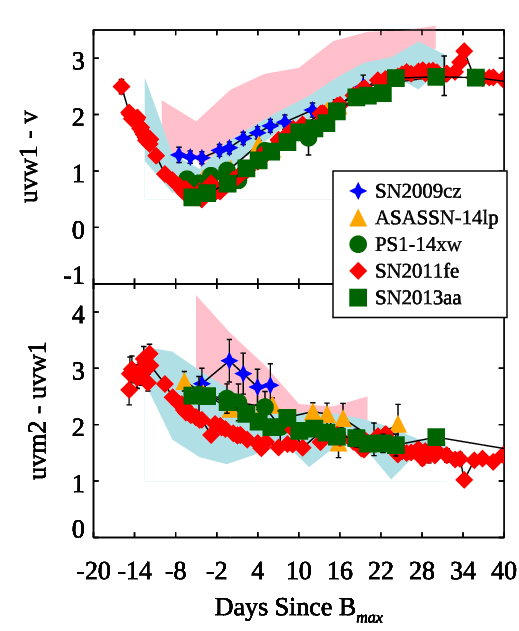}
\caption{Color evolution in $(uvw1-v)$ and $(uvm2-uvw1)$ of the 91T-like (upper-left), 99aa-like (upper-right), and slow-declining CN SNe  (lower).  As detailed in the text, the colors have been 
corrected for both Milky Way and host galaxy dust extinction.  The near-UV red and blue groups defined by \citet{milne13} are indicated by the red and blue regions,
respectively.}
\label{fig:uv_colors}
\end{figure*}

Figure \ref{fig:uv_colors} displays the extinction-corrected color evolution, $(uvw1-v)$ and $(uvm2-uvw1)$,  in the Swift UVOT filters for the 91T-like, 99aa-like, and
slow-declining  Branch CN SNe listed in Table~\ref{tab:host_reddenings}. 
Extinction corrections in the UV filters are complicated because the reddening coefficients
(e.g. $A_{uvw1}=R_{uvw1} \times E(B-V)$) are spectral dependent 
and non-linear with increasing $E(B-V)$ \citep{brown10}.  In order to estimate the UV extinction corrections for the UVOT filters we performed the following steps 
for each SN.  A UV spectrum of the slow-declining SN~2013dy \citep{pan15} near maximum light was first dimmed by the line-of-sight Milky Way reddening from 
\citet{schlafly11} using the \citet{cardelli89} law with $R_V = 3.1$.
Best-fit host galaxy reddening $E(B-V)$ and $R_V$ values
(listed in Table~\ref{tab:host_reddenings})
inferred from the optical and NIR colors of each SN using the method of \citet{burns14} 
were then used to further extinguish the SN~2013dy spectrum.  The magnitude difference 
between the original and extinguished spectra in each filter was taken as the extinction value and used to correct the color.

The red and blue regions in Figure \ref{fig:uv_colors} show the color evolution (uncorrected for reddening) of the near-UV red and blue groups 
from \citep{milne13}.  Unlike the optical colors, the differences in the $(uvw1-v)$ color evolution between the three subgroups are considerably more evident.  All of the 
members of the three subgroups lie within or below the blue edge of the \citet{milne13} blue group.  However, the pre-maximum colors of the 91T-like SNe are, on 
average, considerably bluer than the colors of both the 99aa-like events and the slow-declining Branch CN SNe.
This is particularly true at the earliest epochs observed.  In the case of the 91T-like LSQ12gdj, the $(uvw1-v)$ color grew 
steadily redder from $-0.25$~mag at $-9$~days to a value of $+0.55$~mag at maximum light.  This contrasts with the Branch CN SN~2011fe for which the 
$(uvw1-v)$ color at $-9$~days was $+0.90$~mag, evolving through a minimum value of $+0.55$~mag at $-5$~days, and then growing redder to $+0.82$~mag at maximum.  
Except for SN~2012fr (see below), the $(uvw1-v)$ color evolution of the 99aa-like SNe appears to be intermediate between that of the 91T-like events and the slow-declining 
Branch CN SNe.  The well-studied 99aa-like iPTF14bdn \citep{smitka15}, which was discovered within a few days of explosion, showed a steady evolution to redder $(uvw1-u)$ 
colors from $-16$ to $-9$~days in contrast to the Branch CN SN~2011fe which evolved rapidly from
red to blue over the same period.  Unfortunately, no 91T-like SNe have yet been observed at these early epochs, although the steady evolution  of LSQ12gdj  from bluer to
redder $(uvw1-v)$ colors from $-9$ to $-5$~days is clearly more similar to iPTF14bdn than to SN~2011fe, albeit offset to bluer colors.

Comparison of the evolution of the $(uvm2-uvw1)$ color is limited by the small number of SNe with well-observed $uvm2$ light curves.
In general, the 91T-like and 99aa-like events display $(uvm2-uvw1)$ colors that again fall within or below the blue edge of the \citet{milne13} blue group.
Interestingly, the blue colors of the 99aa-like SNe 2007cq and 2015M, and the 91T-like SN 2016hvl, do not have corollaries among the slow-declining Branch CN
SNe, although small numbers may enter into play here.  \citet{hoogendam23} have suggested that
the $(uvm2-uvw1)$ color can be used to distinguish between normal SNe~Ia and 02es-like 
\citep{ganeshalingam12} and 03fg-like \citep{howell06} events.  It should be noted, however, 
that the evolution of the $(uvm2-uvw1)$ colors of the two 99aa-like SNe, 2015M, and 2016hvl, is
indistinguishable from that of the 02es-like and 03fg-like SNe plotted in Figures~2 and 5 of
\citeauthor{hoogendam23}.

The differences in the pre-maximum $(uvw1-v)$ color evolution of the 91T-like, 99aa-like, and slow-declining Branch CN SNe are likely related to the differences 
observed by \citet{stritzinger18} in the slope of the $(B-V)$ color evolution at early epochs between Branch SS and CN events.  
As will be presented in \S\ref{sec:ifs}, there is no strong evidence that the distributions of the
gas phase metallicities of the local host galaxy environments of the 91T-like, 99aa-like, and 
slow-declining Branch CN SNe are significantly different.  It is unlikely, therefore, that the differences in the pre-maximum $(uvw1-v)$ color evolution seen in
Figure~\ref{fig:uv_colors} are a metallicity effect.  In their study of the 99aa-like iPTF14bdn,
\citet{smitka15} obtained {\it Swift} UVOT grism spectra revealing that the blue $(uvw1-v)$ color of this object at early epochs was a temperature effect 
produced by a 
predominance of doubly-ionized iron-peak elements that created an opacity ``window'' at near-UV wavelengths.  This effect was ascribed by these authors
to the mixing of $^{56}$Ni into the outer ejecta.

Finally, the $(uvw1-v)$ color evolution of the 99aa-like SN~2012fr deserves special mention since, at the earliest epochs, it more closely resembles that of the Branch CN 
SN~2011fe rather than the 99aa-like iPTF14bdn.  The fact that SN~2012fr falls near the border separating the Branch SS and CN subtypes \citep{childress13,contreras18}
may partially explain this.  However, as discussed in detail by \citet{contreras18}, SN~2012fr showed photometric and spectroscopic properties that link
it more closely to the peculiar Branch SS SN~2000cx \citep{li01a,candia03} than with typical Branch SS SNe.

\section{Pseudo-Bolometric Light Curves}
\label{sec:bolometric}

In this section we calculate UV-optical integrated luminosity light curves for three of the best observed examples of 91T-like, 99aa-like,
and slow-declining Branch CN SNe Ia: LSQ12gdj \citep{scalzo14b}, iPTF14bdn \citep{smitka15}, and ASASSN-14lp \citep{shappee16}.
To create these ``pseudo-bolometric'' light curves we employed the following procedure.
Ultraviolet data for the 91T-like SN LSQ12gdj, the 99aa-like iPTF14bdn, and the
slow-declining Branch CN SN ASASSN-14lp were obtained from the Swift Optical Ultraviolet Supernova Archive (SOUSA; \citealp{brown14}), and optical photometry  was obtained from \citet{scalzo14b}, \citet{smitka15}, and \citet{shappee16}, respectively.
A grid of the observation dates was created, and magnitudes from the filters were linearly interpolated so that each epoch had a magnitude in each filter.  At each epoch, the closest spectral epoch (in days since $B_{max}$) from the UV spectral series of ASASSN-14lp was selected.  To mimic the actual path of the photons, this rest-frame spectrum was reddened by the estimated host galaxy dust.  The resulting host-reddened spectrum was then redshifted to the observer frame of the SN and reddened by the Milky Way reddening from \citet{schlafly11}.  That spectrum, which mimics what would reach the Earth, was then color-matched to be consistent with the multi-band photometry of the epoch that was actually observed.  The reddened, redshifted, color-matched spectrum was then dereddened by Milky Way dust, deredshifted into the rest frame and dereddened by the estimated host dust, and converted from observed flux to luminosity using the distance to each object (see Table~\ref{tab:bol_param}). This spectrum was then integrated between the wavelengths of 2000 and 9000 \AA.  Although NIR photometry was not included in the calculation of these pseudo-bolometric light curves since no such data are available for iPTF14bdn, the contribution of the NIR is typically $\leq$20\% over the first three weeks after $B_{max}$ for slow-declining SNe~Ia \citep{scalzo14a,scalzo14b,contreras18}.
Beyond this phase, our light curves will significantly underestimate the total UV-optical-NIR
luminosity.

\begin{deluxetable}{lcccc}
\tabletypesize{\scriptsize}
\tablecolumns{5}
\tablewidth{0pt}
\tablecaption{Bolometric Light Curve Parameters\label{tab:bol_param}}
\tablehead{
\colhead{SN} &
\colhead{$(m-M)$ (mag)\tablenotemark{a}} &
\colhead{$t(B_{max})$\tablenotemark{b}} &
\colhead{$t$(first light)\tablenotemark{c}} &
\colhead{Rise Time (days)\tablenotemark{d}}
}
\startdata
ASASSN-14lp & 30.82 (0.45) & 57015.3 (0.0) & 56998.9 (0.1)\tablenotemark{e} & 16.3 (0.1) \\
iPTF14bdn   & 34.09 (0.14) & 56822.2 (0.1) & 56803.8 (0.5)\tablenotemark{f} & 18.1 (0.5) \\
LSQ12gdj    & 35.48 (0.08) & 56253.6 (0.1) & 56237.1 (1.0)\tablenotemark{g} & 16.0 (1.0) \\
\enddata
\tablenotetext{a}{Distance moduli for iPTF14bdn and LSQ12gdj derived from $z_{CMB}$ 
assuming standard $\Lambda$CDM cosmology and a fixed Hubble constant 
$H_0 = 72~{\rm km~s^{-1}~Mpc^{-1}}$, density parameter $\Omega_m = 0.27$, and cosmological 
constant parameter $\Omega_\Lambda = 0.73$, with errors corresponding to an assumed peculiar
velocity of $300~{\rm km~s^{-1}}$. For ASASSN-14lp, the Tully-Fisher distance modulus of
\citet{tully16} for the host galaxy NGC~4666 was adopted, adjusted for $H_0 = 72~{\rm km~s^{-1}~Mpc^{-1}}$.}
\tablenotetext{b}{MJD of $B$ maximum derived from SNooPy fits to the available photometry. Errors given in parentheses.}
\tablenotetext{c}{MJD of the time of first light. Errors given in parentheses. See text for details.}
\tablenotetext{d}{Rise time to $B$ maximum in days, corrected for time dilation.}
\tablenotetext{e}{\citet{shappee16}}
\tablenotetext{f}{\citet{cao14}}
\tablenotetext{g}{\citet{scalzo14b}}
\end{deluxetable}

\begin{figure}[t]
\epsscale{1.}
\plotone{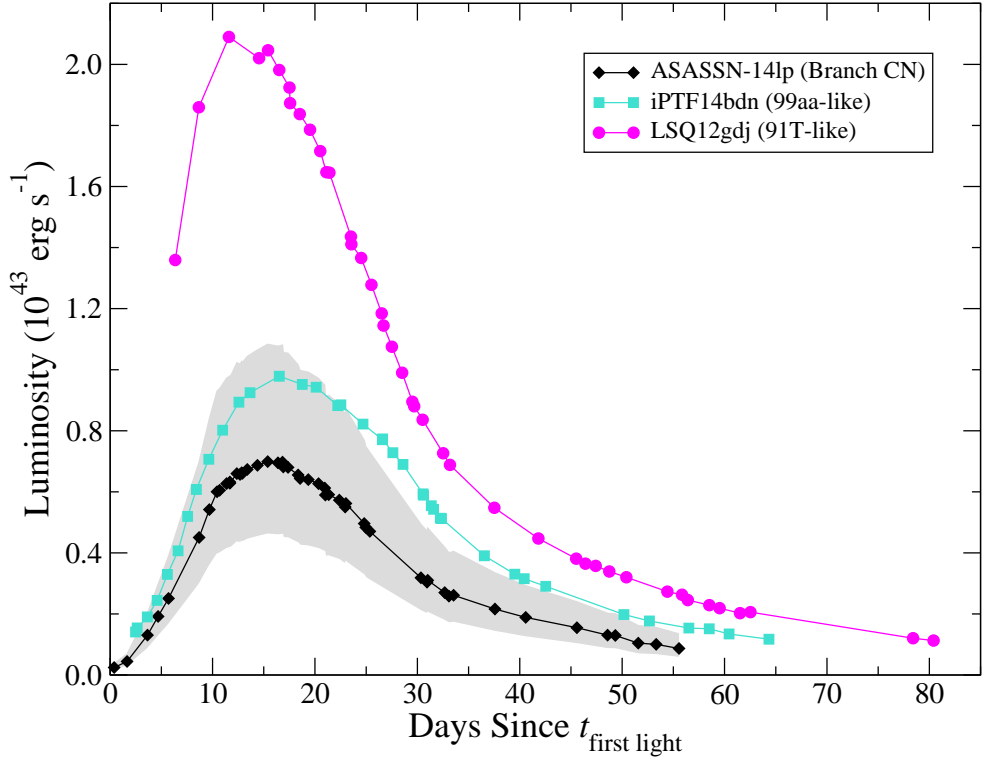} 
\caption{Pseudo-bolometric light curves of the 91T-like SN LSQ12gdj, the 99aa-like iPTF14bdn, and the
slow-declining Branch CN SN ASASSN-14lp integrated between 2000--9000~\AA. The data
are plotted as a function of time in days since first light.  The gray shading indicates
the luminosity uncertainty for ASASSN-14lp due to the large error in its distance
(see Table~\ref{tab:bol_param}).}
\label{fig:bolo_2000_9000}
\end{figure}

Figure~\ref{fig:bolo_2000_9000} displays the results of these calculations where total
luminosity from 2000--9000~\AA\ is plotted versus time since first light. The host galaxy
reddenings adopted for LSQ12gdj and ASASSN-14lp are given in Table~\ref{tab:host_reddenings},
while for iPTF14bdn we assumed zero host reddening. As indicated in Table~\ref{tab:bol_param},
distances for LSQ12adj and iPTF14bdn were calculated using their host galaxy redshifts, 
whereas for ASASSN-14lp we opted to use a Tully-Fisher distance because of the proximity of its host.  Fortunately, the times of first light of these three SNe are known
to $\pm1$~day or better (see Table~\ref{tab:bol_param}).
Note that \citet{scalzo14b} independently computed a UV-optical-NIR bolometric light curve for
LSQ12gdj. After correction to the same Hubble constant, our result compares reasonably well through 20 days past $t(B_{max})$, after which the additional luminosity from the
NIR included in the \citet{scalzo14b} calculation becomes noticeable.

Two things are immediately evident in Figure~\ref{fig:bolo_2000_9000}.  First, and 
most obvious, is the much greater luminosity at maximum of the 91T-like LSQ12gdj, 
$\sim$2--5 times greater than the peak luminosities of iPTF14bdn and ASASSN-14lp.
This is a much larger luminosity difference at peak than is observed at 
optical and NIR wavelengths, and likely reflects the greater mixing of 
$^{56}$Ni into the outer ejecta of the 91T-like SNe.
Secondly, LSQ12gdj reaches maximum approximately four days before the other two
SNe.
Interestingly, the light curve of the 99aa-like iPTF14bdn is also somewhat broader than those
of LSQ12gdj and ASASSN-14lp. Clearly there is a need for more high-quality UV, optical, 
and NIR photometry of other 91T-like, 99aa-like, and slow-declining to explore the
full dispersion of bolometric light curve properties.

\section{Host Galaxies}
\label{sec:hosts}

Luminous, slow-declining SNe~Ia have long been known to preferentially occur in star-forming galaxies \citep{hamuy96a,riess99,hamuy00,sullivan06,neill09}.
Indeed, all but one of the 91T-like SNe in the literature and CSP-II samples with host morphological classifications in the
\citet{ned19}\footnote{NED is operated by the Jet Propulsion Laboratory, California Institute of Technology,
under contract with the National Aeronautics and Space Administration.} occurred 
in spiral galaxies.  In this section, we 
examine and intercompare in more detail the host galaxy properties of the 91T-like, 99aa-like, and slow-declining Branch CN SNe in the CSP-II sample.

\subsection{Masses}
\label{sec:masses}

\begin{figure}[ht]
\epsscale{0.9}
\plotone{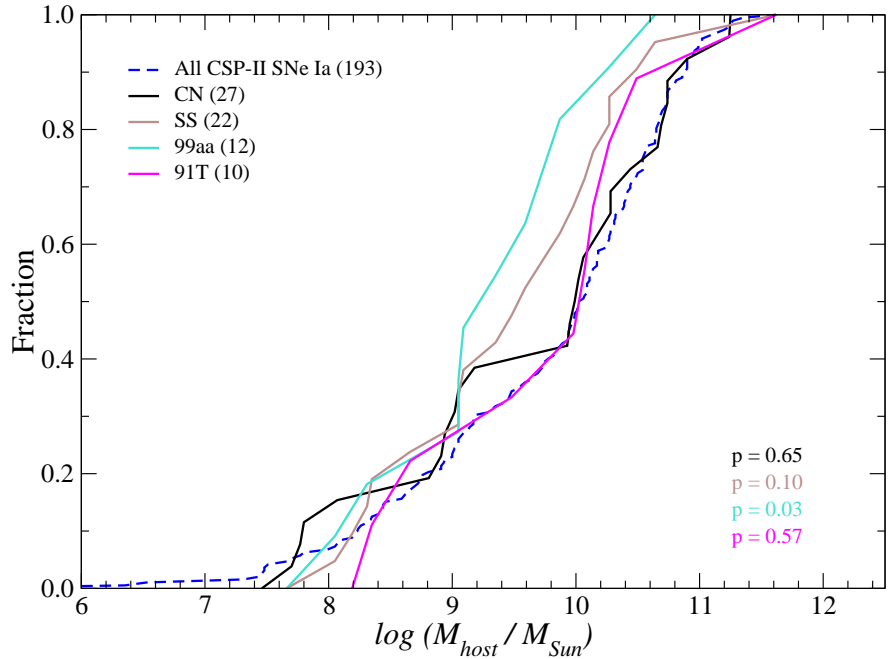}
\caption{Cumulative histograms of the host galaxy masses of 91T-like (magenta), 99aa-like (cyan), and Branch slow-declining CN SNe
(black) observed by the CSP-II.  Also plotted is the host mass distribution of the Branch SS subsample (brown), which is the combination of the 91T-like and 99aa-like SNe.
The number of SNe in each of these subsamples is indicated in the figure legend. The cumulative host mass distribution for the full sample of 193 CSP-II 
SNe Ia is plotted with a dashed blue line.  Shown in the lower-right corner are the $p$~values derived from two-sample K-S tests with the full CSP-II SNe Ia sample 
serving as the reference. These are color-coded to indicate the different CSP-II subsamples (CN = slow-declining Branch CN; 99aa = 99aa-like; 91T = 91T-like; 
SS = 99aa-like + 91T-like).
}
\label{fig:91T_99aa_CN_host_masses}
\end{figure}

Figure~\ref{fig:91T_99aa_CN_host_masses} displays the cumulative histogram of host galaxy masses of the full sample of 193 CSP-II
SNe~Ia for which it was possible to unambiguously assign a host.  
These measurements are taken from \citet{uddin23}, and were derived from $uBgVriYJH$ imaging following the procedures detailed in \citet{uddin20}.
Plotted in the same diagram are the host mass distributions for the CSP-II 91T-like, 99aa-like, and
slow-declining Branch CN subsamples.  Since the numbers of the 91T-like and 99aa-like SNe are small, the histogram of the CSP-II Branch SS SNe --- which 
is the combination of the 91T-like and 99aa-like subsamples --- is also included in the figure.

A Kolmogorov-Smirnov (K-S) test comparing the cumulative distribution of host masses of the slow-declining Branch CN SNe with the full sample
of CSP-II SNe gives a $p$ value of 0.65, indicating
that the K-S test is unable to distinguish the two distributions
at the 95\% confidence level.
A K-S test comparing the host masses
of the Branch SS SNe with the full CSP-II sample gives $p = 0.10$, meaning again that it is not possible to distinguish
between the two distributions at the 2-$\sigma$ level.
A K-S test comparing the 91T-like SNe with the full 
CSP-II sample returns $p = 0.57$, implying a similar conclusion, but the $p$ value of 0.03 returned by a K-S test 
for the 99aa-like subsample is inconsistent with this hypothesis. Thus, we conclude that the distributions of
host masses of the Branch SS and slow-declining Branch CN subsamples are statistically indistinguishable from the distribution for the full CSP-II sample of SNe~Ia, but there is some evidence that 99aa-like SNe explode in host galaxies
with somewhat lower masses.  Larger samples of 91T-like and 99aa-like SNe with
well-determined host masses are needed to reach any stronger conclusions.

\begin{figure}[t]
\epsscale{0.8}
\plotone{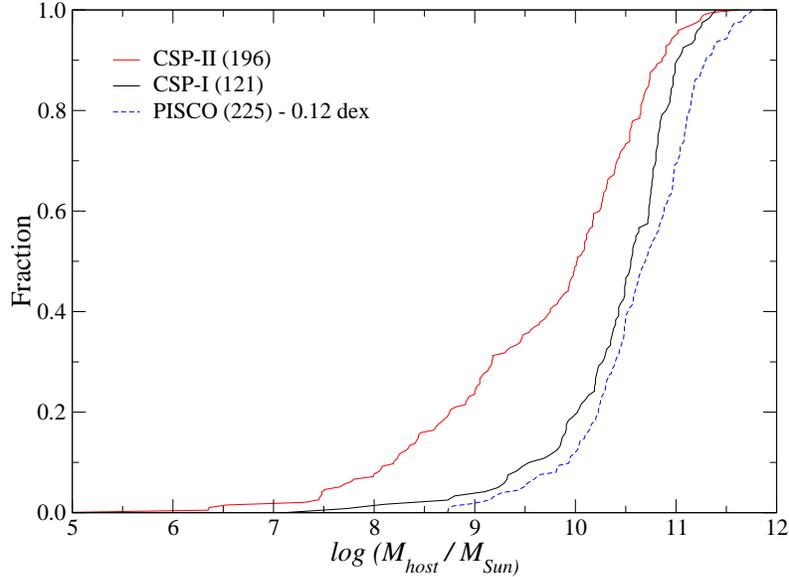}
\caption{Cumulative histograms of the host galaxy masses of the CSP-I (black) and CSP-II (red) SNe Ia samples.  Note the heavy weighting to larger masses of the 
CSP-I sample, which was mostly drawn from {\em targeted} SN searches, compared to the CSP-II sample, which came nearly entirely from {\em untargeted} searches.
Also plotted is the host mass distribution of the updated PISCO sample (see~\ref{sec:ifs}) which has been shifted to lower mass values by 0.12~dex to
account for differences in the Hubble constants and stellar initial mass functions (\citealt{rana92} versus \citealt{salpeter55}) assumed in computing the
masses for the CSP-I and CSP-II samples \citep{uddin20,uddin23} and the PISCO
sample \citep{galbany18}.}
\label{fig:csp1_vs_csp2_host_masses}
\end{figure}

As detailed in \S\ref{sec:cspI}, no true 91T-like events were found among the 121 SNe~Ia observed during the course of the CSP-I.  However, 
nearly 90\% of the SNe~Ia in the CSP-I sample came from {\em targeted} searches that are strongly biased toward luminous (massive) host galaxies,
whereas 96\% of the CSP-II SNe were drawn from {\em untargeted} searches.  In Figure~\ref{fig:csp1_vs_csp2_host_masses}, which compares cumulative
histograms of the host galaxy masses of the CSP-I and CSP-II SNe Ia samples, it is clearly seen that the CSP-I sample is heavily weighted towards 
massive hosts, consistent with other studies comparing the distributions of host masses of SNe~Ia discovered in targeted versus untargeted surveys
\citep[e.g., see Figure~4 of ][]{galbany18}. 
The median host mass of the 91T-like SNe in the CSP-II sample is $log~(M_{host}/M_{\sun}) = 10.0$, and nine out of ten have $log~(M_{host}/M_{\sun})$
values less than the median host mass of the CSP-I sample of $log~(M_{host}/M_{\sun}) = 10.6$. In contrast,
only 20\% of the hosts in the CSP-I sample have $log~(M_{host}/M_{\sun}) < 10.0$.  Thus, it is  not so surprising that the 
CSP-I sample of SNe~Ia does not include any 91T-like events, particularly considering the rarity of these objects compared to the total population of SNe~Ia (see \S\ref{sec:discussion}).

\subsection{Integral Field Unit Spectroscopy}
\label{sec:ifs}

Through integral field unit (IFU) spectroscopy, properties of the local (1~kpc$^2$) SN environment such as age, metallicity, and star formation rate can be studied.
Of the total of 51 SNe comprising the 91T-like, 99aa-like, and slow-declining Branch CN events in the CSP-II sample, IFU observations are available for 45.  Of these,
all but four reveal detectable ionized gas.  The breakdown into subtypes of these 41 SNe is as follows: 7 91T-like SNe, 11 99aa-like events, and 23 
slow-declining Branch CN SNe.  IFU data for 34 of these hosts were obtained as part of the  All-weather MUse Supernova Integral-field of Nearby Galaxies
\citep[AMUSING;][]{galbany16,lopez_coba20} survey which is focused on studying the host environments of different types of SNe.  These observations were acquired using
the Multi-Unit Spectroscopic Explorer \citep[MUSE;][]{bacon10} mounted on the Unit~4 (``Yepun'') telescope at the ESO Very Large Telescope at the the Cerro
Paranal Observatory.  The remaining seven hosts were observed with the Potsdam Multi-Aperture Spectrophotometer \citep[PMAS][]{roth05} in the PPak
configuration \citep{kelz06} on the 3.5~m Calar Alto telescope  as part of the PMAS/PPak Integral-field Supernova hosts COmpilation (PISCO) program \citep{galbany18}.

Analysis of the IFU spectroscopy was carried out following the procedures detailed by \citet{galbany14} and \citet{galbany16}.  Briefly, after correction for Milky Way
dust extinction using the maps of \citet{schlafly11}, the stellar continuum in each spectrum was fit using \texttt{STARLIGHT} \citep{cid05,cid09} which assumes that
a galaxy can be represented as the sum of spectra of a single stellar population with different ages and metallicities.  The \texttt{STARLIGHT} fits were then subtracted 
from the spectra to
obtain pure nebular emission line spectra to allow measurement of fluxes for the strongest emission lines (H$\beta$, [\ion{O}{3}]~$\lambda$5007, H$\alpha$, 
[\ion{N}{2}]~$\lambda\lambda$6548, 6583, and [\ion{S}{2}]~6717, 6731).  These measurements were then corrected for host galaxy extinction using the observed ratio of 
the H$\alpha$ and H$\beta$ fluxes and assuming an intrinsic ratio of 2.86 representative of typical H~II regions.
The star formation rate (SFR) may then be calculated from the extinction-corrected H$\alpha$ flux \citep{kennicutt98} which, in turn, allows the SFR density 
($\Sigma_{{\rm SFR}}$ = SFR/area) and the specific SFR (sSFR = SFR/mass) to be estimated.  The H$\alpha$ equivalent width (EW) provides insight into the
strength of current SFR compared to the past SFR and can be used to estimate the age of the youngest
stellar population \citep[e.g., see][and references therein]{kuncarayakti16}.  Finally, the oxygen abundance of the ionized gas was estimated using an empirical
method (O3N2) based on the extinction-corrected flux ratio of the [\ion{O}{3}]~$\lambda$5007 and H$\beta$ emission lines as calibrated by \citet{marino13}.

\begin{figure*}[t]
\epsscale{1.1}
\plotone{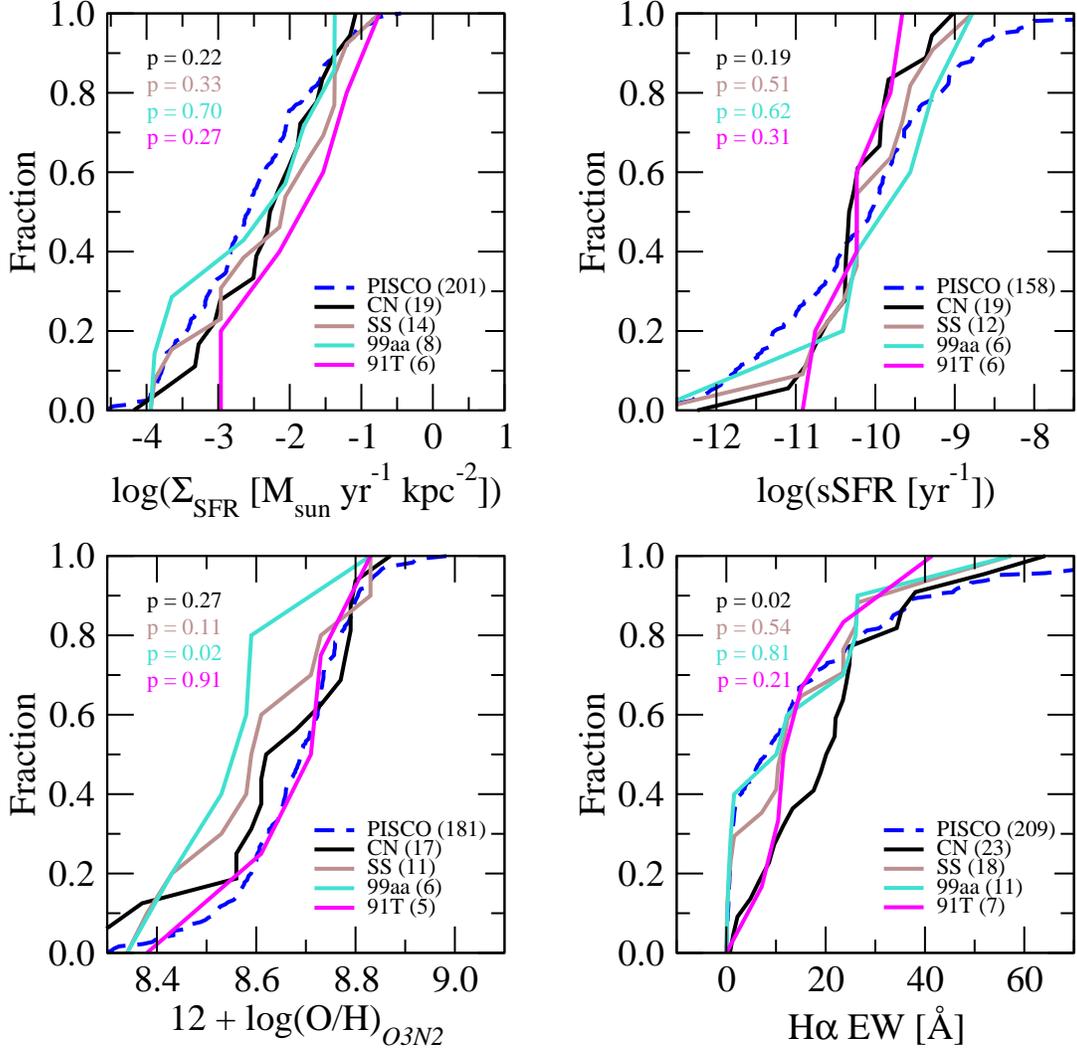} 
\caption{Cumulative histograms of local environmental properties derived from IFU observations.  Shown in the upper-left corner of each
plot are the $p$~values derived from two-sample K-S tests with the updated PISCO sample serving as the reference sample.  These are color-coded to 
indicate the different CSP-II subsamples (CN = slow-declining Branch CN; 99aa = 99aa-like; 91T = 91T-like; SS = 99aa-like + 91T-like).
}
\label{fig:IFU_local}
\end{figure*}

Figure~\ref{fig:IFU_local} displays cumulative histograms of four local parameters derived from the IFU spectroscopy for the CSP-II 91T-like, 99aa-like, Branch SS, and 
slow-declining Branch CN SNe.  Since IFU observations of the full sample of CSP-II SNe~Ia are not yet available, we use an updated version
of the PISCO sample \citep{galbany18} as a reference.  This sample now numbers 
228 SNe~Ia hosts, or double the number published by \citet{galbany18}.  However, an important caveat is that slightly more than half (56\%) of the SNe in the 
PISCO sample were discovered in targeted searches, and so the sample is biased to higher host galaxy masses than is the CSP-II sample.  As
illustrated in Figure~\ref{fig:csp1_vs_csp2_host_masses}, the cumulative distribution of host masses of the updated PISCO sample closely resembles that of
the CSP-I.

The cumulative distributions of the star formation rate parameters, $\Sigma_{{\rm SFR}}$ and sSFR, for the CSP-II 91T-like, 99aa-like, Branch SS, and slow-declining 
Branch CN SNe displayed in Figure~\ref{fig:IFU_local} are all very similar, and also resemble the distributions of these two parameters for the PISCO SNe~Ia sample.
This is born out by 
two-sample K-S tests for each subsample compared with the PISCO sample which all give $p > 0.05$, indicating that they cannot be distinguished at
the 95\% confidence level.

In the case of the H$\alpha$ EW, the histograms of the CSP-II 91T-like, 99aa-like, and Branch SS are, again, all consistent with being indistinguishable from the distribution of the same parameter for
the PISCO sample.  However, this is not the case when comparing the slow-declining Branch CN SNe with the PISCO sample.  Here, Figure~\ref{fig:IFU_local} 
shows that the median H$\alpha$ EW for the slow-declining Branch CN SNe is 20~\AA\ whereas the median for the PISCO sample is 8~\AA.  This is confirmed by the
the two-sample K-S test which gives $p < 0.05$, rejecting at the 95\% confidence level the hypothesis that they are consistent with the same distribution. 

Regarding the local host galaxy oxygen abundance, the 
K-S test $p$ values in the O3N2 diagram imply that the distributions for the 91T-like and slow-declining Branch CN SNe are indistinguishable from the
PISCO sample distribution, but the distribution for the hosts of the 99aa-like SNe may be different.
However, combining the 91T-like, 99aa-like, and 
slow-declining Branch CN SNe and then performing a K-S test comparing the distribution of oxygen abundances with the PISCO sample yields a $p$ value of 0.05, in which case we cannot rule out 
the hypothesis that they are drawn from the same population as the PISCO hosts.

In summary, in most cases, the local environment parameters derived from the analysis of the IFU spectroscopy of the CSP-II 91T-like, 99aa-like, and slow-declining Branch CN SNe hosts
suggest that, overall, they are similar to the general population of SNe~Ia host galaxies as represented by the PISCO sample.  The 99aa-like hosts may have a lower median oxygen metallicity
than the PISCO sample, and the slow-declining Branch CN SNe appear to have a 
larger H$\alpha$ EW compared to the PISCO sample.  It will be interesting to
see if these differences hold up when the full IFU data set for the CSP-II sample
can be used as the reference sample in place of the PISCO sample.

\subsection{Na I D Absorption}
\label{sec:naid}

Optical spectra of SN~1991T showed prominent \ion{Na}{1}~D absorption at the redshift of its host galaxy \citep{hamuy91,sivaraman91,wheeler91}.
The $B$-band light curve was also observed to flatten $\sim$600~days after explosion because of a presumed light echo \citep{schmidt94}, which was 
confirmed by images obtained a few years later with the Hubble Space Telescope by \citet{sparks99}.  \citeauthor{sparks99} interpreted the echo as arising from
dust extending to $\sim$50~pc from the SN, but an unpublished reanalysis suggests that the dust could be much closer (6-9~pc) and might have
originated from the pre-explosion evolution of the progenitor system \citep{thormann09}.
Although it is not clear if the gas that produced the \ion{Na}{1}~D absorption in SN~1991T was associated with the dust that produced the light echo, 
considering the possible link between 91T-like SNe and the Type~Ia-CSM subclass
(see \S\ref{sec:Ia-csm}), it is 
interesting to ask whether 91T-like events preferentially show stronger  \ion{Na}{1}~D lines than do 99aa-like and slow-declining Branch CN SNe.

\begin{figure}[ht]
\epsscale{0.8}
\plotone{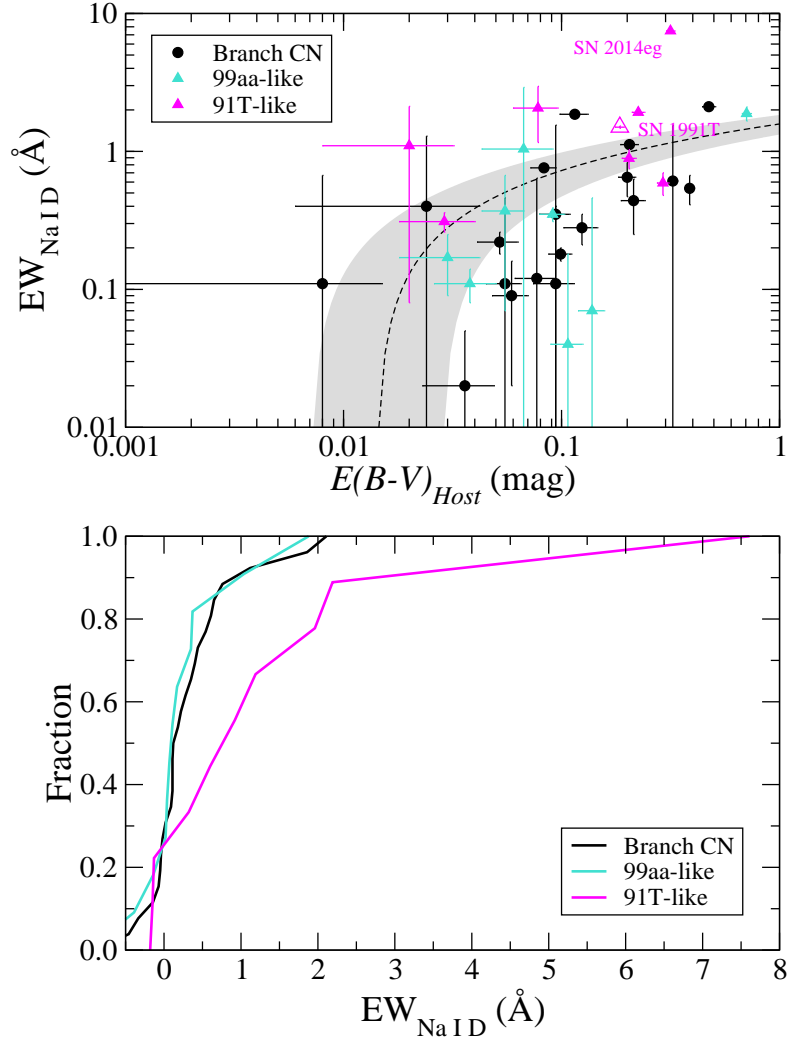}
\caption{(Upper) Equivalent width of the host galaxy component of \ion{Na}{1}~D absorption, EW(\ion{Na}{1}~D), plotted versus the host reddening,
$E(B-V)_{Host}$, for the 91T-like, 99aa-like, and slow-declining Branch CN SNe in the CSP-II sample.  Plotted as a dashed line is the relation given
by \citet{poznanski12}, while the gray area corresponds to the $1\sigma$ dispersion observed for lines of sight through the Milky Way \citep{phillips13}.
(Lower) Cumulative distributions of host galaxy equivalent widths for the 91T-like, 99aa-like, and slow-declining Branch CN SNe in the CSP-II sample.}
\label{fig:NaID}
\end{figure}

In an attempt to answer this question, \ion{Na}{1}~D equivalent widths were measured for the 91T-like, 99aa-like, and slow-declining Branch CN SNe in the CSP-II sample.
The upper panel of Figure~\ref{fig:NaID} shows these measurements plotted versus the host galaxy dust extinction derived from analyses of the light curves using \texttt{SNooPy}. 
One third (9 of 27) of the slow-declining Branch CN SNe, one fourth (3 of 12) of the 99aa-like events, and one half (5 of 10) of the 91T-like SNe  
have $3\sigma$ \ion{Na}{1}~D absorption detections.  The well-studied correlation between \ion{Na}{1}~D equivalent widths and color excess for lines of
sight through the Milky Way is plotted in the top panel of Figure~\ref{fig:NaID}.  The measurements for the SNe also display a loose correlation, albeit with considerably more
scatter, similar to that observed for a sample of 32 SNe~Ia from high-dispersion spectra \citep{phillips13}.  In the lower panel of Figure~\ref{fig:NaID}, cumulative
histograms of all of the equivalent width measurements (including negative values) are plotted for the three subsamples.  The distributions for the 99aa-like and
slow-declining Branch CN SNe are quite similar, whereas that of the 91T-like events is skewed to higher values.  (This effect is also apparent in the upper panel of
Figure~\ref{fig:NaID}.)  However, the numbers of 99aa-like and 91T-like SNe are small, and a K-S test comparing the distribution of the 91T-like events to the 
slow-declining Branch CN SNe does not allow us to reject the null hypothesis that the measurements for both subsamples are drawn from the same 
population at the 95\% confidence level.  Clearly much larger samples will be required to confirm or deny that interstellar \ion{Na}{1}~D absorption lines are
more common in 91T-like SNe.

\section{SNe~Ia-CSM}
\label{sec:Ia-csm}

As mentioned in the introduction to this paper, 91T-like SNe have been linked by some investigators with the rare class of SNe~Ia-CSM
\citep[see][for detailed studies of the properties of these objects]{silverman13b,sharma23}.
In this section, we take a close look at optical spectra of three of the best-observed examples of SNe~Ia-CSM at early phases
when the SN ejecta--CSM interaction was weakest, and compare them to spectra of SN~1991T at comparable epochs.

\subsection{SN~2002ic}
\label{sec:02ic}

SN~2002ic was discovered approximately a week before maximum light by \citet{wood-vasey02} from images taken by the Near-Earth Tracking (NEAT)
program \citep{pravdo99}.  An optical spectrum reported by \citet{hamuy02} near maximum showed strong \ion{Fe}{3} absorption features and weak 
\ion{Si}{2} $\lambda$6355 absorption.  Closer examination of this spectrum revealed prominent H$\alpha$ emission with 
a broad (1,800~\kms\ FWHM) component suggestive of an ongoing SN eject--CSM interaction \citep{hamuy03a}.   In a detailed spectroscopic study
covering the first two months following discovery, \citet{hamuy03b} argued that the evolution of the spectral features of SN~2002ic was very similar to that of
a 91T-like SN, but diluted in strength by continuous emission arising from the ejecta--CSM interaction that also accounted for its enhanced
luminosity ($M_V \sim -20.1$~mag) at maximum light.  These authors suggested that the progenitor system included a massive asymptotic giant branch (AGB)
star that had undergone significant mass loss prior to the explosion of an accompanying white dwarf.  Subsequent modeling by 
\citet{wood-vasey04} suggested the presence of a $\sim$100 a.u. gap between the SN and the CSM, 
consistent with the AGB star having been in the protoplanetary nebula phase at the time of explosion.

\begin{figure}[t]
\epsscale{0.9}
\plotone{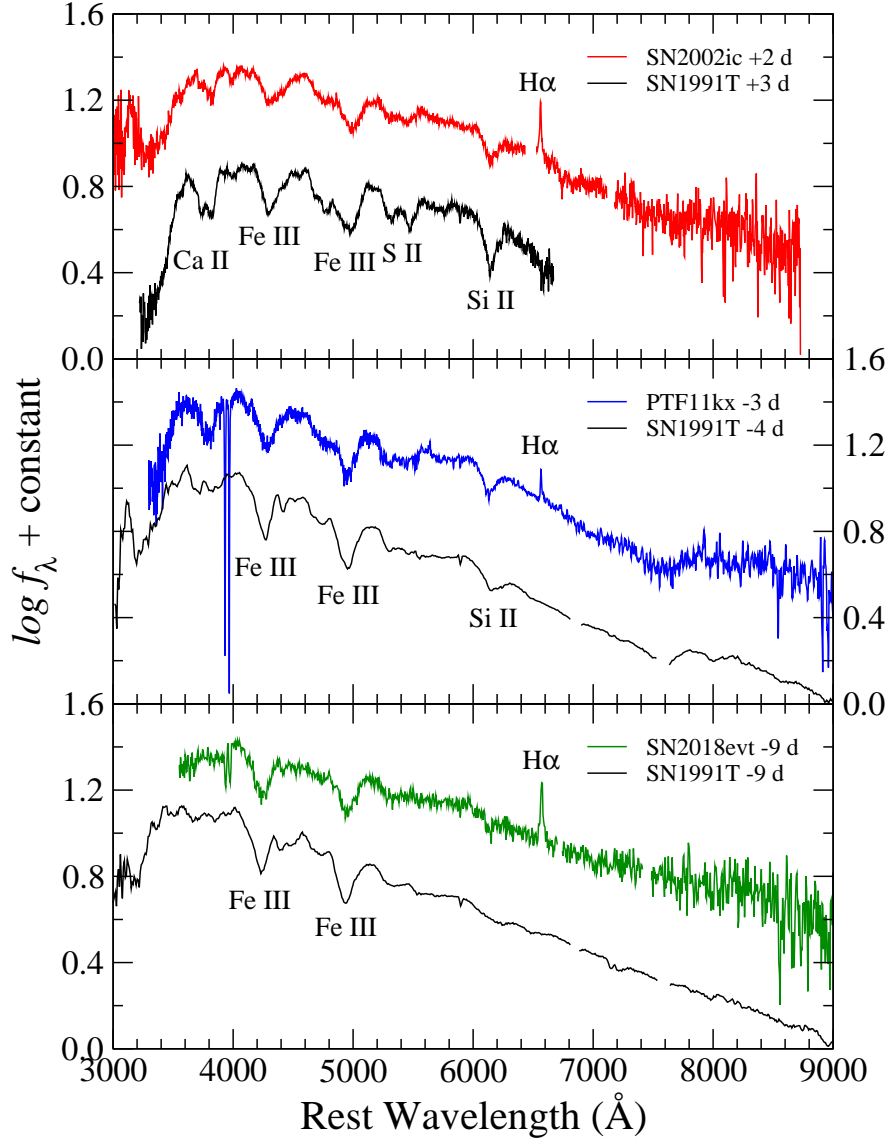}
\caption{Spectra of three SNe~Ia-CSM (SN~2002ic, PTF11kx, and SN~2018evt).  In each panel, comparison is made with a spectrum of
SN~1991T at a comparable epoch with respect to $B$ maximum.}
\label{fig:02ic_11kx_18evt}
\end{figure}

The top panel of Figure~\ref{fig:02ic_11kx_18evt} displays the spectrum of SN~2002ic at a phase of $+2$~days.
For comparison, a spectrum of SN~1991T is shown at a comparable epoch.  The similarity of the two is obvious in spite of the dilution of the 
absorption features in the spectrum of SN~2002ic.  Although this interpretation was questioned by \citet{benetti06} who suggested that 
SN~2002ic might be a Type~Ic supernova surrounded by hydrogen-rich CSM, the cases of PTF11kx and SN~2018evt (see below) argue strongly that SN~2002ic
was, indeed, a 91T-like event.

\subsection{PTF11kx}
\label{sec:11kx}

PTF11kx was discovered approximately two weeks before $B$ maximum by the Palomar Transient Factory.
The middle panel of Figure~\ref{fig:02ic_11kx_18evt} shows a comparison of a spectrum obtained by \cite{dilday12} at a phase of $-3$~days
with the spectrum of SN~1991T at $-4$~days.  Aside from the presence of unusually strong and narrow \ion{Ca}{2}
H~\&~K absorption, the evolution of the optical spectra of PTF11kx closely resembled that of SN~1991T until a strong 
ejecta--CSM interaction began two months after explosion \citep{dilday12}.  Our
measurements of pEW(\ion{Si}{2}~$\lambda6355$) from $-3$~days to $+9$~ days are fully consistent with the evolution
of SN~1991T,  and the $B$/$g$, $R$/$r$, and $I$/$i$ light curves over the first three weeks after maximum were also
consistent with those of a luminous, slow-declining SN~Ia, with $M_B \sim -19.3$~mag at maximum, with the $I$/$i$ light curve showing a clear secondary maximum 
and a primary maximum that appears to peak before $B$~maximum \citep[see Figure~S1 of][]{dilday12}.

If the observations
had ended a month after maximum, PTF11kx would have been classified as a typical 91T-like SN.  However, by day $+41$, the \ion{Ca}{2} H~\&~K
absorption had disappeared, and was replaced by emission lines with a broad ($\sim$1,000~\kms\ FWHM) P-Cygni profile
marking the beginning of a strong ejecta--CSM interaction.  At the same time, H$\alpha$ also developed a broad P-Cygni profile.
The timing of this interaction implies a distance from the SN progenitor to the CSM of $\sim$700 a.u. \citep{dilday12}.
Three months after maximum, the $r$-band light curve had flattened at an absolute magnitude of $-16.3$ \citep{dilday12},
independently signaling a strong CSM interaction, and 
declined only very slowly thereafter in observations obtained up to $\sim$300~days after maximum \citep{silverman13c}.
The spectrum of PTF11kx at these late times also closely resembled that of other SNe~Ia-CSM 
(e.g., SN~2005gj) at similar epochs \citep{silverman13c}.

\subsection{SN~2018evt}
\label{sec:18evt}

SN~2018evt (ASASSN-18ro) was discovered with the All-Sky Automated Survey for Supernovae \citep[ASASSN;][]{shappee14} Cassius telescope on 2018 August 11.0 (UT).
From a PESSTO spectrum obtained two days later, the SN was classified as a 91T-like event approximately nine days before maximum light \citep{brimacombe18}.
Publicly-available ASAS-SN photometry \citep[\url{https://asas-sn.osu.edu/photometry};][]{kochanek17,jayasinghe19}, uncorrected for the contribution of the
host galaxy, shows the $V$-band light curve appearing to approach maximum around 2018 August 22 UT.  Unfortunately, the data end at
this point, presumably since the SN was rapidly becoming inaccessible in the evening twilight sky.  However, when ASAS-SN observations were resumed in
mid-December of 2018, \citet{dong18} reported that the SN was still visible at an absolute $g$-band magnitude of about $-19$~mag --- i.e., nearly four 
magnitudes brighter than expected for a 91T-like SN.  These authors also reported that re-inspection of the PESSTO spectrum on 2018 August 13 UT
revealed prominent H$\alpha$ emission with a FWHM of $\sim$1,000~\kms\ from a likely ejecta--CSM interaction.  Light curves 
obtained by both ZTF (\url{https://lasair.roe.ac.uk/object/ZTF18actuhrs/}) and Gaia (\url{http://gsaweb.ast.cam.ac.uk/alerts/alert/Gaia18dwd/})
show that SN~2018evt declined at a rate of only $\sim$0.005~mag~day$^{-1}$ from mid-December 2018 through mid-August 2019,
confirming an ongoing shock interaction \citep[see also][]{sharma23}.
Extensive optical and infrared observations of this object have been published by \citet{yang23}
and \citet{wang24}.

The lower panel of Figure~\ref{fig:02ic_11kx_18evt} shows the PESSTO spectrum of SN~2018evt is well-matched by a spectrum of
SN~1991T obtained nine days before $B$ maximum, although the \ion{Fe}{3} absorption features in SN~2018evt are somewhat diluted
in comparison.  This is one of the earliest spectra obtained of a SN~Ia-CSM.  Like PTF11kx, strong interstellar \ion{Ca}{2} H~\&~K absorption 
is also clearly visible in the PESSTO spectrum, whereas the interstellar \ion{Na}{1}~D lines are much weaker.  
This is unusual for SNe~Ia, in which the strength of the \ion{Ca}{2}
interstellar lines is normally comparable to, or less than, the strength of the \ion{Na}{1}~D lines, consistent with the lines having been produced in the
cold interstellar medium of the host galaxy.  However, strong
\ion{Ca}{2} compared to \ion{Na}{1} originating in the CSM is expected if the SN explodes in a binary system with a red giant star which has undergone
significant mass loss \citep{chugai08}.

Spectropolarimetry of SN~2018evt was obtained by \citet{yang23} from $\sim$172--219 days after $V$ maximum.  Over this period, the continuum polarization decreased slowly from $0.55 \pm 0.23$\% to $0.33 \pm 0.18$\%.  In contrast, the polarization position angle remained constant to within the errors as the photosphere recessed.  Taken together, these observations suggest SN~2018evt exploded within a massive aspherical CSM, with dust forming three years after explosion in a dense shell
between the shocked CSM and SN ejecta \citep{wang24}.

\section{Discussion and Conclusions}
\label{sec:discussion}

The goal of this research has been to describe the observational properties of 91T-like SNe at UV, optical, and NIR wavelengths, and to attempt to understand how
these events fit into the general scheme of SNe~Ia.  The main findings (including those of Paper~I) may be summarized as follows:

\begin{itemize}

\item 91T-like SNe are the extreme members of the Branch SS group in displaying the weakest \ion{Si}{2} and
\ion{Ca}{2} absorption at maximum light of any SNe~Ia.  In cases where a spectrum at maximum light is unavailable, they
can be identified by plotting the pseudo-equivalent width of the \ion{Si}{2} $\lambda6355$ absorption versus light curve phase for $t-t(B_{max}) \leq +10$~days.  They are
differentiated from 02cx-like and 03fg-like SNe in possessing $i/I$-band light curves that reach maximum before the epoch of $B$ maximum,
and that also display a clear secondary maximum. 

\item 99aa-like SNe are the less extreme members of the Branch SS group.  Lacking knowledge of light curve phase, they can be difficult 
to distinguish from 91T-like SNe.  For example, the spectrum of a 99aa-like SN a week before maximum is essentially identical to that
of a 91T-like SN at maximum.

\item The optical spectra of 91T-like SNe observed at similar phases are quite similar.  At pre-maximum phases, they are dominated by the
\ion{Fe}{3} $\lambda$4404 and $\lambda$5129 absorption features, although the overall strengths of these lines vary from object to object.

\item 91T-like SNe are distinguished from 99aa-like and slow-declining Branch CN SNe by a nearly flat evolution of the \ion{Si}{2} $\lambda6355$ expansion 
velocity over phases ranging from $-10$ to $+10$~days with respect to $B$ maximum.  Nevertheless, over similar phases, the 
\ion{Si}{3} $\lambda$4560 line covers a much larger velocity range (from $-12,500$~\kms\ at $-12$~days to $-5,000$~\kms\ at $+6$~days
for SN~1991T), comparable to what is observed for 99aa-like and slow-declining Branch CN SNe.  This is confirmed by a comparison of the \ion{Si}{2} $\lambda6355$
profiles of SNe~1991T, 1999aa, and the Branch CN SN~2011fe which shows that, in all three cases, the blue wings extend to velocities of $\sim 20,000$~\kms.

\item The \ion{S}{2} $\lambda\lambda$5449,5622 lines of 91T-like SNe are only clearly visible at phases between $-5$ and $+5$~days, but
cover a similar velocity range observed for slow-declining Branch CN SNe over the same period of time.  Taken together with the previous point, the overall
impression is that the IMEs in 91T-like, 99aa-like, and slow-declining members of the Branch CN class are present at similar velocities in
the ejecta.  Hence, the weakness and flat velocity evolution of the \ion{Si}{2} $\lambda$6355 line in 91T-like SNe is explained not by the IMEs being confined
to a small velocity range in the ejecta, but rather by the higher
ionization of the outer ejecta in these objects, which also accounts for the enhanced strength of the \ion{Fe}{3} absorption.

\item The two optical nebular spectra of 91T-like SNe obtained to date
are closely similar to those of 99aa-like and slow-declining Branch CN SNe, with
the exception of the profile of the [\ion{Fe}{2}] and [\ion{Ni}{2}] features 
between 7000 and 7600~\AA\ where strong [\ion{Ca}{2}] $\lambda$7291,7324 emission
appears to also be present in the two 91T-like SNe.  The earlier appearance of
[\ion{Ca}{2}] emission in 91T-like SNe is perhaps explained by greater mixing of 
$^{56}$Ni into the IMEs.

\item In general, the NIR spectra of 91T-like, 99aa-like, and slow-declining Branch CN events such as SN~2011fe are very similar at
comparable epochs.  The major difference is the weakness of the \ion{Ca}{2} triplet and \ion{Mg}{2} $\lambda$1.0927~$\mu$m
absorption at maximum and pre-maximum epochs in the spectra of 91T-like SNe which,
again, is explained by higher ionization conditions in the ejecta.  Velocity measurements of the blue edge of the $H$-band 
break at approximately 10~days past maximum for the 99aa-like and 91T-like SNe observed by the CSP-II range from $-14,000$ to nearly $-16,000$~\kms, 
providing direct evidence of significant mixing of $^{56}$Ni into the outer ejecta.

\item 91T-like and 99aa-like SNe, as well as many slow-declining Branch CN SNe, have strikingly similar $B$ and $V$ 
light curves, and cannot be distinguished on the basis of photometric parameters such as \dm\ and \sbv.  However, 91T-like SNe show a larger diversity 
in the depth of the minimum between primary and secondary maxima of the $i$-band light curves.
This behavior is consistent with model predictions that greater mixing of $^{56}$Ni
into the region of the IMEs has the effect of advancing the secondary maximum by a few
days and to decrease its contrast with respect to the primary maximum. 

\item The $(B-V)$ and $(r-i)$ color evolution of the 91T-like SNe from $-10$ to $+7$~days is generally similar to that of the 99aa-like and slow-declining 
Branch CN SNe.  The largest differences are observed in $(r-i)$ from $+10$ to $+25$~days, reflecting the larger dispersion in the 
depth of the minimum between the primary and secondary maxima of the $i$-band light curves of the 91T-like SNe.

\item Much greater differences in light curve morphology are observed at UV wavelengths, with the 91T-like and 99aa-like events displaying
considerably broader pre-maximum UV light curves compared to the slow-declining Branch CN SNe.  The data also suggest that 91T-like SNe reach maximum
in the UV a few days earlier than do the 99a-like and slow-declining Branch CN SNe.  Post maximum, the differences in the UV light curves are less pronounced.
The broader UV light curves and earlier peaks are indicative of higher temperatures, most likely caused by the greater outward mixing of $^{56}$Ni.

\item As expected, differences in the pre-maximum UV$-$optical color evolution between the 91T-like, 99aa-like, and slow-declining Branch CN SNe are also 
much more dramatic than in $(B-V)$ and $(r-i)$.  The blue UV$-$optical colors of the 91T-like and 99aa-like events compared to the slow-declining Branch CN SNe
are likely related to the differences in the slope of the $(B-V)$ color evolution in the first few days following explosion found by \citet{stritzinger18} in
comparing Branch SS and CN SNe.

\item Hubble diagram residuals for the CSP-II samples of 91T-like, 99aa-like, and slow-declining Branch CN SNe provide clear evidence
that 91T-like SNe are over-luminous by $\sim$0.1--0.6~mag compared to 99aa-like and slow-declining Branch CN SNe.  This difference in luminosity is
remarkably constant from optical to NIR wavelengths, arguing that errant host galaxy dust corrections are not
to blame.  The data further suggest that 99aa-like events may be intermediate in luminosity between 91T-like SNe and slow-declining Branch CN
SNe with similar decline rates.

\item A pseudo-bolometric light curve integrated from $2000$--$9000$~\AA\ 
of the 91T-like LSQ12gdj indicates that the peak luminosity of this event was
$\sim$2--5 times greater, and was reached $\sim$4 days
earlier, than the maximum pseudo-bolometric luminosities of the 99aa-like iPTF14bdn, and the slow-declining Branch
CN SN ASASSSN-14lp.  This is considerably larger than the differences
in peak luminosity observed at optical and NIR wavelengths, providing yet further evidence of the greater extent of mixing of $^{56}$Ni into the outer ejecta of 91T-like SNe.

\item Like most luminous, slow-declining SNe~Ia, 91T-like events occur preferentially
in star-forming galaxies.
The cumulative distributions of the host masses of the 91T-like and slow-declining Branch CN SNe observed by the CSP-II are indistiguishable from that of the full CSP-II sample.
While there is some evidence that 99aa-like SNe may prefer hosts
with somewhat lower masses, the distribution for Branch SS (91T-like + 99aa-like) SNe cannot be distinguished from that of the 
full CSP-II sample.

\item IFU observations of the local host environment indicate that the cumulative distributions of the SFR density,
$\Sigma_{{\rm SFR}}$, and the specific SFR, eSFR, of the CSP-II 91T-like, 99aa-like,
Branch SS, and slow-declining Branch CN SNe are indistinguishable from the distributions of the same parameters for the PISCO 
sample of SNe Ia. The same conclusion applies to the cumulative histograms of the H$\alpha$ EW values of the 
CSP-II 91T-like, 99aa-like, and Branch SS SNe, but the subsample of 
slow-declining Branch CN SNe is characterized by a median H$\alpha$ EW that
is twice that of the PISCO sample.  The local host galaxy oxygen abundances of
the CSP-II 91T-like, Branch SS, and slow-declining Branch CN SNe are all consistent
with the distribution of abundances of the PISCO sample, but this hypothesis is
rejected at the 95\% confidence level when considering the 99aa-like SNe alone.
The strength of these conclusions is unfortunately limited by the small numbers
of SNe for which information on the local host environment is available, as well as the inclusion of a significant number of
targeted SNe in the PISCO sample.

\item Taken together as a group, 91T-like SNe may show somewhat stronger host galaxy \ion{Na}{1}~D absorption in their spectra
than do the 99aa-like and slow-declining Branch CN SNe, but this conclusion is also weakened by the small sample sizes, particularly in
the case of the 91T-like SNe. 

\item Three of the best-observed SNe~Ia-CSM at early phases when the ejecta--CSM interaction was weakest displayed spectra that were
identical to spectra of SN~1991T at similar phases.

\end{itemize}

Based on these findings, it could be argued that 91T-like SNe are not ``peculiar'' in the sense of being an intrinsically distinct phenomenon, 
but rather are connected to the larger population of luminous, slow-declining SNe~Ia.
The main properties that distinguish the 91T-like SNe
from the 99aa-like and slow-declining Branch CN events are the higher ionization observed in their outer layers, their greater peak luminosities, and their broader UV light curves, all of which are related to the amount of $^{56}$Ni produced in the explosion and the extent to which it is mixed outward into the ejecta.

Spectropolarimetric observations provide further evidence that the Branch SS SNe (i.e., 91T-like and 99aa-like events) are closely related to 
the slow-declining Branch CN SNe.  The continuum polarization observed for most SNe~Ia is low \citep[$\sim$0.3\%;][]{cikota19}, indicating global 
asphericities of 10\% or less \citep{chornock08}.
Nevertheless, polarization in lines such as \ion{Ca}{2} and \ion{Si}{2} can be significant, especially before maximum light, reaching values as large as 
$\sim$1\% \citep[e.g., see][]{wang08,patat17,cikota19}\footnote{As paraphrased by \citet{patat17}, SNe~Ia ``are globally spherical explosions 
[that] show chemical asymmetries''.}.
From a study of 17 SNe~Ia, \citet{wang07} showed that the polarization of the \ion{Si}{2} $\lambda$6355 line corrected to $t_{B_{max}} = -5$~d was highly
correlated with the decline rate parameter \dm\ for ``spectroscopically normal'' events, with the slowest-declining (Branch CN and SS) SNe showing the least
 \ion{Si}{2} polarization.  In the context of DDT models, these authors interpreted this result as evidence that these SNe have suffered more complete nuclear burning,
 thus erasing most of the chemical inhomogeneities produced by the initial deflagration phase.

 Viewing angle effects can be considered using early time \ion{Si}{2} velocity gradients and nebular phase line velocity shifts
to examine the distribution of material along the observer’s line of sight.
The \ion{Si}{2} $\lambda6355$ ``velocity plateau'' \citep{scalzo12,scalzo14b} and low-velocity gradient
classification \citep{benetti05} at early times are inconsistent with observing the outer layers of the ejecta at 
a ``steep'' viewing angle. As shown in Figure~\ref{fig:nebular_IR}, the nebular phase spectrum of 
SN~1991T has a fairly symmetric [\ion{Fe}{2}] 1.644~$\mu$m and only slightly blueshifted emission line profile indicating
that the iron-group elements located around the geometric center of the SN ejecta.
The combination of low polarization with early- and late-time line velocity evolution suggests that 91T-like SNe Ia do not exhibit the asymmetries expected from a significantly off-center ignition \citep{maeda10}. 

In spite of the large amount of data presented in this paper
in support of the idea that 91T-like SNe are simply the extreme members of the
general population of luminous SNe~Ia, 
the association of 91T-like events with the
rare SNe~Ia-CSM may be difficult to reconcile with the original \citet{nugent95} conclusion 
that the progression from normal to 91T-like SNe is purely a temperature effect.  
In the majority of the 
SNe~Ia-CSM discovered to date, the CSM interaction began too early to be able to
clearly detect spectral features of the underlying SN 
\citep[e.g., see][]{silverman13b}.
However, in a recent study by \citet{sharma23} of 12 SNe~Ia-CSM discovered in the 
course of the Bright Transient Survey of the Zwicky Transit Factory (ZTF),
underlying SN~Ia absorption features were observed for three SNe.  
One of the three is SN~2018evt, whose observations were discussed in
\S\ref{sec:11kx}, while the other two are SN~2020qxz and SN~2020aekp.
In all three cases, the first spectrum obtained revealed strong \ion{Fe}{3}
absorption and weak or missing \ion{Si}{2} $\lambda$6355 consistent with a 91T-like
classification.

The host galaxies of SNe~Ia-CSM have a similar range of masses and absolute
magnitudes as the star-forming hosts of the general population of luminous 
SNe~Ia \citep{sharma23}.
However, 91T-like events are rare in the local universe as shown in 
a recent study by \citet{desai24}.  From the $V$-band light curves
of 404 SNe~Ia discovered by ASAS-SN from 2014-2017, these authors derived a volumetric rate for 91T-like SNe of $8.5~_{-1.6}^{+1.7} \times 
10^2$~yr$^{-1}$~Gpc$^{-3}$~h$^3_{70}$, which translates to 
$\sim4$\% of all
SNe~Ia that obey the peak luminosity versus decline rate relation.
In the same paper, a rate for SNe~Ia-CSM of 
$10~_{-7}^{+7}$~yr$^{-1}$~Gpc$^{-3}$~h$^3_{70}$ was estimated.
These numbers imply that 91T-like SNe 
outnumber SNe~Ia-CSM by a factor of $85\pm62$, and we thus
might conclude that only a small fraction of 91T-like SNe have CSM interactions.
However, the case of the 91T-like SN~2015cp, for which
the ejecta--CSM interaction was not initially observed but was visible two years 
after maximum  \citep{graham19}, begs the question of
whether a CSM interaction might eventually be detectable for many (or even all)
91T-like SNe if observations were to continue indefinitely\footnote{Unfortunately, only a single optical spectrum of SN~2015cp 
was obtained at a phase of $+45$~days, precluding a comparison of its spectral evolution at earlier epochs with
other 91T-like SNe.  Indeed, because of the late phase of this spectrum, \citeauthor{graham19} could not completely rule out a match with a ``normal SN Ia.''}.
\citeauthor{graham19}
estimated that the percentage of luminous SNe~Ia with late-onset CSM interaction
was 6\%.  More recently, \citet{dubay22} derived rates 
from Galaxy Evolution Explorer \citep[GALEX;][]{martin05} light
curves of a much larger sample of 1,080 SNe~Ia covering a range of luminosities.  These authors
concluded that $\lesssim 5.1$\% of SNe~Ia undergo late-onset (between 0 and 500 days 
after discovery) CSM interactions as strong as that
observed for PTF11kx, and that weaker interactions 
such as observed for SN~2015cp occur $\lesssim 16$\%.
Thus, it is conceivable (although far from proven) that the ``91T-like'' and ``Ia-CSM'' subtypes 
refer to one and the same population of SNe~Ia.
Nevertheless, \citet{badenes08} concluded that the observed X-ray emission of the 
supernova remnant 0509--67.5 in the Large Magellanic Cloud, whose light echo spectrum
is best matched by a 91T-like SN \citep{rest08}, is
consistent with interaction with an assumed uniform ambient medium of density $10^{-24}~{\rm g~cm}^{-3}$,
placing strong constraints on the density and location of any CSM associated with this SN.

The purpose of this paper has been to present a comprehensive review 
of the UV-optical-NIR properties of 91T-like SNe.  In a companion paper 
(Hoeflich et al., in preparation), these observations will be used to study possible 
models for the progenitors and explosion mechanism(s).  It is worth remarking, however, 
that the accumulation of evidence to date points to the progenitors of the SNe~Ia-CSM being
single-degenerate systems --- i.e., with a main-sequence or evolved,
non-degenerate companion\footnote{\citet{shen13} have pointed out that hydrogen-rich material is also
ejected prior to explosion in double-degenerate systems consisting of a He white dwarf and a
C-O white dwarf, but this most likely occurs hundreds or thousands of years before the SN 
explodes.}.
Except in the case of the recently-described 
SN~2020eyj which showed a helium-rich CSM \citep{kool23}, the CSM for all
other SNe~Ia-CSM discovered to date appears to have been hydrogen-rich.
If 91T-like SNe actually
are the extreme members of luminous SNe Ia, then the implication is that
a significant fraction of, or perhaps even all, luminous SNe~Ia might also be the product of single-degenerate binary systems.
A number of studies have attempted to address the progenitor system question through observations and modelling of early-time light curves
with mixed conclusions \citep[e.g.,][]{fausnaugh21,burke22a,burke22b,magee22,deckers22,fausnaugh23}.  Some of these have suggested that
as many as $\sim$30\% of SNe~Ia show excess blue flux at very early epochs that may be consistent with interaction with a 
non-degenerate companion, and two of the best-observed of these SNe,
2017cbv \citep{hosseinzadeh17} and 2018oh \citep{dimitriadis19,shappee19}, were luminous, slow-declining events.
However, for the nearby, well-observed Branch CN SN~2011fe, early observations at optical and UV wavelengths have been used to place tight
constraints on the presence of a non-degenerate companion \citep{bloom12,brown12}.
Mixing of $^{56}$Ni into the outer ejecta or the presence of circumstellar material could also produce an early blue excess \citep{piro16},
complicating the interpretation of early light curves.
Perhaps the strongest argument against the single-degenerate scenario for normal, luminous SNe~Ia is the
absence of narrow ($\sim1000$~\kms) stripped-companion \ion{H}{1} emission in nebular-phase spectra \citep[e.g.,][]{mattila05,leonard07,shappee13,sand16,tucker20,sand21}. 

Thus, it is still not clear what the relationship is between normal luminous 
SNe~Ia and the 91T-like events.  This is an important question to resolve, not
only from the point of view of understanding the progenitor systems and
explosion mechanisms of SNe~Ia, but also for SN~Ia cosmology.  The association
of luminous SNe~Ia with host galaxies undergoing significant star formation 
implies that the decline rates of SNe~Ia are a function of the ages of the 
progenitor systems.  Indeed, the distribution of decline rates (or stretch) in
SN Ia populations has been shown to shift to events with brighter, slower light 
curves at high redshift \citep{howell07,nicolas21}.  While 91T-like SNe represent
a relatively small fraction of the {\em total} SN~Ia population in the local universe,
in the CSP-II sample they amounted to $\sim$~25\% of the luminous
SNe with (\dm~$\la 1.1$~mag).  If the C-O white dwarf progenitors of 91T-like SNe have
evolved from the high end of the possible range of main-sequence masses, then
91T-like SNe may well dominate the population of SNe~Ia in the early universe
($z \sim$~4--11) \citep{chakraborty23}. Either optical spectroscopy, UV light curves
(or spectra), or perhaps narrow-band light curves of features such as \ion{Si}{2} $\lambda6355$
or \ion{Ca}{2}~H~\&~K as \citep{nordin18,boone21} will be required to recognize them since, as emphasized in Paper~I and
by \citet{boone21}, \citet{yang22}, and, most recently, by \citet{bi24}, 91T-like SNe cannot
be identified on the basis of decline rate (or light curve shape) as measured at optical wavelengths.

From statistics and spectra extracted from the Transient
Name Server (TNS)\footnote{\url{https://www.wis-tns.org/}} between 2021--2023, when the ASAS-SN \citep{shappee14},
ATLAS \citep{tonry11},
and ZTF \citep{masci19} surveys were all active, we estimate that 91T-like SNe 
explode within 100~Mpc at a rate of only $\sim$2 per year.
The aforementioned volumetric rate for 91T-like SNe derived
from ASAS-SN alone by \citet{desai24} implies a similar
value of $\sim$1 per year within this
distance.
It is incumbent, therefore, to obtain
detailed X-ray, UV, optical, and infrared observations --- from the earliest moments
following explosion to nebular-phase epochs --- of future nearby 91T-like SNe
whenever the opportunity presents itself.
Photometric monitoring at very late epochs with the Vera Rubin Observatory 
to search for late-onset CSM interactions in these 91T-like events, as well as for all nearby, luminous SNe~Ia, will 
also be key to understanding the interrelations and  progenitors systems of these cosmologically-important objects.

\begin{acknowledgments}
The work of the CSP-I and CSP-II has been generously supported by the National Science Foundation under 
grants AST-0306969, AST-0607438, AST-1008343, AST-1613426, AST-1613455, and AST-1613472.
The CSP-II was also supported in part by the Danish Agency for Science and Technology and Innovation through a 
Sapere Aude Level 2 grant. M. Stritzinger acknowledges  funding by a research 
grant (13261) from VILLUM FONDEN, and a grant from the Independent Research Fund Denmark (IRFD, grant number  10.46540/2032-00022B).
E.B. is supported in part by NASA grant 80NSSC20K0538.
L.G. acknowledges financial support from the Spanish Ministerio de Ciencia e Innovaci\'on (MCIN) and the Agencia Estatal de Investigaci\'on (AEI) 10.13039/501100011033 under the PID2020-115253GA-I00 HOSTFLOWS project, from Centro Superior de Investigaciones Cient\'ificas (CSIC) under the PIE project 20215AT016 and the program Unidad de Excelencia Mar\'ia de Maeztu CEX2020-001058-M, and from the Departament de Recerca i Universitats de la Generalitat de Catalunya through the 2021-SGR-01270 grant.
We gratefully acknowledge the use of WISeREP (\url{https://wiserep.weizmann.ac.il}) and TNS (\url{https://www.wis-tns.org)},
and are especially thankful to Peter Meikle for providing us with the NIR
spectra of SN~1991T included in this paper.
Thanks also to Carles Badenes for reminding us of SNR~0509--67.5.
This research has made use of the \citet{ned19}, which is funded by the National 
Aeronautics and Space Administration and operated by the California Institute of Technology.
\end{acknowledgments}


\vspace{5mm}
\facilities{Magellan:Baade (IMACS imaging spectrograph, FourStar wide-field near-infrared camera, 
FIRE near-infrared echellette), Magellan:Clay (LDSS3 imaging spectrograph), Swope (SITe3 CCD imager, 
e2v 4K x 4K CCD imager), du~Pont (SITe2 CCD imager, Tek5 CCD imager, WFCCD imaging spectrograph, RetroCam 
near-infrared imager), Gemini:North (GNIRS near-infrared spectrograph), Gemini:South (FLAMINGOS2),
VLT (ISAAC, MUSE), IRTF (SpeX near-infrared spectrograph), NOT (ALFOSC), Calar Alto 3.5~m (PMAS/PPak),
La Silla-QUEST, CRTS, PTF, iPTF, OGLE, ASAS-SN, PS1, KISS, ISSP, MASTER, SMT}


\software{IRAF \citep{tody86},
SNID \citep{blondin07},
% GELATO \citep{harutyunyan08},
SNooPy \citep{burns11}}



\clearpage

\appendix

\restartappendixnumbering

\section{91T-like Supernovae from the Literature}
\label{sec:historical}

In this appendix, we consider a sample of nearby 91T-like SNe derived from the literature.  This sample is undoubtedly incomplete as the number
of discovered SNe has grown exponentially since the 1990's \citep[e.g., see Figure~2 of][]{stritzinger18}.  However, as far as we are aware, 
it includes most of the best-observed examples of the 91T-like phenomenon discovered through
2019.

\subsection{CfA}
\label{sec:cfa}

\begin{figure*}[t]
\epsscale{1.}
\plottwo{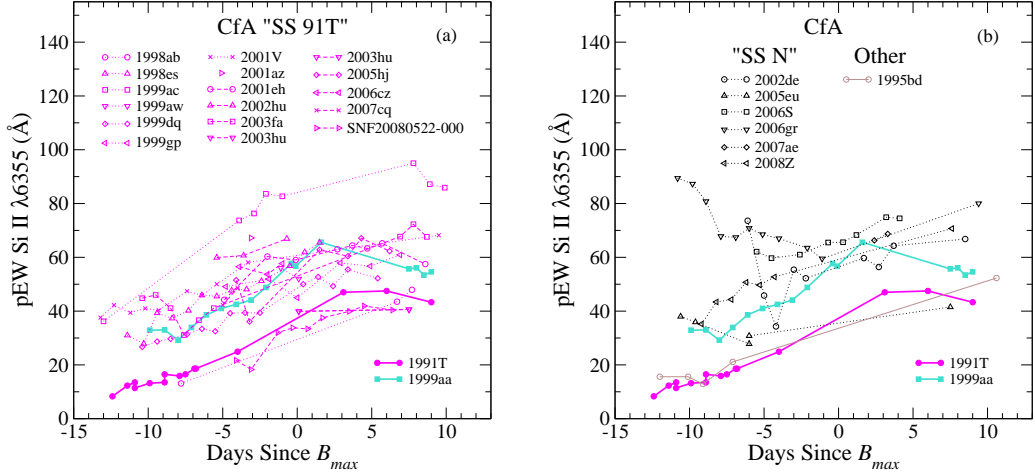}{FigA1b.eps}
\caption{Evolution of the pEW(\ion{Si}{2}~$\lambda6355$) measurements for Branch SS-class SNe~Ia in the \citet{blondin12} spectroscopic sample 
classified as Wang-class ``91T'' (a) and ``N'' (b).  The trajectories of SN~1991T and SN~1999aa in these diagram are plotted for reference.}
\label{fig:CfA}
\end{figure*}

One of the largest samples of SNe~Ia spectra published to date is from the Harvard-Smithsonian Center for Astrophysics (CfA) 
Supernova Group \citep{blondin12}.  These authors provided \citet{branch06} classifications for each supernova having a spectrum 
obtained near maximum, and also independently classified the SNe in the \citet{wang09} system consisting of four classes --- ``Normal (N)'', 
``HV'', ``91T-like'', and ``91bg-like''.  The Wang classes correspond reasonably closely to the \citet{branch06} CN, BL, SS, and CN 
subgroups, respectively, although the criterion used by \citet{wang09} to classify 91T-like events of ``weak \ion{Si}{2} absorption and prominent 
\ion{Fe}{3} lines in the near-maximum spectra'' is non-quantitative.  As a case in point, all SNe listed by \citet{blondin12} as 
Wang 91T-like also belong to the Branch~SS class, but some SNe classified as Branch~SS events were 
categorized as Wang~Normal.

Figure~\ref{fig:CfA} displays the time evolution of 
pEW(\ion{Si}{2}~$\lambda6355$) for the SNe~Ia classified as ``SS 91T'' and ``SS N'' in the \citet{blondin12} sample.  Shown for 
comparison are the trajectories of SN~1991T and SN~1999aa.  Clearly only a few SNe  --- SN~1999ab, SN~2003hu, and
SNF20080522-000 ---  are as extreme as SN~1991T.  The published $I$-band light curve of SNF20080522-000 shows a  
secondary maximum and a primary maximum that peaked before $B$ maximum \citep{scalzo12}, consistent with a 91T-like event.  
Unfortunately, the $I$-band photometry of SN~1999ab \citep{jha06} and SN~2003hu \citep{ganeshalingam10} does not allow us to 
determine if the primary maxima peaked before $B$ maximum, but we strongly suspect that both of these SNe were 91T-like.

\begin{figure*}[t]
\epsscale{1.}
\plottwo{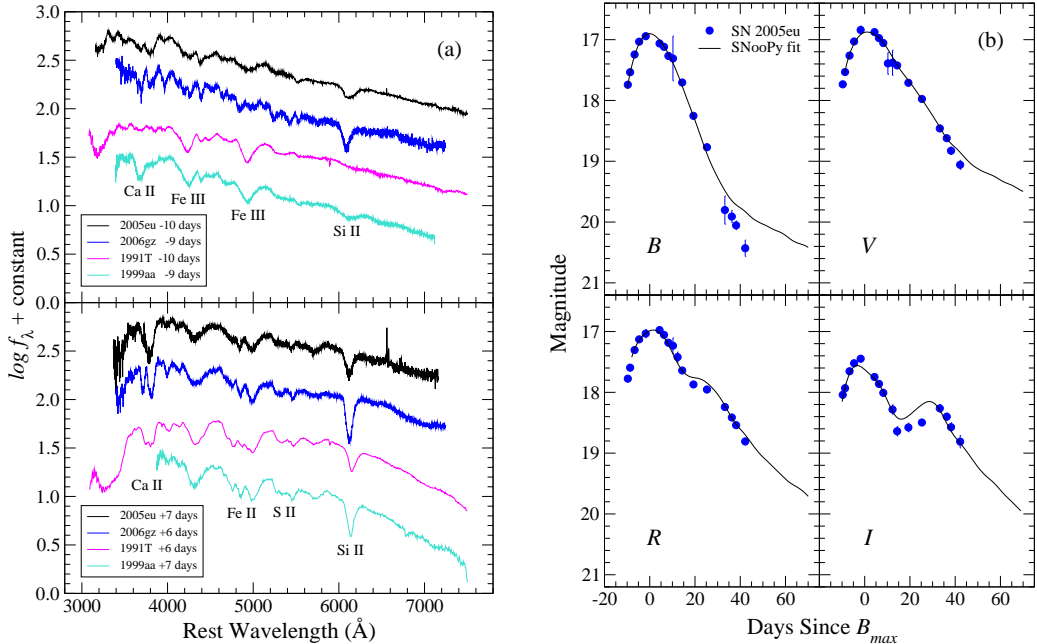}{FigA2b.eps}
\caption{(a) The spectrum of SN~2005eu at $-10$~days is compared with spectra at similar epochs of SN~1991T, SN~1999aa, and the 03fg-like 
SN~2006gz in the upper panel. In the lower panel, a similar comparison at approximately $+7$~days is shown.  (b) The $BVRI$ photometry of
SN~2005eu published by \citet{ganeshalingam10} is plotted along with best \texttt{SNooPy} fits.}
\label{fig:05eu_comparison}
\end{figure*}

The rest of the objects in the CfA sample are mostly consistent with SN~1999aa events with the exception of SN~2005eu.  
Pre-maximum pEW(\ion{Si}{2} $\lambda6355$) measurements of this SN resemble a 99aa-like event, but the 
single post-maximum spectrum obtained at a phase of $+7.5$ days has a pEW(\ion{Si}{2}~$\lambda6355$) consistent with 91T-like SNe.
Figure~\ref{fig:05eu_comparison}a shows spectra of SN~2005eu at phases of $-10$~days and $+7$~days with respect to the epoch
of $B$ maximum.  For comparison, spectra of SN~1991T and SN~1999aa at similar phases are plotted in the same figure.  At $-10$~days,
neither is a good match to SN~2005eu due to the absence of strong \ion{Fe}{3} absorption in the latter SN, while at $+7$~days,
SN~1991T provides a better match.  Also included in this figure are spectra of the 03fg-like SN~2006gz \citep{hicken07}, which resemble
the spectra of SN~2005eu at both epochs, except for somewhat stronger \ion{Si}{2}~$\lambda6355$ absorption.  
Light curves of SN~2005eu in $BVRI$ are displayed in Figure~\ref{fig:05eu_comparison}b.  The best-fitting \texttt{SNooPy} templates 
 do a poor job of matching the $B$ photometry beyond $+30$~days, and fail to fit the $I$-band light curve in general.  There is a hint that the $I$
light curve may have peaked after $B$ maximum; nevertheless, it displays a clear secondary maximum, 
whereas SN~2006gz --- like most other 03fg-like SNe --- was characterized by a broad $i$-band light curve with no clear secondary maximum.
Although it is unclear how to classify SN~2005eu, we conclude that it was most likely neither a 91T- or 99aa-like event. 

We also checked for additional 91T-like events among the SNe for which \citet{blondin12} 
did not provide Branch and Wang classes because no spectra were obtained within a few days of maximum.   One object, SN~1995bd, for 
which four spectra were obtained by \citeauthor{blondin12} between $-12$ and $-7$~days is clearly consistent with the 91T-like class 
(see Figure~\ref{fig:CfA}) and also meets the photometric requirement.  

\subsection{CSP-I}
\label{sec:cspI}

\begin{figure*}[t]
\epsscale{1.}
\plottwo{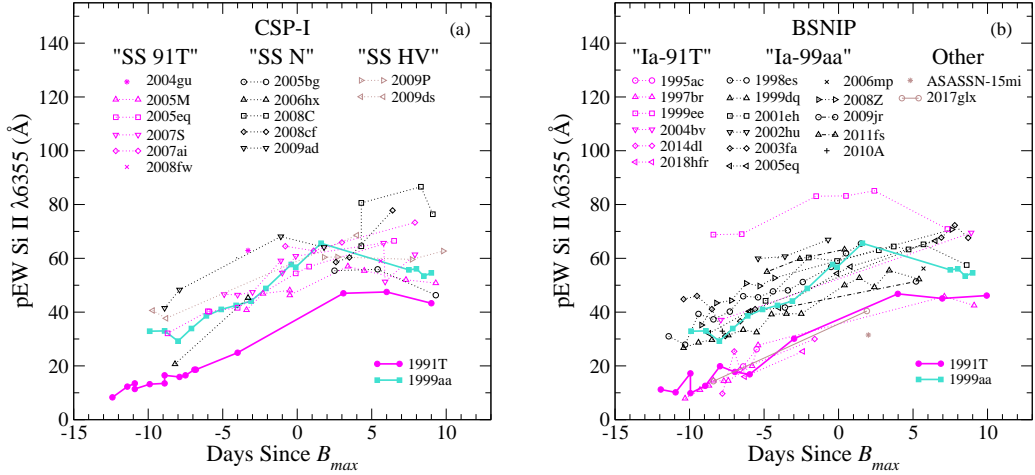}{FigA3b.eps}
\caption{(a) Same as Figure~\ref{fig:CfA}, but for Branch SS-class SNe~Ia in the \citet{folatelli13} spectroscopic sample.  (b)
Same as Figure~\ref{fig:CfA}, but for SNe~Ia in the \citet{silverman12} spectroscopic sample which these authors classified 
as 91T-like and 99aa-like.}
\label{fig:CSPI_BSNIP}
\end{figure*}

Spectra of a sample of 126 SNe~Ia observed by the CSP-I have been published by \citet{folatelli13} and \citet{morrell24}. Like 
\citet{blondin12},  these authors provided both \citet{branch06} and \citet{wang09} classifications
for each supernova having a spectrum obtained near maximum.  The temporal evolution of pEW(\ion{Si}{2}~$\lambda6355$) for the 
SNe~Ia classified as Branch~SS in the CSP-I sample is shown in Figure~\ref{fig:CSPI_BSNIP}a.  Note
that two of the CSP-I SNe are classified as both ``SS'' and \citet{wang09} ``HV'', demonstrating further that there is not a perfect 
one-to-one correspondence between the Branch ``SS'' and \citeauthor{wang09} ``91T'' classes.  None
of the SNe in the CSP-I sample are as extreme as SN~1991T, although for one object --- SN~2006hx --- the classification is
ambiguous.  The pEW(\ion{Si}{2}~$\lambda6355$) measurement obtained from the $-8$~day spectrum of this SN lies squarely 
in the 91T-like portion of the diagram, whereas a spectrum taken five days later gives a measurement that is consistent with
a 99aa-like SN.  The signal-to-noise ratio of the first spectrum is low, and may suffer some host galaxy contamination.  The second
spectrum obtained at $-3$~days shows strong \ion{Ca}{2} absorption more typical of a 99aa-like event.  We therefore do not consider
this SN to be a bona fide 91T-like SN.
 
We also checked for additional 91T-like events among the SNe in the \citet{folatelli13} and \citet{morrell24} samples for which 
Branch and Wang classes were not provided, but none were identified. 

\subsection{BSNIP}
\label{sec:bsnip}

A large sample of SNe~Ia spectra has been published by the Berkeley Supernova~Ia Program \citep[BSNIP;][]{silverman12,stahl20}.
These authors used SNID \citep{blondin07} to provide automated spectral classification of the SNe in their sample.  The 
standard SNID subgroups for SNe~Ia are ``Ia-norm'', ``Ia-91T'', ``Ia-91bg'', ``Ia-csm'', and ``Ia-pec''.  However, \citet{silverman12}
recognized that the default SNID spectral templates for the Ia-91T category includes both 91T-like and 99aa-like events, and so in
order to separate these two classes, they created their own custom set of spectral templates for use with SNID.  In 
Figure~\ref{fig:CSPI_BSNIP}b, the time evolution of pEW(\ion{Si}{2}~$\lambda6355$) is plotted for the SNe~Ia in the
\citet{silverman12} sample classified as Ia-99aa and Ia-91T.  Here the discrimination between the two
subtypes is much better than for the CfA and CSP-I samples, with two events (SN~1995ac and 
SN~1997br\footnote{\citet{li99} published extensive spectroscopic and photometric data for SN~1997br.  This object was the first 
91T-like SN to be recognized after SN~1991T itself.  The primary maximum of the $I$-band light curve also peaked before the epoch of 
$B$~maximum, and a clear secondary maximum in this filter was also observed \citep{li99}.}) clearly qualifying as 91T-like.  
Nevertheless, two events classified by \citet{silverman12} as Ia-91T --- SN~1999ee and SN~2004bv --- do not resemble SN~1991T in this 
diagram.  These authors also classified two further SNe --- SN~2001eu and SN~2003K --- as Ia-91T events.  
However, the signal-to-noise ratio of their single spectrum of SN~2001eu is low and the epoch is unknown since there is no light curve 
available.  In the case of SN~2003K, the first spectrum was obtained 13.4~days after $B$~maximum when the differences between 91T-like, 
99aa-like, and slow-declining Branch CN SNe such as SN~1999ee are subtle.  Hence, we consider the Ia-91T 
classifications assigned by these authors to these two SNe to be uncertain.

 \citet{stahl20} classified three SNe --- 2013dj, 2014dl, and 2018hfr --- as Ia-91T using the same custom SNID templates 
 as \citet{silverman12}.  SN~2013dj was discovered on 2013~June~10.7 (UT) \citep{lipunov13}, with no object visible at the same 
 position $\sim$1~month before.  We measured pseudo-equivalent widths of 32~\AA\ and 39~\AA, respectively, for the 
 \ion{Si}{2}~$\lambda6355$ and \ion{Ca}{2}~H~\&~K
 absorption features in the spectrum obtained by \citet{stahl20} on 2013-June~14.4 (UT), that 
  are consistent with either SN~1991T at $\sim$4~days before
 $B$~maximum, or SN~1999aa at ~$\sim$10~days before maximum.  A light curve is therefore required to resolve this ambiguity.  Fortunately,
 SN~2013dj was independently discovered by the La Silla-QUEST survey \citep{baltay13} as LSQ13avx, and was observed to 
 reach maximum light on approximately 2013~June~24 (UT) (D. Rabinowitz, private communication).  The \citet{stahl20}
 spectrum therefore corresponds to an epoch of $-10$~days, plus or minus a few days, in which case SN~2013dj was likely a 99aa-like 
 event.  An unpublished spectrum obtained near the date of maximum confirms this conclusion (P. Nugent, private communication).  SN~2014dl
 is a bona fide 91T-like event as confirmed by CSP-II photometry (see Paper~I).   \citet{stahl20} obtained three spectra of 
 SN~2018hfr (ASASSN-18xn), all of which were best matched by spectra of the 91T-like template, SN~1997br.  The date of maximum implied by the three
 SNID fits of 2018~October~19.5 (UT) is consistent with the publicly-available ASAS-SN \citep{shappee14,kochanek17} $g$ and $V$
 light curves, confirming that SN~2018hfr is likely to have been a 91T-like event.  However, since an $i/I$-band light is unavailable,
 we cannot rule out the possibility that it was a 03fg-like SN.

We also checked those SNe in the \cite{silverman12}
sample classified as Ia-norm.  This search yielded only one event, SN~2004br, which should correctly be classified as 91T-like.  The 
photometry of this SN obtained by \citet{ganeshalingam10} shows that the $I$-band light curve had a clear secondary maximum and 
a primary maximum that peaked a few days before the $B$ band.  The single spectrum published by \citeauthor{silverman12} was taken 
at a phase of $+3.5$~days, and closely resembles a spectrum of SN~1991T obtained at $+3.1$~days. 

The pEW(\ion{Si}{2} $\lambda6355$) measurements given in Table~4 of \citet{stahl20} were searched for unrecognized 
91T-like SNe.  We found two candidates --- ASASSN-15mi and SN~2017glx --- which meet our criteria for
classification as 91T-like events.  \citet{stahl20} obtained a single spectrum of ASASSN-15mi at a phase of $+2$~days that
is a good match for the $+3.1$ day spectrum of SN~1991T and the light curves published
by \citet{foley18} meet the photometric requirement.  The spectrum of SN~2017glx taken at $+1.9$~days as well as the spectrum
of \citet{zhang17} obtained at $-8.5$~days are both consistent with the 91T-like classification,
as are the light curves published by \citet{stahl19}.

\subsection{Other Sources}
\label{sec:others}

\begin{figure*}[t]
\epsscale{0.8}
\plotone{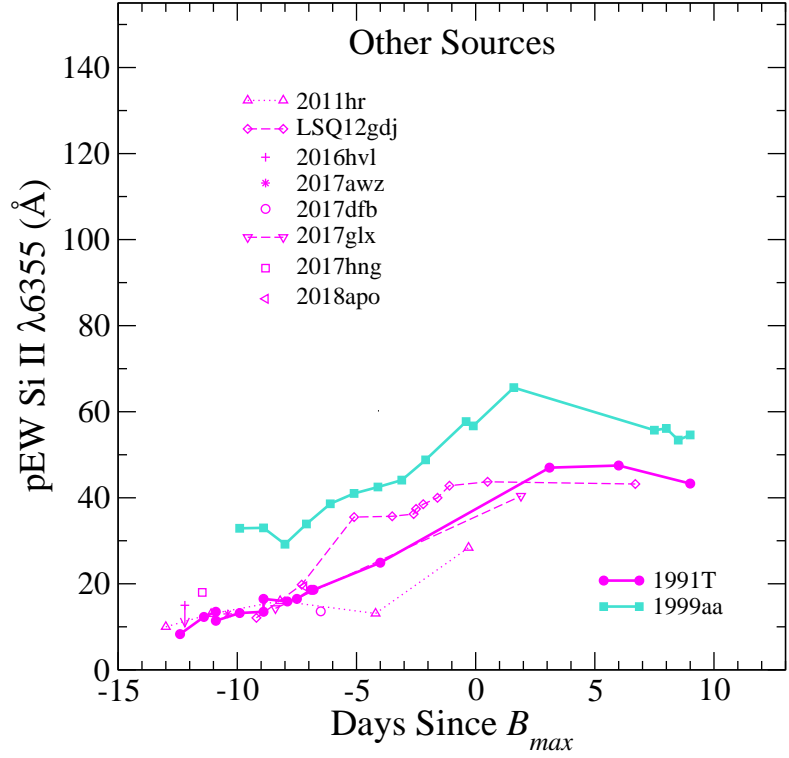}
\caption{Same as Figure~\ref{fig:CfA}, but for 91T-like events taken from other sources in the literature.}
\label{fig:literature_other}
\end{figure*}

A number of other SNe~Ia have been identified as possible members of the 91T-like class from classification spectra or in the literature.  In many cases, 
these SNe should more properly be classified as 99aa-like events based on the pseudo-equivalent widths of the \ion{Ca}{2}~H~\&~K and 
\ion{Si}{2} $\lambda6355$ absorption features.  However, we have been able to confirm  91T-like classifications for ten additional SNe, and for two that we suspect were also
91T-like events. 
The temporal evolution of pEW(\ion{Si}{2}~$\lambda6355$) for eight of these SNe is shown in Figure~\ref{fig:literature_other}, and 
each is briefly described in the remainder of this section.

\begin{itemize}

\item SNF20070803-005 and SNF20080723-012:  \citet{scalzo12} published spectra and light curves of these two SNe~Ia which they suspected to be
03fg-like events.  Unfortunately, the spectral data are not publicly available, and so we are unable to measure the evolution of the \ion{Si}{2}~$\lambda6355$ 
pseudo-equivalent widths.  However, as \citet{scalzo12} showed in Figure~3 of their paper, the spectra of these two SNe obtained approximately a 
week before and a week after maximum closely resembled spectra of SN~1991T and the 91T-like SNF20080522+000 (also observed by the CfA group) 
 at similar epochs.  As opposed to 03fg-like SNe, the $I$-band light curves of SNF20070803-005 and SNF20080723-012 plotted in Figure~1 of  
\citet{scalzo12} displayed strong secondary maxima, and primary maxima that peaked before $B$ maximum.  Hence, these two SNe qualify 
 as 91T-like events.  Two other objects in the \citet{scalzo12} paper --- SNF20070528-003 and SNF20070912-000 --- also have spectra that closely
 resemble 91T-like SNe.  Unfortunately, $I$-band light curves are not available for either SN, and so although we suspect that they also
 were 91T-like events, this cannot be confirmed with the data in hand.

\item SN~2011hr:  Observations of this object were presented by \citet{zhang16a} who pointed out its resemblance to SN~1991T, but who also suggested
that it might be a transitional object between 91T-like and 03fg-like SNe.  Our measurements of pEW(\ion{Si}{2}~$\lambda6355$) from $-13$~days to 
maximum, plotted in 
Figure~\ref{fig:literature_other}, are consistent with the trajectory of SN~1991T.  The photometry published by \citet{zhang16a} reveals
a clear secondary maximum in the $I$~filter light curve, and is also consistent with the primary maximum in $I$ peaking before $B$-band maximum.

\item LSQ12gdj:  This SN was observed extensively by \citet{scalzo14b}, who concluded that it was ``spectroscopically 91T-like'', and was
included in the CSP-II sample of 91T-like events (see Paper~I).  Figure~A1 of Paper~I
shows that the evolution of the pseudo-equivalent width of \ion{Si}{2}~$\lambda6355$ was more similar to SN~1991T than SN~1999aa.  The light curves
plotted in Figure~2 of \citet{scalzo14b} also clearly show that LSQ12gdj had a strong secondary maximum in the $i$ band, and that primary maximum
in $i$ was reached before $B$~maximum. 

\item SN~2016hvl:  This SN was discovered by ATLAS and classified by PESSTO \citep{smartt15} as ``similar to SN 1999aa'' 
 at a phase of $-10$ to $-8$~days \citep{dimitriadis16}.  Published light curves provide a date of $B$~maximum of MJD~57711 \citep{stahl19},
implying that the PESSTO spectrum was obtained at a phase of $-12$~days.  \ion{Si}{2}~$\lambda6355$ absorption is too weak to be detected
 in this spectrum, from which we estimate an upper limit of pEW(\ion{Si}{2}~$\lambda6355$)~$\leq$~15~\AA~(see Figure~\ref{fig:literature_other}).  \ion{Ca}{2}~H~\&~K
absorption is also undetected, consistent with a 91T-like event.  Two more spectra of this SN are available on WISeREP \citep{yaron12}, but 
unfortunately at phases of $+15$ and $+103$ when differentiating between 91T-like and 99aa-like SNe is unreliable.  The photometry of
\citep{stahl19} is of excellent quality, and shows that there was a clear secondary maximum in $I$, with primary maximum occurring before
$B$~maximum.  It is highly probable, therefore, that SN~2016hvl was a 91T-like SN.

\item SN~2017awz: This is another SN discovered by ATLAS and classified by PESSTO with the comment ``best SNID match with SN~Ia 91T-like 
1998es at $-10$~d'' \citep{barbarino17}\footnote{As demonstrated in \S\ref{sec:cfa}, SN~1998es was actually a 99aa-like event.}.  Photometry published by \citet{yang22}
confirms that the $i$-band light curve had an obvious secondary maximum, with the primary maximum occurring 
before $B$~maximum.  The PESSTO spectrum, which shows very weak 
\ion{Si}{2}~$\lambda6355$ absorption (see Figure~\ref{fig:literature_other}) and little, if any, absorption of \ion{Ca}{2}~H~\&~K, is fully consistent with that of a 91T-like event.

\item SN~2017dfb: Discovered by the ASAS-SN group \citep{stanek17}, this SN was classified
by \citet{tomasella17} as being consistent with a 91T-like event from a spectrum
obtained with the Asiago 1.82~m Copernico Telescope.  These authors called attention
to the strong interstellar \ion{Na}{1}~D absorption in the spectrum (see Figure~\ref{fig:historical_91T_spec}) suggesting significant host galaxy reddening.

\item SN~2017hng: This SN was discovered by the PMO-Tsinghua Transient Survey survey
\citep{xu17}, and confirmed spectroscopically by \citet{zhang17b} who remarked that the
spectrum ``[matched] that of a young 91T-like SN Ia at about -9 days after maximum.''
Light curves published by \citet{yang22} show that the $i$-band light curve
peaked before $B$~maximum and displayed a clear secondary maximum.  The pEW(\ion{Si}{2} $\lambda6355$) measurement is also fully consistent with the 91T-like
classification (see Figure~\ref{fig:literature_other}).

\item SN~2018apo: Discovered by ASAS-SN \citep{chen18}, this SN was classified by the PESSTO
collaboration as a ``young SNe Ia of the bright sub-types 91T and 99aa'' \citep{malesani18}.
The pEW(\ion{Si}{2} $\lambda6355$) measurement obtained from this spectrum is fully consistent with a 91T-like
classification (see Figure~\ref{fig:literature_other}).  \citet{yang22} published $gri$ photometry, with the
$i$ filter clearly displaying a secondary maximum. However, a $B$-band light curve is not available, and
so we cannot confirm that the primary $i$-band maximum occurred before $B$ maximum.  Hence, we classify
 this SN as a suspected 91T-like event.

 \item SN~2018eay: This SN was discovered by ZTF \citep{fremling18}.  \citet{yin18}
 reported that the spectrum was that of a young SN~Ia, similar to SN~1991T, with 
 $E(B-V) \sim 0.5$~mag of host galaxy reddening.  Light curves obtained by ZTF
 indicate that this spectrum was obtained approximately a week before $g$ maximum.  The
 weak \ion{Ca}{2}~H~\&~K and \ion{Si}{2}~$\lambda6355$ observed in the spectrum at
 this epoch confirms the 91T-like classification. The ZTF $r$-band light curve
shows a clear post-maximum "shoulder" as would be expected for a 91T-like event, but 
since a $B$-band light curve is not available, we list this object as a suspected 91T-like
SN.

\item SN~2019gwa: This SN was also discovered by ZTF \citep{nordin19} and classified by the Spectroscopic Classification of Astronomical Transients \citep[SCAT;][]{tucker22b} Survey in
\citet{do19} as a 91T-like SN with an age of $-7.4 \pm 2$~days. A 
spectrum obtained at nearly the same epoch by the PESSTO collaboration \citep{frohmaier19}
shows weak or absent \ion{Ca}{2}~H~\&~K and \ion{Si}{2}~$\lambda6355$ absorption with
pEW $\sim 16$~\AA, consistent with a 91T-like SN caught approximately one week before $B$~maximum.  Unfortunately, the photometry published
by \citet{yang22} does not include the $i$ filter, and so we are unable to rule out
the possibility that this SN was an 03fg-like event.  However, the ZTF $r$-band light curve
shows a clear post-maximum "shoulder" as would be expected for a 91T-like event, and so we
strongly suspect that this SN was, indeed, a true 91T-like object.

\end{itemize}

\subsection{Final List of 91T-like SNe from the Literature}
\label{sec:final_list}

Table~\ref{tab:historical_91T}  provides host galaxy names, 
redshifts, the decline rate parameters, \dm\ and \sbv,
and references to the published spectroscopy and photometry of the literature sample of 91T-like SNe discussed in this appendix.

\clearpage

\begin{deluxetable*}{lllllll}
\tabletypesize{\scriptsize}
\tablecolumns{7}
\tablewidth{0pt}
\tablecaption{1991T-like SNe from the Literature\label{tab:historical_91T}}
\tablehead{
 \colhead{SN Name} &
\colhead{Host Galaxy} &
\colhead{$z_{\rm{helio}}$\tablenotemark{a}} &
\colhead{\dm \tablenotemark{b}} &
\colhead{\sbv \tablenotemark{c}} &
\colhead{Spectroscopy\tablenotemark{d}} &
\colhead{Photometry\tablenotemark{e}}
}
\startdata
\sidehead {{\bf Confirmed}}
\hline
SN1991T                     &   NGC 4527                                        &   $0.0058$   &  0.93 (0.02)        &  1.21 (0.01)   & 1,2,3     & 1$^\dagger$,2$^\dagger$,4    \\
SN1995ac                   &   2MFGC 17122                                 &   $0.0499$   &  0.86 (0.04)         &  1.19 (0.03)   &  5,6       & 7    \\
SN1995bd                   &   UGC 3151                                        &  $0.0146$   &  0.88 (0.02)         &   1.10 (0.04)   &  5,6        & 7    \\
SN1997br                    &    ESO 576-40                                    &   $0.0070$   &  1.12 (0.03)         &  1.19 (0.04)  &   5,6,8,9 & 8,10,11    \\
SN2004br                    &  NGC 4493                                         &   $0.0231$   &  0.77 (0.04)         &  1.12 (0.02)    &  9           & 12 \\
SNF20070803-005      &  GALEXASC J222623.66+211506.0  &   $0.0317$   &  0.85 (0.02)        &  1.17 (0.02)  &  13       & 13  \\
SNF20080522-000      &  SDSS J133647.59+050833.0           &   $0.0453$   &  0.82 (0.04)          &  1.08 (0.06)  &  5,6,13    & 13,14$^\dagger$,15$^\dagger$  \\
SNF20080723-012      &   anonymous                                      &   $0.0745$   &   1.38 (0.03)     &  1.21 (0.03)  &  13       & 13 \\
%PTF11kx                      &  WISEA J080913.19+461843.5         &   $0.0466$   &   \nodata                &   \nodata        &  16      &  16  \\
SN2011hr                    &   NGC 2691                                       &   $0.0134$   &  0.83 (0.02)   &   1.20 (0.07)  &  16       & 16  \\
LSQ12gdj                    &   ESO 472-G007                                &   $0.0303$   &  0.73 (0.03)         &  1.14 (0.05)   &  17,18   &  19$^\dagger$,20  \\
SN2014dl                    &   UGC 10414                                      &   $0.0330$  &   1.05 (0.03)$^*$  &  1.22 (0.05)  &  20,21,22    & 20   \\
ASASSN-15mi            &  MRK 0283a                                       &  $0.0345$   &   1.04 (0.10)$^*$   &  1.19 (0.05)   & 22        & 23 \\
%SN2015bq                   &   KUG 1232+315                                &   $0.0282$   &  0.88 (0.04)         &  1.23 (0.03)  &  20         &  21   \\
SN2016hvl                  &   UGC 3524                                        &   $0.0131$   &  1.04 (0.01)         &  0.90 (0.03)   &  24        &  20$^\dagger$,25,26,27  \\
SN2017awz                &  SDSS J110735.46+225104.2           &   $0.0219$   &  0.94 (0.04)   &  1.19 (0.04)  &  28         &  19$^\dagger$,23,27  \\
SN2017dfb                &  ARK 481                            &   $0.0259$   &    1.01 (0.03)$^*$              &  0.98 (0.03)         &  29           & 27                \\
SN2017glx                  &   NGC 6824                                        &   $0.0118$   &  0.87 (0.03)      &   1.19 (0.03)  &  22,30    &  19$^\dagger$,25,27    \\
SN2017hng                &  2MASX J04214029-0332267        &   $0.0445$  &  0.80 (0.03)$^*$     &   1.26 (0.03)        &        31,32       &  27               \\
\hline
\sidehead {{\bf Suspected}}
\hline
SN1998ab &  NGC 4704  &  0.0272  &  1.06 (0.03)$^*$ &  1.19 (0.03)  &  5  &  11  \\
SN2003hu                   &   2MASX J19113272+7753382          &   $0.0750$   &   0.80 (0.08)$^*$   &   1.29 (0.09)   &   4,5      & 33  \\
SNF20070528-003      &   anonymous                                      &   $0.1170$    &  1.04 (0.05)         &  1.32 (0.04)  &  13       & 13   \\ 
SNF20070912-000      &  GALEXASC J000436.74+180912.2  &   $0.1230$   &  1.09 (0.09)       &  1.33 (0.06)  &  13       & 13  \\ 
SN2018apo                &  ESO 268-G037                   &   $0.0163$       &  0.82 (0.03)$^*$      &   1.13 (0.03)          &  34           &  27               \\
SN2018eay               &  IC 1286                          &  $0.0185$        &  \nodata             &  \nodata          &  35             &  36   \\
SN2018hfr                  &  2MASX J09305509-0434173             &   $0.0226$   &  \nodata            &  \nodata        &  22,37       &  38  \\
SN2019gwa                &  SDSS J155841.10+111425.5                  &   $0.0550$       &   0.93 (0.01)   &   1.10 (0.05)            &   39        &  27               \\
%\hline
%\sidehead {{\bf Ambiguous}}
%\hline
%%SN2005eu                   &   2MASS J02274330+2810378          &   $0.0349$   &  0.90 (0.05)      &  1.08 (0.04)   &   5,6,9  & 11,12$^\dagger$,13  \\
%SN2013dj                    &   UGC 10535                                      &   $0.0253$    &  \nodata           &  \nodata        &  21       & \nodata   \\ 
\enddata
\tablenotetext{a}{Heliocentric redshifts are from the \citet{ned19}  except for the hosts of SNF20080723-012 and SNF20070528-003 whose redshifts are from \citet{childress13b}.}
\tablenotetext{b}{\dm~decline rate in magnitudes \citep{phillips93} as measured with SNooPy. The 1$\sigma$ error is given in parentheses.
An asterisk indicates values derived from template fits only.}
\tablenotetext{c}{\sbv~color stretch \citep{burns14} as measured with SNooPy template fits. The 1$\sigma$ error is given in between parentheses.}
\tablenotetext{d}{Spectroscopy references.}
\tablenotetext{e}{Photometry references.
The $\dagger$~symbol indicates data not used in the SNooPy fits.
}
\tablerefs{
(1) \citet{filippenko92b}; 
(2) \citet{phillips92};
(3) \citet{ruiz-lapuente92};
(4) \citet{lira98};
(5) \citet{blondin12};
(6) \citet{matheson08};
(7) \citet{riess99};
(8) \citet{li99};
(9) \citet{silverman12};
(10) \citet{altavilla04}
(11) \citet{jha06};
(12) \citet{ganeshalingam10};
(13) \citet{scalzo12};
(14) \citet{friedman15};
(15) \citet{hicken12};
(16) \citet{zhang16a};
(17) \citet{maguire12};
(18) \citet{scalzo14b};
%(20) \citet{zhang15};
(19) Swift Optical/Ultraviolet Supernova Archive \citep{brown14};
(20) This paper;
(21) \citet{drake14};
(22) \citet{stahl20}
(23) \citet{foley18};
%(24) CPCS Alert 26431;
(24) \citet{dimitriadis16};
(25) \citet{stahl19};
(26) \citet{baltay21};
(27) \citet{yang22};
(28) \citet{barbarino17};
(29) \citet{tomasella17};
(30) \citet{zhang17};
(31) \citet{zhang17b};
(32) \citet{floers17};
(33) \citet{hicken09};
(34) \citet{malesani18};
(35) \citet{yin18}
(36) \url{https://alerce.online/object/ZTF18abgmcmv}
(37) \citet{kuncarayakti18};
(38) ASASSN Photometry Database \citep{shappee14,kochanek17};
(39) \citet{frohmaier19}
}
\end{deluxetable*}

\clearpage

\bibliography{main}{}

\begin{thebibliography}{}
\expandafter\ifx\csname natexlab\endcsname\relax\def\natexlab#1{#1}\fi
\providecommand{\url}[1]{\href{#1}{#1}}
\providecommand{\dodoi}[1]{doi:~\href{http://doi.org/#1}{\nolinkurl{#1}}}
\providecommand{\doeprint}[1]{\href{http://ascl.net/#1}{\nolinkurl{http://ascl.net/#1}}}
\providecommand{\doarXiv}[1]{\href{https://arxiv.org/abs/#1}{\nolinkurl{https://arxiv.org/abs/#1}}}

\bibitem[{{Aldering} {et~al.}(2006){Aldering}, {Antilogus}, {Bailey}, {Baltay},
  {Bauer}, {Blanc}, {Bongard}, {Copin}, {Gangler}, {Gilles}, {Kessler},
  {Kocevski}, {Lee}, {Loken}, {Nugent}, {Pain}, {P{\'e}contal}, {Pereira},
  {Perlmutter}, {Rabinowitz}, {Rigaudier}, {Scalzo}, {Smadja}, {Thomas},
  {Wang}, {Weaver}, \& {Nearby Supernova Factory}}]{aldering06}
{Aldering}, G., {Antilogus}, P., {Bailey}, S., {et~al.} 2006, \apj, 650, 510,
  \dodoi{10.1086/507020}

\bibitem[{{Altavilla} {et~al.}(2004){Altavilla}, {Fiorentino}, {Marconi},
  {Musella}, {Cappellaro}, {Barbon}, {Benetti}, {Pastorello}, {Riello},
  {Turatto}, \& {Zampieri}}]{altavilla04}
{Altavilla}, G., {Fiorentino}, G., {Marconi}, M., {et~al.} 2004, \mnras, 349,
  1344, \dodoi{10.1111/j.1365-2966.2004.07616.x}

\bibitem[{{Aouad} {et~al.}(2022){Aouad}, {Mazzali}, {Hachinger}, {Teffs},
  {Pian}, {Ashall}, {Benetti}, {Filippenko}, \& {Tanaka}}]{aouad22}
{Aouad}, C.~J., {Mazzali}, P.~A., {Hachinger}, S., {et~al.} 2022, \mnras, 515,
  4445, \dodoi{10.1093/mnras/stac2024}

\bibitem[{{Ashall} {et~al.}(2019){Ashall}, {Hsiao}, {Hoeflich}, {Stritzinger},
  {Phillips}, {Morrell}, {Davis}, {Baron}, {Piro}, {Burns}, {Contreras},
  {Galbany}, {Holmbo}, {Kirshner}, {Krisciunas}, {Marion}, {Sand},
  {Shahbandeh}, {Suntzeff}, \& {Taddia}}]{ashall19}
{Ashall}, C., {Hsiao}, E.~Y., {Hoeflich}, P., {et~al.} 2019, \apjl, 875, L14,
  \dodoi{10.3847/2041-8213/ab1654}

\bibitem[{{Ashall} {et~al.}(2020){Ashall}, {Lu}, {Burns}, {Hsiao},
  {Stritzinger}, {Suntzeff}, {Phillips}, {Baron}, {Contreras}, {Davis},
  {Galbany}, {Hoeflich}, {Holmbo}, {Morrell}, {Karamehmetoglu}, {Krisciunas},
  {Kumar}, {Shahbandeh}, \& {Uddin}}]{ashall20}
{Ashall}, C., {Lu}, J., {Burns}, C., {et~al.} 2020, \apjl, 895, L3,
  \dodoi{10.3847/2041-8213/ab8e37}

\bibitem[{{Bacon} {et~al.}(2010){Bacon}, {Accardo}, {Adjali}, {Anwand},
  {Bauer}, {Biswas}, {Blaizot}, {Boudon}, {Brau-Nogue}, {Brinchmann},
  {Caillier}, {Capoani}, {Carollo}, {Contini}, {Couderc}, {Daguis{\'e}},
  {Deiries}, {Delabre}, {Dreizler}, {Dubois}, {Dupieux}, {Dupuy}, {Emsellem},
  {Fechner}, {Fleischmann}, {Fran{\c{c}}ois}, {Gallou}, {Gharsa}, {Glindemann},
  {Gojak}, {Guiderdoni}, {Hansali}, {Hahn}, {Jarno}, {Kelz}, {Koehler},
  {Kosmalski}, {Laurent}, {Le Floch}, {Lilly}, {Lizon}, {Loupias}, {Manescau},
  {Monstein}, {Nicklas}, {Olaya}, {Pares}, {Pasquini}, {P{\'e}contal-Rousset},
  {Pell{\'o}}, {Petit}, {Popow}, {Reiss}, {Remillieux}, {Renault}, {Roth},
  {Rupprecht}, {Serre}, {Schaye}, {Soucail}, {Steinmetz}, {Streicher}, {Stuik},
  {Valentin}, {Vernet}, {Weilbacher}, {Wisotzki}, \& {Yerle}}]{bacon10}
{Bacon}, R., {Accardo}, M., {Adjali}, L., {et~al.} 2010, in Society of
  Photo-Optical Instrumentation Engineers (SPIE) Conference Series, Vol. 7735,
  Ground-based and Airborne Instrumentation for Astronomy III, ed. I.~S.
  {McLean}, S.~K. {Ramsay}, \& H.~{Takami}, 773508, \dodoi{10.1117/12.856027}

\bibitem[{{Badenes} {et~al.}(2008){Badenes}, {Hughes}, {Cassam-Chena{\"\i}}, \&
  {Bravo}}]{badenes08}
{Badenes}, C., {Hughes}, J.~P., {Cassam-Chena{\"\i}}, G., \& {Bravo}, E. 2008,
  \apj, 680, 1149, \dodoi{10.1086/524700}

\bibitem[{{Baltay} {et~al.}(2013){Baltay}, {Rabinowitz}, {Hadjiyska}, {Walker},
  {Nugent}, {Coppi}, {Ellman}, {Feindt}, {McKinnon}, {Horowitz}, \&
  {Effron}}]{baltay13}
{Baltay}, C., {Rabinowitz}, D., {Hadjiyska}, E., {et~al.} 2013, \pasp, 125,
  683, \dodoi{10.1086/671198}

\bibitem[{{Baltay} {et~al.}(2021){Baltay}, {Grossman}, {Howard}, {Rabinowitz},
  {Arcavi}, {Barbour}, {Burke}, {Contreras}, {Dilday}, {Graham}, {Hiramatsu},
  {Hossenzadeh}, {Howell}, {McCully}, {McKinnon}, {Ment}, {Montesi},
  {Pellegrino}, \& {Valenti}}]{baltay21}
{Baltay}, C., {Grossman}, L., {Howard}, R., {et~al.} 2021, \pasp, 133, 044002,
  \dodoi{10.1088/1538-3873/abd417}

\bibitem[{{Barbarino} {et~al.}(2017){Barbarino}, {Nyholm}, {Taddia},
  {Fremling}, {Sollerman}, {Botticella}, {Fraser}, {Inserra}, {Kankare},
  {Maguire}, {Smartt}, {Smith}, {Wright}, {Young}, {Sullivan}, {Valenti},
  {Yaron}, {Manulis}, {Stalder}, {Chambers}, {Denneau}, {Flewelling}, {Heinze},
  {Huber}, {Magnier}, {Tonry}, {Waters}, {Wainscoat}, {Weiland}, \&
  {Rest}}]{barbarino17}
{Barbarino}, C., {Nyholm}, A., {Taddia}, F., {et~al.} 2017, The Astronomer's
  Telegram, 10094, 1

\bibitem[{{Baron} {et~al.}(2008){Baron}, {Jeffery}, {Branch}, {Bravo},
  {Garc{\'\i}a-Senz}, \& {Hauschildt}}]{baron08}
{Baron}, E., {Jeffery}, D.~J., {Branch}, D., {et~al.} 2008, \apj, 672, 1038,
  \dodoi{10.1086/524009}

\bibitem[{{Benetti} {et~al.}(2006){Benetti}, {Cappellaro}, {Turatto},
  {Taubenberger}, {Harutyunyan}, \& {Valenti}}]{benetti06}
{Benetti}, S., {Cappellaro}, E., {Turatto}, M., {et~al.} 2006, \apjl, 653,
  L129, \dodoi{10.1086/510667}

\bibitem[{{Benetti} {et~al.}(2005){Benetti}, {Cappellaro}, {Mazzali},
  {Turatto}, {Altavilla}, {Bufano}, {Elias-Rosa}, {Kotak}, {Pignata}, {Salvo},
  \& {Stanishev}}]{benetti05}
{Benetti}, S., {Cappellaro}, E., {Mazzali}, P.~A., {et~al.} 2005, \apj, 623,
  1011, \dodoi{10.1086/428608}

\bibitem[{{Bi} {et~al.}(2024){Bi}, {Woods}, \& {Fabbro}}]{bi24}
{Bi}, C., {Woods}, T.~E., \& {Fabbro}, S. 2024, arXiv e-prints,
  arXiv:2401.06087, \dodoi{10.48550/arXiv.2401.06087}

\bibitem[{{Blondin} \& {Tonry}(2007)}]{blondin07}
{Blondin}, S., \& {Tonry}, J.~L. 2007, \apj, 666, 1024, \dodoi{10.1086/520494}

\bibitem[{{Blondin} {et~al.}(2006){Blondin}, {Dessart}, {Leibundgut}, {Branch},
  {H{\"o}flich}, {Tonry}, {Matheson}, {Foley}, {Chornock}, {Filippenko},
  {Sollerman}, {Spyromilio}, {Kirshner}, {Wood-Vasey}, {Clocchiatti},
  {Aguilera}, {Barris}, {Becker}, {Challis}, {Covarrubias}, {Davis},
  {Garnavich}, {Hicken}, {Jha}, {Krisciunas}, {Li}, {Miceli}, {Miknaitis},
  {Pignata}, {Prieto}, {Rest}, {Riess}, {Salvo}, {Schmidt}, {Smith}, {Stubbs},
  \& {Suntzeff}}]{blondin06}
{Blondin}, S., {Dessart}, L., {Leibundgut}, B., {et~al.} 2006, \aj, 131, 1648,
  \dodoi{10.1086/498724}

\bibitem[{{Blondin} {et~al.}(2012){Blondin}, {Matheson}, {Kirshner}, {Mandel},
  {Berlind}, {Calkins}, {Challis}, {Garnavich}, {Jha}, {Modjaz}, {Riess}, \&
  {Schmidt}}]{blondin12}
{Blondin}, S., {Matheson}, T., {Kirshner}, R.~P., {et~al.} 2012, \aj, 143, 126,
  \dodoi{10.1088/0004-6256/143/5/126}

\bibitem[{{Bloom} {et~al.}(2012){Bloom}, {Kasen}, {Shen}, {Nugent}, {Butler},
  {Graham}, {Howell}, {Kolb}, {Holmes}, {Haswell}, {Burwitz}, {Rodriguez}, \&
  {Sullivan}}]{bloom12}
{Bloom}, J.~S., {Kasen}, D., {Shen}, K.~J., {et~al.} 2012, \apjl, 744, L17,
  \dodoi{10.1088/2041-8205/744/2/L17}

\bibitem[{{Boone} {et~al.}(2021){Boone}, {Aldering}, {Antilogus}, {Aragon},
  {Bailey}, {Baltay}, {Bongard}, {Buton}, {Copin}, {Dixon}, {Fouchez},
  {Gangler}, {Gupta}, {Hayden}, {Hillebrandt}, {Kim}, {Kowalski},
  {K{\"u}sters}, {L{\'e}get}, {Mondon}, {Nordin}, {Pain}, {Pecontal},
  {Pereira}, {Perlmutter}, {Ponder}, {Rabinowitz}, {Rigault}, {Rubin}, {Runge},
  {Saunders}, {Smadja}, {Suzuki}, {Tao}, {Taubenberger}, {Thomas}, \&
  {Vincenzi}}]{boone21}
{Boone}, K., {Aldering}, G., {Antilogus}, P., {et~al.} 2021, \apj, 912, 71,
  \dodoi{10.3847/1538-4357/abec3b}

\bibitem[{{Bowers} {et~al.}(1997){Bowers}, {Meikle}, {Geballe}, {Walton},
  {Pinto}, {Dhillon}, {Howell}, \& {Harrop-Allin}}]{bowers97}
{Bowers}, E.~J.~C., {Meikle}, W.~P.~S., {Geballe}, T.~R., {et~al.} 1997,
  \mnras, 290, 663, \dodoi{10.1093/mnras/290.4.663}

\bibitem[{{Branch}(1981)}]{branch81}
{Branch}, D. 1981, \apj, 248, 1076, \dodoi{10.1086/159237}

\bibitem[{{Branch}(1987)}]{branch87}
---. 1987, \apjl, 316, L81, \dodoi{10.1086/184897}

\bibitem[{{Branch}(2001)}]{branch01}
---. 2001, \pasp, 113, 169, \dodoi{10.1086/318614}

\bibitem[{{Branch} {et~al.}(2006){Branch}, {Dang}, {Hall}, {Ketchum},
  {Melakayil}, {Parrent}, {Troxel}, {Casebeer}, {Jeffery}, \&
  {Baron}}]{branch06}
{Branch}, D., {Dang}, L.~C., {Hall}, N., {et~al.} 2006, \pasp, 118, 560,
  \dodoi{10.1086/502778}

\bibitem[{{Brimacombe} {et~al.}(2018){Brimacombe}, {Castro}, {Clocchiatti},
  {Nicholls}, {Vallely}, {Stanek}, {Kochanek}, {Brown}, {Shields}, {Thompson},
  {Shappee}, {Holoien}, {Prieto}, {Bersier}, {Dong}, {Bose}, {Chen},
  {Stritzinger}, {Holmbo}, {Bock}, {Cacella}, {Krannich}, {Carballo}, {Farfan},
  \& {Stone}}]{brimacombe18}
{Brimacombe}, J., {Castro}, N., {Clocchiatti}, A., {et~al.} 2018, The
  Astronomer's Telegram, 11963, 1

\bibitem[{{Brown} {et~al.}(2014){Brown}, {Breeveld}, {Holland}, {Kuin}, \&
  {Pritchard}}]{brown14}
{Brown}, P.~J., {Breeveld}, A.~A., {Holland}, S., {Kuin}, P., \& {Pritchard},
  T. 2014, \apss, 354, 89, \dodoi{10.1007/s10509-014-2059-8}

\bibitem[{{Brown} \& {Crumpler}(2020)}]{brown20}
{Brown}, P.~J., \& {Crumpler}, N.~R. 2020, \apj, 890, 45,
  \dodoi{10.3847/1538-4357/ab66b3}

\bibitem[{{Brown} {et~al.}(2015){Brown}, {Roming}, \& {Milne}}]{brown15}
{Brown}, P.~J., {Roming}, P. W.~A., \& {Milne}, P.~A. 2015, Journal of High
  Energy Astrophysics, 7, 111, \dodoi{10.1016/j.jheap.2015.04.007}

\bibitem[{{Brown} {et~al.}(2010){Brown}, {Roming}, {Milne}, {Bufano},
  {Ciardullo}, {Elias-Rosa}, {Filippenko}, {Foley}, {Gehrels}, {Gronwall},
  {Hicken}, {Holland}, {Hoversten}, {Immler}, {Kirshner}, {Li}, {Mazzali},
  {Phillips}, {Pritchard}, {Still}, {Turatto}, \& {Vanden Berk}}]{brown10}
{Brown}, P.~J., {Roming}, P. W.~A., {Milne}, P., {et~al.} 2010, \apj, 721,
  1608, \dodoi{10.1088/0004-637X/721/2/1608}

\bibitem[{{Brown} {et~al.}(2012){Brown}, {Dawson}, {de Pasquale}, {Gronwall},
  {Holland}, {Immler}, {Kuin}, {Mazzali}, {Milne}, {Oates}, \&
  {Siegel}}]{brown12}
{Brown}, P.~J., {Dawson}, K.~S., {de Pasquale}, M., {et~al.} 2012, \apj, 753,
  22, \dodoi{10.1088/0004-637X/753/1/22}

\bibitem[{{Bulla} {et~al.}(2020){Bulla}, {Miller}, {Yao}, {Dessart}, {Dhawan},
  {Papadogiannakis}, {Biswas}, {Goobar}, {Kulkarni}, {Nordin}, {Nugent},
  {Polin}, {Sollerman}, {Bellm}, {Coughlin}, {Dekany}, {Golkhou}, {Graham},
  {Kasliwal}, {Kupfer}, {Laher}, {Masci}, {Porter}, {Rusholme}, \&
  {Shupe}}]{bulla20}
{Bulla}, M., {Miller}, A.~A., {Yao}, Y., {et~al.} 2020, \apj, 902, 48,
  \dodoi{10.3847/1538-4357/abb13c}

\bibitem[{{Burke} {et~al.}(2022{\natexlab{a}}){Burke}, {Howell}, {Sand}, \&
  {Hosseinzadeh}}]{burke22b}
{Burke}, J., {Howell}, D.~A., {Sand}, D.~J., \& {Hosseinzadeh}, G.
  2022{\natexlab{a}}, arXiv e-prints, arXiv:2208.11201,
  \dodoi{10.48550/arXiv.2208.11201}

\bibitem[{{Burke} {et~al.}(2022{\natexlab{b}}){Burke}, {Howell}, {Sand},
  {Amaro}, {Brown}, {Andrews}, {Bostroem}, {Dong}, {Haislip}, {Hiramatsu},
  {Hosseinzadeh}, {Kouprianov}, {Lundquist}, {McCully}, {Pellegrino},
  {Reichart}, {Tartaglia}, {Valenti}, \& {Yang}}]{burke22a}
{Burke}, J., {Howell}, D.~A., {Sand}, D.~J., {et~al.} 2022{\natexlab{b}}, arXiv
  e-prints, arXiv:2207.07681, \dodoi{10.48550/arXiv.2207.07681}

\bibitem[{{Burns} {et~al.}(2011){Burns}, {Stritzinger}, {Phillips}, {Kattner},
  {Persson}, {Madore}, {Freedman}, {Boldt}, {Campillay}, {Contreras},
  {Folatelli}, {Gonzalez}, {Krzeminski}, {Morrell}, {Salgado}, \&
  {Suntzeff}}]{burns11}
{Burns}, C.~R., {Stritzinger}, M., {Phillips}, M.~M., {et~al.} 2011, \aj, 141,
  19, \dodoi{10.1088/0004-6256/141/1/19}

\bibitem[{{Burns} {et~al.}(2014){Burns}, {Stritzinger}, {Phillips}, {Hsiao},
  {Contreras}, {Persson}, {Folatelli}, {Boldt}, {Campillay}, {Castell{\'o}n},
  {Freedman}, {Madore}, {Morrell}, {Salgado}, \& {Suntzeff}}]{burns14}
---. 2014, \apj, 789, 32, \dodoi{10.1088/0004-637X/789/1/32}

\bibitem[{{Cadonau} {et~al.}(1985){Cadonau}, {Sandage}, \&
  {Tammann}}]{cadonau85}
{Cadonau}, R., {Sandage}, A., \& {Tammann}, G.~A. 1985, in Supernovae as
  Distance Indicators, ed. N.~{Bartel}, Vol. 224, 151,
  \dodoi{10.1007/3-540-15206-7_56}

\bibitem[{{Candia} {et~al.}(2003){Candia}, {Krisciunas}, {Suntzeff},
  {Gonz{\'a}lez}, {Espinoza}, {Leiton}, {Rest}, {Smith}, {Cuadra}, {Tavenner},
  {Logan}, {Snider}, {Thomas}, {West}, {Gonz{\'a}lez}, {Gonz{\'a}lez},
  {Phillips}, {Hastings}, \& {McMillan}}]{candia03}
{Candia}, P., {Krisciunas}, K., {Suntzeff}, N.~B., {et~al.} 2003, \pasp, 115,
  277, \dodoi{10.1086/368229}

\bibitem[{{Cao} {et~al.}(2014){Cao}, {Perley}, {Kasliwal}, {Johansson},
  {Goobar}, {Horesh}, {Sagiv}, {Ofer}, {Ofek}, {Nugent}, \& {Fox}}]{cao14}
{Cao}, Y., {Perley}, D., {Kasliwal}, M., {et~al.} 2014, The Astronomer's
  Telegram, 6175, 1

\bibitem[{{Cardelli} {et~al.}(1989){Cardelli}, {Clayton}, \&
  {Mathis}}]{cardelli89}
{Cardelli}, J.~A., {Clayton}, G.~C., \& {Mathis}, J.~S. 1989, \apj, 345, 245,
  \dodoi{10.1086/167900}

\bibitem[{{Chakraborty} {et~al.}(2023){Chakraborty}, {Sadler}, {Hoeflich},
  {Hsiao}, {Phillips}, {Burns}, {Diamond}, {Dominguez}, {Galbany}, {Uddin},
  {Ashall}, {Krisciunas}, {Kumar}, {Mera}, {Morrell}, {Baron}, {Contreras},
  {Stritzinger}, \& {Suntzeff}}]{chakraborty23}
{Chakraborty}, S., {Sadler}, B., {Hoeflich}, P., {et~al.} 2023, arXiv e-prints,
  arXiv:2311.03473, \dodoi{10.48550/arXiv.2311.03473}

\bibitem[{{Chen} {et~al.}(2018){Chen}, {Dong}, \& {Stanek}}]{chen18}
{Chen}, P., {Dong}, S., \& {Stanek}, K. 2018, Transient Name Server Discovery
  Report, 2018-440, 1

\bibitem[{{Childress} {et~al.}(2013{\natexlab{a}}){Childress}, {Aldering},
  {Antilogus}, {Aragon}, {Bailey}, {Baltay}, {Bongard}, {Buton}, {Canto},
  {Cellier-Holzem}, {Chotard}, {Copin}, {Fakhouri}, {Gangler}, {Guy}, {Hsiao},
  {Kerschhaggl}, {Kim}, {Kowalski}, {Loken}, {Nugent}, {Paech}, {Pain},
  {Pecontal}, {Pereira}, {Perlmutter}, {Rabinowitz}, {Rigault}, {Runge},
  {Scalzo}, {Smadja}, {Tao}, {Thomas}, {Weaver}, \& {Wu}}]{childress13b}
{Childress}, M., {Aldering}, G., {Antilogus}, P., {et~al.} 2013{\natexlab{a}},
  \apj, 770, 107, \dodoi{10.1088/0004-637X/770/2/107}

\bibitem[{{Childress} {et~al.}(2013{\natexlab{b}}){Childress}, {Scalzo}, {Sim},
  {Tucker}, {Yuan}, {Schmidt}, {Cenko}, {Silverman}, {Contreras}, {Hsiao},
  {Phillips}, {Morrell}, {Jha}, {McCully}, {Filippenko}, {Anderson}, {Benetti},
  {Bufano}, {de Jaeger}, {Forster}, {Gal-Yam}, {Le Guillou}, {Maguire},
  {Maund}, {Mazzali}, {Pignata}, {Smartt}, {Spyromilio}, {Sullivan}, {Taddia},
  {Valenti}, {Bayliss}, {Bessell}, {Blanc}, {Carson}, {Clubb}, {de Burgh-Day},
  {Desjardins}, {Fang}, {Fox}, {Gates}, {Ho}, {Keller}, {Kelly}, {Lidman},
  {Loaring}, {Mould}, {Owers}, {Ozbilgen}, {Pei}, {Pickering}, {Pracy}, {Rich},
  {Schaefer}, {Scott}, {Stritzinger}, {Vogt}, \& {Zhou}}]{childress13}
{Childress}, M.~J., {Scalzo}, R.~A., {Sim}, S.~A., {et~al.} 2013{\natexlab{b}},
  \apj, 770, 29, \dodoi{10.1088/0004-637X/770/1/29}

\bibitem[{{Chornock} \& {Filippenko}(2008)}]{chornock08}
{Chornock}, R., \& {Filippenko}, A.~V. 2008, \aj, 136, 2227,
  \dodoi{10.1088/0004-6256/136/6/2227}

\bibitem[{{Chugai}(2008)}]{chugai08}
{Chugai}, N.~N. 2008, Astronomy Letters, 34, 389,
  \dodoi{10.1134/S1063773708060030}

\bibitem[{{Cid Fernandes} {et~al.}(2005){Cid Fernandes}, {Mateus}, {Sodr{\'e}},
  {Stasi{\'n}ska}, \& {Gomes}}]{cid05}
{Cid Fernandes}, R., {Mateus}, A., {Sodr{\'e}}, L., {Stasi{\'n}ska}, G., \&
  {Gomes}, J.~M. 2005, \mnras, 358, 363,
  \dodoi{10.1111/j.1365-2966.2005.08752.x}

\bibitem[{{Cid Fernandes} {et~al.}(2009){Cid Fernandes}, {Schoenell}, {Gomes},
  {Asari}, {Schlickmann}, {Mateus}, {Stasinska}, {Sodr{\'e}}, {Torres-Papaqui},
  \& {Seagal Collaboration}}]{cid09}
{Cid Fernandes}, R., {Schoenell}, W., {Gomes}, J.~M., {et~al.} 2009, in Revista
  Mexicana de Astronomia y Astrofisica Conference Series, Vol.~35, Revista
  Mexicana de Astronomia y Astrofisica Conference Series, 127--132,
  \dodoi{10.48550/arXiv.0802.0849}

\bibitem[{{Cikota} {et~al.}(2019){Cikota}, {Patat}, {Wang}, {Wheeler}, {Bulla},
  {Baade}, {H{\"o}flich}, {Cikota}, {Clocchiatti}, {Maund}, {Stevance}, \&
  {Yang}}]{cikota19}
{Cikota}, A., {Patat}, F., {Wang}, L., {et~al.} 2019, \mnras, 490, 578,
  \dodoi{10.1093/mnras/stz2322}

\bibitem[{{Contreras} {et~al.}(2018){Contreras}, {Phillips}, {Burns}, {Piro},
  {Shappee}, {Stritzinger}, {Baltay}, {Brown}, {Conseil}, {Klotz}, {Nugent},
  {Turpin}, {Parker}, {Rabinowitz}, {Hsiao}, {Morrell}, {Campillay},
  {Castell{\'o}n}, {Corco}, {Gonz{\'a}lez}, {Krisciunas}, {Ser{\'o}n},
  {Tucker}, {Walker}, {Baron}, {Cain}, {Childress}, {Folatelli}, {Freedman},
  {Hamuy}, {Hoeflich}, {Persson}, {Scalzo}, {Schmidt}, \&
  {Suntzeff}}]{contreras18}
{Contreras}, C., {Phillips}, M.~M., {Burns}, C.~R., {et~al.} 2018, \apj, 859,
  24, \dodoi{10.3847/1538-4357/aabaf8}

\bibitem[{{Cooke} {et~al.}(2011){Cooke}, {Ellis}, {Sullivan}, {Nugent},
  {Howell}, {Gal-Yam}, {Lidman}, {Bloom}, {Cenko}, {Kasliwal}, {Kulkarni},
  {Law}, {Ofek}, \& {Quimby}}]{cooke11}
{Cooke}, J., {Ellis}, R.~S., {Sullivan}, M., {et~al.} 2011, \apjl, 727, L35,
  \dodoi{10.1088/2041-8205/727/2/L35}

\bibitem[{{Deckers} {et~al.}(2022){Deckers}, {Maguire}, {Magee}, {Dimitriadis},
  {Smith}, {Sainz de Murieta}, {Miller}, {Goobar}, {Nordin}, {Rigault},
  {Bellm}, {Coughlin}, {Laher}, {Shupe}, {Graham}, {Kasliwal}, \&
  {Walters}}]{deckers22}
{Deckers}, M., {Maguire}, K., {Magee}, M.~R., {et~al.} 2022, {Constraining Type
  Ia supernova explosions and early flux excesses with the Zwicky Transient
  Factory}, \dodoi{10.1093/mnras/stac558}

\bibitem[{{DerKacy} {et~al.}(2020){DerKacy}, {Baron}, {Branch}, {Hoeflich},
  {Hauschildt}, {Brown}, \& {Wang}}]{derkacy20}
{DerKacy}, J.~M., {Baron}, E., {Branch}, D., {et~al.} 2020, \apj, 901, 86,
  \dodoi{10.3847/1538-4357/abae67}

\bibitem[{{Desai} {et~al.}(2024){Desai}, {Kochanek}, {Shappee}, {Jayasinghe},
  {Stanek}, {Holoien}, {Thompson}, {Ashall}, {Beacom}, {Do}, {Dong}, \&
  {Prieto}}]{desai24}
{Desai}, D.~D., {Kochanek}, C.~S., {Shappee}, B.~J., {et~al.} 2024, \mnras,
  \dodoi{10.1093/mnras/stae606}

\bibitem[{{Dhawan} {et~al.}(2015){Dhawan}, {Leibundgut}, {Spyromilio}, \&
  {Maguire}}]{dhawan15}
{Dhawan}, S., {Leibundgut}, B., {Spyromilio}, J., \& {Maguire}, K. 2015,
  \mnras, 448, 1345, \dodoi{10.1093/mnras/stu2716}

\bibitem[{{Diamond} {et~al.}(2015){Diamond}, {Hoeflich}, \&
  {Gerardy}}]{diamond15}
{Diamond}, T.~R., {Hoeflich}, P., \& {Gerardy}, C.~L. 2015, \apj, 806, 107,
  \dodoi{10.1088/0004-637X/806/1/107}

\bibitem[{{Diamond} {et~al.}(2018){Diamond}, {Hoeflich}, {Hsiao}, {Sand},
  {Sonneborn}, {Phillips}, {Hristov}, {Collins}, {Ashall}, {Marion},
  {Stritzinger}, {Morrell}, {Gerardy}, \& {Penney}}]{diamond18}
{Diamond}, T.~R., {Hoeflich}, P., {Hsiao}, E.~Y., {et~al.} 2018, \apj, 861,
  119, \dodoi{10.3847/1538-4357/aac434}

\bibitem[{{Dilday} {et~al.}(2012){Dilday}, {Howell}, {Cenko}, {Silverman},
  {Nugent}, {Sullivan}, {Ben-Ami}, {Bildsten}, {Bolte}, {Endl}, {Filippenko},
  {Gnat}, {Horesh}, {Hsiao}, {Kasliwal}, {Kirkman}, {Maguire}, {Marcy},
  {Moore}, {Pan}, {Parrent}, {Podsiadlowski}, {Quimby}, {Sternberg}, {Suzuki},
  {Tytler}, {Xu}, {Bloom}, {Gal-Yam}, {Hook}, {Kulkarni}, {Law}, {Ofek},
  {Polishook}, \& {Poznanski}}]{dilday12}
{Dilday}, B., {Howell}, D.~A., {Cenko}, S.~B., {et~al.} 2012, Science, 337,
  942, \dodoi{10.1126/science.1219164}

\bibitem[{{Dimitriadis} {et~al.}(2016){Dimitriadis}, {Pursiainen}, {Cartier},
  {Lyman}, {Inserra}, {Kankare}, {Maguire}, {Smartt}, {Smith}, {Sullivan},
  {Valenti}, {Yaron}, {Young}, {Manulis}, {Tonry}, {Stalder}, {Denneau},
  {Heinze}, {Sherstyuk}, \& {Rest}}]{dimitriadis16}
{Dimitriadis}, G., {Pursiainen}, M., {Cartier}, R., {et~al.} 2016, The
  Astronomer's Telegram, 9720, 1

\bibitem[{{Dimitriadis} {et~al.}(2019){Dimitriadis}, {Foley}, {Rest}, {Kasen},
  {Piro}, {Polin}, {Jones}, {Villar}, {Narayan}, {Coulter}, {Kilpatrick},
  {Pan}, {Rojas-Bravo}, {Fox}, {Jha}, {Nugent}, {Riess}, {Scolnic}, {Drout},
  {K2 Mission Team}, {Barentsen}, {Dotson}, {Gully-Santiago}, {Hedges}, {Cody},
  {Barclay}, {Howell}, {KEGS}, {Garnavich}, {Tucker}, {Shaya}, {Mushotzky},
  {Olling}, {Margheim}, {Zenteno}, {Kepler spacecraft Team}, {Coughlin}, {Van
  Cleve}, {Cardoso}, {Larson}, {McCalmont-Everton}, {Peterson}, {Ross},
  {Reedy}, {Osborne}, {McGinn}, {Kohnert}, {Migliorini}, {Wheaton}, {Spencer},
  {Labonde}, {Castillo}, {Beerman}, {Steward}, {Hanley}, {Larsen},
  {Gangopadhyay}, {Kloetzel}, {Weschler}, {Nystrom}, {Moffatt}, {Redick},
  {Griest}, {Packard}, {Muszynski}, {Kampmeier}, {Bjella}, {Flynn},
  {Elsaesser}, {Pan-STARRS}, {Chambers}, {Flewelling}, {Huber}, {Magnier},
  {Waters}, {Schultz}, {Bulger}, {Lowe}, {Willman}, {Smartt}, {Smith}, {DECam},
  {Points}, {Strampelli}, {ASAS-SN}, {Brimacombe}, {Chen}, {Mu{\~n}oz},
  {Mutel}, {Shields}, {Vallely}, {Villanueva}, {PTSS/TNTS}, {Li}, {Wang},
  {Zhang}, {Lin}, {Mo}, {Zhao}, {Sai}, {Zhang}, {Zhang}, {Zhang}, {Wang},
  {Zhang}, {Baron}, {DerKacy}, {Li}, {Chen}, {Xiang}, {Rui}, {Wang}, {Huang},
  {Li}, {Cumbres Observatory}, {Hosseinzadeh}, {Howell}, {Arcavi}, {Hiramatsu},
  {Burke}, {Valenti}, {ATLAS}, {Tonry}, {Denneau}, {Heinze}, {Weiland},
  {Stalder}, {Konkoly}, {Vink{\'o}}, {S{\'a}rneczky}, {P{\'a}l}, {B{\'o}di},
  {Bogn{\'a}r}, {Cs{\'a}k}, {Cseh}, {Cs{\"o}rnyei}, {Hanyecz}, {Ign{\'a}cz},
  {Kalup}, {K{\"o}nyves-T{\'o}th}, {Kriskovics}, {Ordasi}, {Rajmon},
  {S{\'o}dor}, {Szab{\'o}}, {Szak{\'a}ts}, {Zsidi}, {ePESSTO}, {Williams},
  {Nordin}, {Cartier}, {Frohmaier}, {Galbany}, {Guti{\'e}rrez}, {Hook},
  {Inserra}, {Smith}, {Arizona}, {Sand}, {Andrews}, {Smith}, \&
  {Bilinski}}]{dimitriadis19}
{Dimitriadis}, G., {Foley}, R.~J., {Rest}, A., {et~al.} 2019, \apjl, 870, L1,
  \dodoi{10.3847/2041-8213/aaedb0}

\bibitem[{{Do} {et~al.}(2019){Do}, {Tucker}, {Payne}, {Huber}, \&
  {Shappee}}]{do19}
{Do}, A., {Tucker}, M.~A., {Payne}, A.~V., {Huber}, M.~E., \& {Shappee}, B.~J.
  2019, Transient Name Server Classification Report, 2019-965, 1

\bibitem[{{Dong} {et~al.}(2018){Dong}, {Bose}, {Stritzinger}, {Stanek},
  {Kochanek}, {Thompson}, {Prieto}, {Shappee}, \& {Holoien}}]{dong18}
{Dong}, S., {Bose}, S., {Stritzinger}, M., {et~al.} 2018, The Astronomer's
  Telegram, 12325, 1

\bibitem[{{Drake} {et~al.}(2014){Drake}, {Djorgovski}, {Graham}, {Mahabal},
  {Williams}, {Catelan}, {Larson}, {Christensen}, {Nascimbeni}, {Tomasella},
  {Pastorello}, {Benetti}, {Cappellaro}, {Elias-Rosa}, {Ochner}, {Turatto},
  {Morrell}, {Contreras}, {Gonzalez}, {Phillips}, {Hsiao}, {Stritzinger},
  {Gall}, {Holmbo}, {Sanchez}, \& {Lira}}]{drake14}
{Drake}, A.~J., {Djorgovski}, S.~G., {Graham}, M.~J., {et~al.} 2014, Central
  Bureau Electronic Telegrams, 3995, 1

\bibitem[{{Dubay} {et~al.}(2022){Dubay}, {Tucker}, {Do}, {Shappee}, \&
  {Anand}}]{dubay22}
{Dubay}, L.~O., {Tucker}, M.~A., {Do}, A., {Shappee}, B.~J., \& {Anand}, G.~S.
  2022, \apj, 926, 98, \dodoi{10.3847/1538-4357/ac3bb4}

\bibitem[{{Ellis} {et~al.}(2008){Ellis}, {Sullivan}, {Nugent}, {Howell},
  {Gal-Yam}, {Astier}, {Balam}, {Balland}, {Basa}, {Carlberg}, {Conley},
  {Fouchez}, {Guy}, {Hardin}, {Hook}, {Pain}, {Perrett}, {Pritchet}, \&
  {Regnault}}]{ellis08}
{Ellis}, R.~S., {Sullivan}, M., {Nugent}, P.~E., {et~al.} 2008, \apj, 674, 51,
  \dodoi{10.1086/524981}

\bibitem[{{Fausnaugh} {et~al.}(2021){Fausnaugh}, {Vallely}, {Kochanek},
  {Shappee}, {Stanek}, {Tucker}, {Ricker}, {Vanderspek}, {Latham}, {Seager},
  {Winn}, {Jenkins}, {Berta-Thompson}, {Daylan}, {Doty}, {F{\H{u}}r{\'e}sz},
  {Levine}, {Morris}, {P{\'a}l}, {Sha}, {Ting}, \& {Wohler}}]{fausnaugh21}
{Fausnaugh}, M.~M., {Vallely}, P.~J., {Kochanek}, C.~S., {et~al.} 2021, \apj,
  908, 51, \dodoi{10.3847/1538-4357/abcd42}

\bibitem[{{Fausnaugh} {et~al.}(2023){Fausnaugh}, {Vallely}, {Tucker},
  {Kochanek}, {Shappee}, {Stanek}, {Ricker}, {Vanderspek}, {Agarwal}, {Daylan},
  {Jayaraman}, {Hounsell}, \& {Muthukrishna}}]{fausnaugh23}
{Fausnaugh}, M.~M., {Vallely}, P.~J., {Tucker}, M.~A., {et~al.} 2023, \apj,
  956, 108, \dodoi{10.3847/1538-4357/aceaef}

\bibitem[{{Filippenko} {et~al.}(1999){Filippenko}, {Li}, \&
  {Leonard}}]{filippenko99}
{Filippenko}, A.~V., {Li}, W.~D., \& {Leonard}, D.~C. 1999, \iaucirc, 7108, 2

\bibitem[{{Filippenko} {et~al.}(1992{\natexlab{a}}){Filippenko}, {Richmond},
  {Matheson}, {Shields}, {Burbidge}, {Cohen}, {Dickinson}, {Malkan}, {Nelson},
  {Pietz}, {Schlegel}, {Schmeer}, {Spinrad}, {Steidel}, {Tran}, \&
  {Wren}}]{filippenko92b}
{Filippenko}, A.~V., {Richmond}, M.~W., {Matheson}, T., {et~al.}
  1992{\natexlab{a}}, \apjl, 384, L15, \dodoi{10.1086/186252}

\bibitem[{{Filippenko} {et~al.}(1992{\natexlab{b}}){Filippenko}, {Richmond},
  {Branch}, {Gaskell}, {Herbst}, {Ford}, {Treffers}, {Matheson}, {Ho}, {Dey},
  {Sargent}, {Small}, \& {van Breugel}}]{filippenko92a}
{Filippenko}, A.~V., {Richmond}, M.~W., {Branch}, D., {et~al.}
  1992{\natexlab{b}}, \aj, 104, 1543, \dodoi{10.1086/116339}

\bibitem[{{Fisher} {et~al.}(1999){Fisher}, {Branch}, {Hatano}, \&
  {Baron}}]{fisher99}
{Fisher}, A., {Branch}, D., {Hatano}, K., \& {Baron}, E. 1999, \mnras, 304, 67,
  \dodoi{10.1046/j.1365-8711.1999.02299.x}

\bibitem[{{Fisher} \& {Jumper}(2015)}]{fisher15}
{Fisher}, R., \& {Jumper}, K. 2015, \apj, 805, 150,
  \dodoi{10.1088/0004-637X/805/2/150}

\bibitem[{{Floers} {et~al.}(2017){Floers}, {Taubenberger}, {Vogl}, {Tomasella},
  {Benetti}, \& {Cappellaro}}]{floers17}
{Floers}, A., {Taubenberger}, S., {Vogl}, C., {et~al.} 2017, The Astronomer's
  Telegram, 10896, 1

\bibitem[{{Folatelli} {et~al.}(2013){Folatelli}, {Morrell}, {Phillips},
  {Hsiao}, {Campillay}, {Contreras}, {Castell{\'o}n}, {Hamuy}, {Krzeminski},
  {Roth}, {Stritzinger}, {Burns}, {Freedman}, {Madore}, {Murphy}, {Persson},
  {Prieto}, {Suntzeff}, {Krisciunas}, {Anderson}, {F{\"o}rster}, {Maza},
  {Pignata}, {Rojas}, {Boldt}, {Salgado}, {Wyatt}, {Olivares E.}, {Gal-Yam}, \&
  {Sako}}]{folatelli13}
{Folatelli}, G., {Morrell}, N., {Phillips}, M.~M., {et~al.} 2013, \apj, 773,
  53, \dodoi{10.1088/0004-637X/773/1/53}

\bibitem[{{Foley} {et~al.}(2016){Foley}, {Pan}, {Brown}, {Filippenko}, {Fox},
  {Hillebrandt}, {Kirshner}, {Marion}, {Milne}, {Parrent}, {Pignata}, \&
  {Stritzinger}}]{foley16}
{Foley}, R.~J., {Pan}, Y.-C., {Brown}, P., {et~al.} 2016, \mnras, 461, 1308,
  \dodoi{10.1093/mnras/stw1440}

\bibitem[{{Foley} {et~al.}(2018){Foley}, {Scolnic}, {Rest}, {Jha}, {Pan},
  {Riess}, {Challis}, {Chambers}, {Coulter}, {Dettman}, {Foley}, {Fox},
  {Huber}, {Jones}, {Kilpatrick}, {Kirshner}, {Schultz}, {Siebert},
  {Flewelling}, {Gibson}, {Magnier}, {Miller}, {Primak}, {Smartt}, {Smith},
  {Wainscoat}, {Waters}, \& {Willman}}]{foley18}
{Foley}, R.~J., {Scolnic}, D., {Rest}, A., {et~al.} 2018, \mnras, 475, 193,
  \dodoi{10.1093/mnras/stx3136}

\bibitem[{{Fremling}(2018)}]{fremling18}
{Fremling}, C. 2018, Transient Name Server Discovery Report, 2018-1005, 1

\bibitem[{{Friedman} {et~al.}(2015){Friedman}, {Wood-Vasey}, {Marion},
  {Challis}, {Mandel}, {Bloom}, {Modjaz}, {Narayan}, {Hicken}, {Foley},
  {Klein}, {Starr}, {Morgan}, {Rest}, {Blake}, {Miller}, {Falco}, {Wyatt},
  {Mink}, {Skrutskie}, \& {Kirshner}}]{friedman15}
{Friedman}, A.~S., {Wood-Vasey}, W.~M., {Marion}, G.~H., {et~al.} 2015, \apjs,
  220, 9, \dodoi{10.1088/0067-0049/220/1/9}

\bibitem[{{Frohmaier} {et~al.}(2019){Frohmaier}, {Swann}, {Short}, {Nicholl},
  {Dennefeld}, {Williams}, {Pessi}, {Pastorello}, {Anderson}, {Bravo}, {Chen},
  {Gromadzki}, {Inserra}, {Kankare}, {Yaron}, {Young}, {Manulis}, {Tonry},
  {Denneau}, {Heinze}, {Weiland}, {Stalder}, {Rest}, {Smith}, {Smartt},
  {McBrien}, \& {Srivastav}}]{frohmaier19}
{Frohmaier}, C., {Swann}, E., {Short}, P., {et~al.} 2019, Transient Name Server
  AstroNote, 27, 1

\bibitem[{{Galbany} {et~al.}(2014){Galbany}, {Stanishev}, {Mour{\~a}o},
  {Rodrigues}, {Flores}, {Garc{\'\i}a-Benito}, {Mast}, {Mendoza},
  {S{\'a}nchez}, {Badenes}, {Barrera-Ballesteros}, {Bland-Hawthorn},
  {Falc{\'o}n-Barroso}, {Garc{\'\i}a-Lorenzo}, {Gomes}, {Gonz{\'a}lez Delgado},
  {Kehrig}, {Lyubenova}, {L{\'o}pez-S{\'a}nchez}, {de Lorenzo-C{\'a}ceres},
  {Marino}, {Meidt}, {Moll{\'a}}, {Papaderos}, {P{\'e}rez-Torres},
  {Rosales-Ortega}, \& {van de Ven}}]{galbany14}
{Galbany}, L., {Stanishev}, V., {Mour{\~a}o}, A.~M., {et~al.} 2014, \aap, 572,
  A38, \dodoi{10.1051/0004-6361/201424717}

\bibitem[{{Galbany} {et~al.}(2016){Galbany}, {Anderson}, {Rosales-Ortega},
  {Kuncarayakti}, {Kr{\"u}hler}, {S{\'a}nchez}, {Falc{\'o}n-Barroso},
  {P{\'e}rez}, {Maureira}, {Hamuy}, {Gonz{\'a}lez-Gait{\'a}n}, {F{\"o}rster},
  \& {Moral}}]{galbany16}
{Galbany}, L., {Anderson}, J.~P., {Rosales-Ortega}, F.~F., {et~al.} 2016,
  \mnras, 455, 4087, \dodoi{10.1093/mnras/stv2620}

\bibitem[{{Galbany} {et~al.}(2018){Galbany}, {Anderson}, {S{\'a}nchez},
  {Kuncarayakti}, {Pedraz}, {Gonz{\'a}lez-Gait{\'a}n}, {Stanishev},
  {Dom{\'\i}nguez}, {Moreno-Raya}, {Wood-Vasey}, {Mour{\~a}o}, {Ponder},
  {Badenes}, {Moll{\'a}}, {L{\'o}pez-S{\'a}nchez}, {Rosales-Ortega},
  {V{\'\i}lchez}, {Garc{\'\i}a-Benito}, \& {Marino}}]{galbany18}
{Galbany}, L., {Anderson}, J.~P., {S{\'a}nchez}, S.~F., {et~al.} 2018, \apj,
  855, 107, \dodoi{10.3847/1538-4357/aaaf20}

\bibitem[{{Ganeshalingam} {et~al.}(2010){Ganeshalingam}, {Li}, {Filippenko},
  {Anderson}, {Foster}, {Gates}, {Griffith}, {Grigsby}, {Joubert}, {Leja},
  {Lowe}, {Macomber}, {Pritchard}, {Thrasher}, \& {Winslow}}]{ganeshalingam10}
{Ganeshalingam}, M., {Li}, W., {Filippenko}, A.~V., {et~al.} 2010, \apjs, 190,
  418, \dodoi{10.1088/0067-0049/190/2/418}

\bibitem[{{Ganeshalingam} {et~al.}(2012){Ganeshalingam}, {Li}, {Filippenko},
  {Silverman}, {Chornock}, {Foley}, {Matheson}, {Kirshner}, {Milne}, {Calkins},
  \& {Shen}}]{ganeshalingam12}
---. 2012, \apj, 751, 142, \dodoi{10.1088/0004-637X/751/2/142}

\bibitem[{{Garavini} {et~al.}(2004){Garavini}, {Folatelli}, {Goobar}, {Nobili},
  {Aldering}, {Amadon}, {Amanullah}, {Astier}, {Balland}, {Blanc}, {Burns},
  {Conley}, {Dahl{\'e}n}, {Deustua}, {Ellis}, {Fabbro}, {Fan}, {Frye}, {Gates},
  {Gibbons}, {Goldhaber}, {Goldman}, {Groom}, {Haissinski}, {Hardin}, {Hook},
  {Howell}, {Kasen}, {Kent}, {Kim}, {Knop}, {Lee}, {Lidman}, {Mendez},
  {Miller}, {Moniez}, {Mour{\~a}o}, {Newberg}, {Nugent}, {Pain}, {Perdereau},
  {Perlmutter}, {Prasad}, {Quimby}, {Raux}, {Regnault}, {Rich}, {Richards},
  {Ruiz-Lapuente}, {Sainton}, {Schaefer}, {Schahmaneche}, {Smith}, {Spadafora},
  {Stanishev}, {Walton}, {Wang}, {Wood-Vasey}, \& {Supernova Cosmology
  Project}}]{garavini04}
{Garavini}, G., {Folatelli}, G., {Goobar}, A., {et~al.} 2004, \aj, 128, 387,
  \dodoi{10.1086/421747}

\bibitem[{{Gehrels} {et~al.}(2004){Gehrels}, {Chincarini}, {Giommi}, {Mason},
  {Nousek}, {Wells}, {White}, {Barthelmy}, {Burrows}, {Cominsky}, {Hurley},
  {Marshall}, {M{\'e}sz{\'a}ros}, {Roming}, {Angelini}, {Barbier}, {Belloni},
  {Campana}, {Caraveo}, {Chester}, {Citterio}, {Cline}, {Cropper}, {Cummings},
  {Dean}, {Feigelson}, {Fenimore}, {Frail}, {Fruchter}, {Garmire}, {Gendreau},
  {Ghisellini}, {Greiner}, {Hill}, {Hunsberger}, {Krimm}, {Kulkarni}, {Kumar},
  {Lebrun}, {Lloyd-Ronning}, {Markwardt}, {Mattson}, {Mushotzky}, {Norris},
  {Osborne}, {Paczynski}, {Palmer}, {Park}, {Parsons}, {Paul}, {Rees},
  {Reynolds}, {Rhoads}, {Sasseen}, {Schaefer}, {Short}, {Smale}, {Smith},
  {Stella}, {Tagliaferri}, {Takahashi}, {Tashiro}, {Townsley}, {Tueller},
  {Turner}, {Vietri}, {Voges}, {Ward}, {Willingale}, {Zerbi}, \&
  {Zhang}}]{gehrels04}
{Gehrels}, N., {Chincarini}, G., {Giommi}, P., {et~al.} 2004, \apj, 611, 1005,
  \dodoi{10.1086/422091}

\bibitem[{{G{\'o}mez} \& {L{\'o}pez}(1998)}]{gomez98}
{G{\'o}mez}, G., \& {L{\'o}pez}, R. 1998, \apss, 263, 295,
  \dodoi{10.1023/A:1002176109402}

\bibitem[{{Gomez} {et~al.}(1996){Gomez}, {Lopez}, \& {Sanchez}}]{gomez96}
{Gomez}, G., {Lopez}, R., \& {Sanchez}, F. 1996, \aj, 112, 2094,
  \dodoi{10.1086/118166}

\bibitem[{{Gonz{\'a}lez-Gait{\'a}n} {et~al.}(2014){Gonz{\'a}lez-Gait{\'a}n},
  {Hsiao}, {Pignata}, {F{\"o}rster}, {Guti{\'e}rrez}, {Bufano}, {Galbany},
  {Folatelli}, {Phillips}, {Hamuy}, {Anderson}, \& {de Jaeger}}]{gonzalez14}
{Gonz{\'a}lez-Gait{\'a}n}, S., {Hsiao}, E.~Y., {Pignata}, G., {et~al.} 2014,
  \apj, 795, 142, \dodoi{10.1088/0004-637X/795/2/142}

\bibitem[{{Graham} {et~al.}(2019){Graham}, {Harris}, {Nugent}, {Maguire},
  {Sullivan}, {Smith}, {Valenti}, {Goobar}, {Fox}, {Shen}, {Kelly}, {McCully},
  {Brink}, \& {Filippenko}}]{graham19}
{Graham}, M.~L., {Harris}, C.~E., {Nugent}, P.~E., {et~al.} 2019, \apj, 871,
  62, \dodoi{10.3847/1538-4357/aaf41e}

\bibitem[{{Graham} {et~al.}(2022){Graham}, {Kennedy}, {Kumar}, {Amaro}, {Sand},
  {Jha}, {Galbany}, {Vinko}, {Wheeler}, {Hsiao}, {Bostroem}, {Burke},
  {Hiramatsu}, {Hosseinzadeh}, {McCully}, {Howell}, {Diamond}, {Hoeflich},
  {Wang}, \& {Li}}]{graham22}
{Graham}, M.~L., {Kennedy}, T.~D., {Kumar}, S., {et~al.} 2022, \mnras, 511,
  3682, \dodoi{10.1093/mnras/stac192}

\bibitem[{{Hamuy} {et~al.}(2002){Hamuy}, {Maza}, \& {Phillips}}]{hamuy02}
{Hamuy}, M., {Maza}, J., \& {Phillips}, M. 2002, \iaucirc, 8028, 2

\bibitem[{{Hamuy} {et~al.}(2003{\natexlab{a}}){Hamuy}, {Phillips}, {Suntzeff},
  \& {Maza}}]{hamuy03a}
{Hamuy}, M., {Phillips}, M., {Suntzeff}, N., \& {Maza}, J. 2003{\natexlab{a}},
  \iaucirc, 8151, 2

\bibitem[{{Hamuy} {et~al.}(1991){Hamuy}, {Phillips}, {Silva}, {Lubcke}, \&
  {Steffey}}]{hamuy91}
{Hamuy}, M., {Phillips}, M.~M., {Silva}, D., {Lubcke}, G., \& {Steffey}, P.
  1991, \iaucirc, 5251, 1

\bibitem[{{Hamuy} {et~al.}(1996{\natexlab{a}}){Hamuy}, {Phillips}, {Suntzeff},
  {Schommer}, {Maza}, \& {Aviles}}]{hamuy96a}
{Hamuy}, M., {Phillips}, M.~M., {Suntzeff}, N.~B., {et~al.} 1996{\natexlab{a}},
  \aj, 112, 2391, \dodoi{10.1086/118190}

\bibitem[{{Hamuy} {et~al.}(1996{\natexlab{b}}){Hamuy}, {Phillips}, {Suntzeff},
  {Schommer}, {Maza}, {Smith}, {Lira}, \& {Aviles}}]{hamuy96b}
---. 1996{\natexlab{b}}, \aj, 112, 2438, \dodoi{10.1086/118193}

\bibitem[{{Hamuy} {et~al.}(2000){Hamuy}, {Trager}, {Pinto}, {Phillips},
  {Schommer}, {Ivanov}, \& {Suntzeff}}]{hamuy00}
{Hamuy}, M., {Trager}, S.~C., {Pinto}, P.~A., {et~al.} 2000, \aj, 120, 1479,
  \dodoi{10.1086/301527}

\bibitem[{{Hamuy} {et~al.}(2003{\natexlab{b}}){Hamuy}, {Phillips}, {Suntzeff},
  {Maza}, {Gonz{\'a}lez}, {Roth}, {Krisciunas}, {Morrell}, {Green}, {Persson},
  \& {McCarthy}}]{hamuy03b}
{Hamuy}, M., {Phillips}, M.~M., {Suntzeff}, N.~B., {et~al.} 2003{\natexlab{b}},
  \nat, 424, 651, \dodoi{10.1038/nature01854}

\bibitem[{{Hicken} {et~al.}(2007){Hicken}, {Garnavich}, {Prieto}, {Blondin},
  {DePoy}, {Kirshner}, \& {Parrent}}]{hicken07}
{Hicken}, M., {Garnavich}, P.~M., {Prieto}, J.~L., {et~al.} 2007, \apjl, 669,
  L17, \dodoi{10.1086/523301}

\bibitem[{{Hicken} {et~al.}(2009){Hicken}, {Challis}, {Jha}, {Kirshner},
  {Matheson}, {Modjaz}, {Rest}, {Wood-Vasey}, {Bakos}, {Barton}, {Berlind},
  {Bragg}, {Brice{\~n}o}, {Brown}, {Caldwell}, {Calkins}, {Cho}, {Ciupik},
  {Contreras}, {Dendy}, {Dosaj}, {Durham}, {Eriksen}, {Esquerdo}, {Everett},
  {Falco}, {Fernandez}, {Gaba}, {Garnavich}, {Graves}, {Green}, {Groner},
  {Hergenrother}, {Holman}, {Hradecky}, {Huchra}, {Hutchison}, {Jerius},
  {Jordan}, {Kilgard}, {Krauss}, {Luhman}, {Macri}, {Marrone}, {McDowell},
  {McIntosh}, {McNamara}, {Megeath}, {Mochejska}, {Munoz}, {Muzerolle},
  {Naranjo}, {Narayan}, {Pahre}, {Peters}, {Peterson}, {Rines}, {Ripman},
  {Roussanova}, {Schild}, {Sicilia-Aguilar}, {Sokoloski}, {Smalley}, {Smith},
  {Spahr}, {Stanek}, {Barmby}, {Blondin}, {Stubbs}, {Szentgyorgyi}, {Torres},
  {Vaz}, {Vikhlinin}, {Wang}, {Westover}, {Woods}, \& {Zhao}}]{hicken09}
{Hicken}, M., {Challis}, P., {Jha}, S., {et~al.} 2009, \apj, 700, 331,
  \dodoi{10.1088/0004-637X/700/1/331}

\bibitem[{{Hicken} {et~al.}(2012){Hicken}, {Challis}, {Kirshner}, {Rest},
  {Cramer}, {Wood-Vasey}, {Bakos}, {Berlind}, {Brown}, {Caldwell}, {Calkins},
  {Currie}, {de Kleer}, {Esquerdo}, {Everett}, {Falco}, {Fernandez},
  {Friedman}, {Groner}, {Hartman}, {Holman}, {Hutchins}, {Keys}, {Kipping},
  {Latham}, {Marion}, {Narayan}, {Pahre}, {Pal}, {Peters}, {Perumpilly},
  {Ripman}, {Sipocz}, {Szentgyorgyi}, {Tang}, {Torres}, {Vaz}, {Wolk}, \&
  {Zezas}}]{hicken12}
{Hicken}, M., {Challis}, P., {Kirshner}, R.~P., {et~al.} 2012, \apjs, 200, 12,
  \dodoi{10.1088/0067-0049/200/2/12}

\bibitem[{{Hoeflich} \& {Khokhlov}(1996)}]{hoeflich96}
{Hoeflich}, P., \& {Khokhlov}, A. 1996, \apj, 457, 500, \dodoi{10.1086/176748}

\bibitem[{{Hoeflich} {et~al.}(1994){Hoeflich}, {Khokhlov}, \&
  {Mueller}}]{hoeflich94}
{Hoeflich}, P., {Khokhlov}, A., \& {Mueller}, E. 1994, \apjs, 92, 501,
  \dodoi{10.1086/192004}

\bibitem[{{Hoeflich} {et~al.}(1996){Hoeflich}, {Khokhlov}, {Wheeler},
  {Phillips}, {Suntzeff}, \& {Hamuy}}]{hoeflich_etal96}
{Hoeflich}, P., {Khokhlov}, A., {Wheeler}, J.~C., {et~al.} 1996, \apjl, 472,
  L81, \dodoi{10.1086/310363}

\bibitem[{{Hoeflich} {et~al.}(2017){Hoeflich}, {Hsiao}, {Ashall}, {Burns},
  {Diamond}, {Phillips}, {Sand}, {Stritzinger}, {Suntzeff}, {Contreras},
  {Krisciunas}, {Morrell}, \& {Wang}}]{hoeflich17}
{Hoeflich}, P., {Hsiao}, E.~Y., {Ashall}, C., {et~al.} 2017, \apj, 846, 58,
  \dodoi{10.3847/1538-4357/aa84b2}

\bibitem[{{H{\"o}flich} {et~al.}(2002){H{\"o}flich}, {Gerardy}, {Fesen}, \&
  {Sakai}}]{hoeflich02}
{H{\"o}flich}, P., {Gerardy}, C.~L., {Fesen}, R.~A., \& {Sakai}, S. 2002, \apj,
  568, 791, \dodoi{10.1086/339063}

\bibitem[{{H{\"o}flich} {et~al.}(1998){H{\"o}flich}, {Wheeler}, \&
  {Thielemann}}]{hoeflich98}
{H{\"o}flich}, P., {Wheeler}, J.~C., \& {Thielemann}, F.~K. 1998, \apj, 495,
  617, \dodoi{10.1086/305327}

\bibitem[{{Hoogendam} {et~al.}(2023){Hoogendam}, {Shappee}, {Brown}, {Tucker},
  {Ashall}, \& {Piro}}]{hoogendam23}
{Hoogendam}, W.~B., {Shappee}, B.~J., {Brown}, P.~J., {et~al.} 2023, arXiv
  e-prints, arXiv:2309.11563, \dodoi{10.48550/arXiv.2309.11563}

\bibitem[{{Hosseinzadeh} {et~al.}(2017){Hosseinzadeh}, {Sand}, {Valenti},
  {Brown}, {Howell}, {McCully}, {Kasen}, {Arcavi}, {Bostroem}, {Tartaglia},
  {Hsiao}, {Davis}, {Shahbandeh}, \& {Stritzinger}}]{hosseinzadeh17}
{Hosseinzadeh}, G., {Sand}, D.~J., {Valenti}, S., {et~al.} 2017, \apjl, 845,
  L11, \dodoi{10.3847/2041-8213/aa8402}

\bibitem[{{Howell} {et~al.}(2007){Howell}, {Sullivan}, {Conley}, \&
  {Carlberg}}]{howell07}
{Howell}, D.~A., {Sullivan}, M., {Conley}, A., \& {Carlberg}, R. 2007, \apjl,
  667, L37, \dodoi{10.1086/522030}

\bibitem[{{Howell} {et~al.}(2006){Howell}, {Sullivan}, {Nugent}, {Ellis},
  {Conley}, {Le Borgne}, {Carlberg}, {Guy}, {Balam}, {Basa}, {Fouchez}, {Hook},
  {Hsiao}, {Neill}, {Pain}, {Perrett}, \& {Pritchet}}]{howell06}
{Howell}, D.~A., {Sullivan}, M., {Nugent}, P.~E., {et~al.} 2006, \nat, 443,
  308, \dodoi{10.1038/nature05103}

\bibitem[{{Hristov} {et~al.}(2021){Hristov}, {Hoeflich}, \&
  {Collins}}]{hristov21}
{Hristov}, B., {Hoeflich}, P., \& {Collins}, D.~C. 2021, \apj, 923, 210,
  \dodoi{10.3847/1538-4357/ac0ef8}

\bibitem[{{Hsiao} {et~al.}(2007){Hsiao}, {Conley}, {Howell}, {Sullivan},
  {Pritchet}, {Carlberg}, {Nugent}, \& {Phillips}}]{hsiao07}
{Hsiao}, E.~Y., {Conley}, A., {Howell}, D.~A., {et~al.} 2007, \apj, 663, 1187,
  \dodoi{10.1086/518232}

\bibitem[{{Hsiao} {et~al.}(2013){Hsiao}, {Marion}, {Phillips}, {Burns},
  {Winge}, {Morrell}, {Contreras}, {Freedman}, {Kromer}, {Gall}, {Gerardy},
  {H{\"o}flich}, {Im}, {Jeon}, {Kirshner}, {Nugent}, {Persson}, {Pignata},
  {Roth}, {Stanishev}, {Stritzinger}, \& {Suntzeff}}]{hsiao13}
{Hsiao}, E.~Y., {Marion}, G.~H., {Phillips}, M.~M., {et~al.} 2013, \apj, 766,
  72, \dodoi{10.1088/0004-637X/766/2/72}

\bibitem[{{Hsiao} {et~al.}(2019){Hsiao}, {Phillips}, {Marion}, {Kirshner},
  {Morrell}, {Sand}, {Burns}, {Contreras}, {Hoeflich}, {Stritzinger},
  {Valenti}, {Anderson}, {Ashall}, {Baltay}, {Baron}, {Banerjee}, {Davis},
  {Diamond}, {Folatelli}, {Freedman}, {F{\"o}rster}, {Galbany}, {Gall},
  {Gonz{\'a}lez-Gait{\'a}n}, {Goobar}, {Hamuy}, {Holmbo}, {Kasliwal},
  {Krisciunas}, {Kumar}, {Lidman}, {Lu}, {Nugent}, {Perlmutter}, {Persson},
  {Piro}, {Rabinowitz}, {Roth}, {Ryder}, {Schmidt}, {Shahbandeh}, {Suntzeff},
  {Taddia}, {Uddin}, \& {Wang}}]{hsiao19}
{Hsiao}, E.~Y., {Phillips}, M.~M., {Marion}, G.~H., {et~al.} 2019, \pasp, 131,
  014002, \dodoi{10.1088/1538-3873/aae961}

\bibitem[{{Jack} {et~al.}(2015){Jack}, {Baron}, \& {Hauschildt}}]{jack15}
{Jack}, D., {Baron}, E., \& {Hauschildt}, P.~H. 2015, \mnras, 449, 3581,
  \dodoi{10.1093/mnras/stv474}

\bibitem[{{Jayasinghe} {et~al.}(2019){Jayasinghe}, {Stanek}, {Kochanek},
  {Shappee}, {Holoien}, {Thompson}, {Prieto}, {Dong}, {Pawlak}, {Pejcha},
  {Shields}, {Pojmanski}, {Otero}, {Hurst}, {Britt}, \& {Will}}]{jayasinghe19}
{Jayasinghe}, T., {Stanek}, K.~Z., {Kochanek}, C.~S., {et~al.} 2019, \mnras,
  485, 961, \dodoi{10.1093/mnras/stz444}

\bibitem[{{Jeffery} {et~al.}(1992){Jeffery}, {Leibundgut}, {Kirshner},
  {Benetti}, {Branch}, \& {Sonneborn}}]{jeffery92}
{Jeffery}, D.~J., {Leibundgut}, B., {Kirshner}, R.~P., {et~al.} 1992, \apj,
  397, 304, \dodoi{10.1086/171787}

\bibitem[{{Jha} {et~al.}(2006){Jha}, {Kirshner}, {Challis}, {Garnavich},
  {Matheson}, {Soderberg}, {Graves}, {Hicken}, {Alves}, {Arce}, {Balog},
  {Barmby}, {Barton}, {Berlind}, {Bragg}, {Brice{\~n}o}, {Brown}, {Buckley},
  {Caldwell}, {Calkins}, {Carter}, {Concannon}, {Donnelly}, {Eriksen},
  {Fabricant}, {Falco}, {Fiore}, {Garcia}, {G{\'o}mez}, {Grogin}, {Groner},
  {Groot}, {Haisch}, {Hartmann}, {Hergenrother}, {Holman}, {Huchra},
  {Jayawardhana}, {Jerius}, {Kannappan}, {Kim}, {Kleyna}, {Kochanek},
  {Koranyi}, {Krockenberger}, {Lada}, {Luhman}, {Luu}, {Macri}, {Mader},
  {Mahdavi}, {Marengo}, {Marsden}, {McLeod}, {McNamara}, {Megeath}, {Moraru},
  {Mossman}, {Muench}, {Mu{\~n}oz}, {Muzerolle}, {Naranjo}, {Nelson-Patel},
  {Pahre}, {Patten}, {Peters}, {Peters}, {Raymond}, {Rines}, {Schild},
  {Sobczak}, {Spahr}, {Stauffer}, {Stefanik}, {Szentgyorgyi}, {Tollestrup},
  {V{\"a}is{\"a}nen}, {Vikhlinin}, {Wang}, {Willner}, {Wolk}, {Zajac}, {Zhao},
  \& {Stanek}}]{jha06}
{Jha}, S., {Kirshner}, R.~P., {Challis}, P., {et~al.} 2006, \aj, 131, 527,
  \dodoi{10.1086/497989}

\bibitem[{{Jiang} {et~al.}(2018){Jiang}, {Doi}, {Maeda}, \&
  {Shigeyama}}]{jiang18}
{Jiang}, J.-a., {Doi}, M., {Maeda}, K., \& {Shigeyama}, T. 2018, \apj, 865,
  149, \dodoi{10.3847/1538-4357/aadb9a}

\bibitem[{{Kasen}(2006)}]{kasen06}
{Kasen}, D. 2006, \apj, 649, 939, \dodoi{10.1086/506588}

\bibitem[{{Kelz} {et~al.}(2006){Kelz}, {Verheijen}, {Roth}, {Bauer}, {Becker},
  {Paschke}, {Popow}, {S{\'a}nchez}, \& {Laux}}]{kelz06}
{Kelz}, A., {Verheijen}, M. A.~W., {Roth}, M.~M., {et~al.} 2006, \pasp, 118,
  129, \dodoi{10.1086/497455}

\bibitem[{{Kennicutt}(1998)}]{kennicutt98}
{Kennicutt}, Robert~C., J. 1998, \apj, 498, 541, \dodoi{10.1086/305588}

\bibitem[{{Khokhlov}(1991{\natexlab{a}})}]{khokhlov91a}
{Khokhlov}, A.~M. 1991{\natexlab{a}}, \aap, 245, 114

\bibitem[{{Khokhlov}(1991{\natexlab{b}})}]{khokhlov91b}
---. 1991{\natexlab{b}}, \aap, 245, L25

\bibitem[{{Kochanek} {et~al.}(2017){Kochanek}, {Shappee}, {Stanek}, {Holoien},
  {Thompson}, {Prieto}, {Dong}, {Shields}, {Will}, {Britt}, {Perzanowski}, \&
  {Pojma{\'n}ski}}]{kochanek17}
{Kochanek}, C.~S., {Shappee}, B.~J., {Stanek}, K.~Z., {et~al.} 2017, \pasp,
  129, 104502, \dodoi{10.1088/1538-3873/aa80d9}

\bibitem[{{Kool} {et~al.}(2023){Kool}, {Johansson}, {Sollerman}, {Mold{\'o}n},
  {Moriya}, {Mattila}, {Schulze}, {Chomiuk}, {P{\'e}rez-Torres}, {Harris},
  {Lundqvist}, {Graham}, {Yang}, {Perley}, {Strotjohann}, {Fremling},
  {Gal-Yam}, {Lezmy}, {Maguire}, {Omand}, {Smith}, {Andreoni}, {Bellm},
  {Bloom}, {De}, {Groom}, {Kasliwal}, {Masci}, {Medford}, {Park}, {Purdum},
  {Reynolds}, {Riddle}, {Robert}, {Ryder}, {Sharma}, \& {Stern}}]{kool23}
{Kool}, E.~C., {Johansson}, J., {Sollerman}, J., {et~al.} 2023, \nat, 617, 477,
  \dodoi{10.1038/s41586-023-05916-w}

\bibitem[{{Krisciunas} {et~al.}(2001){Krisciunas}, {Phillips}, {Stubbs},
  {Rest}, {Miknaitis}, {Riess}, {Suntzeff}, {Roth}, {Persson}, \&
  {Freedman}}]{krisciunas01}
{Krisciunas}, K., {Phillips}, M.~M., {Stubbs}, C., {et~al.} 2001, \aj, 122,
  1616, \dodoi{10.1086/322120}

\bibitem[{{Kumar} {et~al.}(2023){Kumar}, {Hsiao}, {Ashall}, {Phillips},
  {Morrell}, {Hoeflich}, {Burns}, {Galbany}, {Baron}, {Contreras}, {Davis},
  {Diamond}, {F{\"o}rster}, {Graham}, {Karamehmetoglu}, {Kirshner},
  {Koribalski}, {Krisciunas}, {Lu}, {Marion}, {Pessi}, {Piro}, {Shahbandeh},
  {Stritzinger}, {Suntzeff}, \& {Uddin}}]{kumar23}
{Kumar}, S., {Hsiao}, E.~Y., {Ashall}, C., {et~al.} 2023, \apj, 945, 27,
  \dodoi{10.3847/1538-4357/acad73}

\bibitem[{{Kuncarayakti} {et~al.}(2016){Kuncarayakti}, {Galbany}, {Anderson},
  {Kr{\"u}hler}, \& {Hamuy}}]{kuncarayakti16}
{Kuncarayakti}, H., {Galbany}, L., {Anderson}, J.~P., {Kr{\"u}hler}, T., \&
  {Hamuy}, M. 2016, \aap, 593, A78, \dodoi{10.1051/0004-6361/201628813}

\bibitem[{{Kuncarayakti} {et~al.}(2018){Kuncarayakti}, {Reynolds}, {Moran},
  {Kankare}, {Harmanen}, {Wiik}, {Cartier}, {Dennefeld}, {Prentice},
  {Taubenberger}, {Inserra}, {Maguire}, {Smartt}, {Yaron}, {Young}, {Manulis},
  {Tonry}, {Denneau}, {Heinze}, {Weiland}, {Stalder}, {Rest}, {McBrien}, \&
  {Wright}}]{kuncarayakti18}
{Kuncarayakti}, H., {Reynolds}, T., {Moran}, S., {et~al.} 2018, The
  Astronomer's Telegram, 12107, 1

\bibitem[{{Leibundgut} {et~al.}(1993){Leibundgut}, {Kirshner}, {Phillips},
  {Wells}, {Suntzeff}, {Hamuy}, {Schommer}, {Walker}, {Gonzalez}, {Ugarte},
  {Williams}, {Williger}, {Gomez}, {Marzke}, {Schmidt}, {Whitney}, {Caldwell},
  {Peters}, {Chaffee}, {Foltz}, {Rehner}, {Siciliano}, {Barnes}, {Cheng},
  {Hintzen}, {Kim}, {Maza}, {Parker}, {Porter}, {Schmidtke}, \&
  {Sonneborn}}]{leibundgut93}
{Leibundgut}, B., {Kirshner}, R.~P., {Phillips}, M.~M., {et~al.} 1993, \aj,
  105, 301, \dodoi{10.1086/116427}

\bibitem[{{Lentz} {et~al.}(2000){Lentz}, {Baron}, {Branch}, {Hauschildt}, \&
  {Nugent}}]{lentz00}
{Lentz}, E.~J., {Baron}, E., {Branch}, D., {Hauschildt}, P.~H., \& {Nugent},
  P.~E. 2000, \apj, 530, 966, \dodoi{10.1086/308400}

\bibitem[{{Leonard}(2007)}]{leonard07}
{Leonard}, D.~C. 2007, \apj, 670, 1275, \dodoi{10.1086/522367}

\bibitem[{{Li} {et~al.}(2001{\natexlab{a}}){Li}, {Filippenko}, {Treffers},
  {Riess}, {Hu}, \& {Qiu}}]{li01b}
{Li}, W., {Filippenko}, A.~V., {Treffers}, R.~R., {et~al.} 2001{\natexlab{a}},
  \apj, 546, 734, \dodoi{10.1086/318299}

\bibitem[{{Li} {et~al.}(2001{\natexlab{b}}){Li}, {Filippenko}, {Gates},
  {Chornock}, {Gal-Yam}, {Ofek}, {Leonard}, {Modjaz}, {Rich}, {Riess}, \&
  {Treffers}}]{li01a}
{Li}, W., {Filippenko}, A.~V., {Gates}, E., {et~al.} 2001{\natexlab{b}}, \pasp,
  113, 1178, \dodoi{10.1086/323355}

\bibitem[{{Li} {et~al.}(1999){Li}, {Qiu}, {Qiao}, {Zhu}, {Hu}, {Richmond},
  {Filippenko}, {Treffers}, {Peng}, \& {Leonard}}]{li99}
{Li}, W.~D., {Qiu}, Y.~L., {Qiao}, Q.~Y., {et~al.} 1999, \aj, 117, 2709,
  \dodoi{10.1086/300895}

\bibitem[{{Lipunov} {et~al.}(2013){Lipunov}, {Tyurina}, {Denisenko},
  {Gorbovskoy}, {Cenko}, {Tucker}, {Shivvers}, {Clubb}, \&
  {Filippenko}}]{lipunov13}
{Lipunov}, V., {Tyurina}, N., {Denisenko}, D., {et~al.} 2013, Central Bureau
  Electronic Telegrams, 3563, 1

\bibitem[{{Lira} {et~al.}(1998){Lira}, {Suntzeff}, {Phillips}, {Hamuy}, {Maza},
  {Schommer}, {Smith}, {Wells}, {Avil{\'e}s}, {Baldwin}, {Elias},
  {Gonz{\'a}lez}, {Layden}, {Navarrete}, {Ugarte}, {Walker}, {Williger},
  {Baganoff}, {Crotts}, {Rich}, {Tyson}, {Dey}, {Guhathakurta}, {Hibbard},
  {Kim}, {Rehner}, {Siciliano}, {Roth}, {Seitzer}, \& {Williams}}]{lira98}
{Lira}, P., {Suntzeff}, N.~B., {Phillips}, M.~M., {et~al.} 1998, \aj, 115, 234,
  \dodoi{10.1086/300175}

\bibitem[{{L{\'o}pez-Cob{\'a}} {et~al.}(2020){L{\'o}pez-Cob{\'a}},
  {S{\'a}nchez}, {Anderson}, {Cruz-Gonz{\'a}lez}, {Galbany}, {Ruiz-Lara},
  {Barrera-Ballesteros}, {Prieto}, \& {Kuncarayakti}}]{lopez_coba20}
{L{\'o}pez-Cob{\'a}}, C., {S{\'a}nchez}, S.~F., {Anderson}, J.~P., {et~al.}
  2020, \aj, 159, 167, \dodoi{10.3847/1538-3881/ab7848}

\bibitem[{{Lu} {et~al.}(2023){Lu}, {Hsiao}, {Phillips}, {Burns}, {Ashall},
  {Morrell}, {Ng}, {Kumar}, {Shahbandeh}, {Hoeflich}, {Baron}, {Uddin},
  {Stritzinger}, {Suntzeff}, {Baltay}, {Davis}, {Diamond}, {Folatelli},
  {F{\"o}rster}, {Gagn{\'e}}, {Galbany}, {Gall}, {Gonz{\'a}lez-Gait{\'a}n},
  {Holmbo}, {Kirshner}, {Krisciunas}, {Marion}, {Perlmutter}, {Pessi}, {Piro},
  {Rabinowitz}, {Ryder}, \& {Sand}}]{lu23}
{Lu}, J., {Hsiao}, E.~Y., {Phillips}, M.~M., {et~al.} 2023, \apj, 948, 27,
  \dodoi{10.3847/1538-4357/acc100}

\bibitem[{{Maeda} {et~al.}(2010){Maeda}, {Benetti}, {Stritzinger}, {R{\"o}pke},
  {Folatelli}, {Sollerman}, {Taubenberger}, {Nomoto}, {Leloudas}, {Hamuy},
  {Tanaka}, {Mazzali}, \& {Elias-Rosa}}]{maeda10}
{Maeda}, K., {Benetti}, S., {Stritzinger}, M., {et~al.} 2010, \nat, 466, 82,
  \dodoi{10.1038/nature09122}

\bibitem[{{Magee} {et~al.}(2022){Magee}, {Cuddy}, {Maguire}, {Deckers},
  {Dhawan}, {Frohmaier}, {Miller}, {Nordin}, {Coughlin}, {Feinstein}, \&
  {Riddle}}]{magee22}
{Magee}, M.~R., {Cuddy}, C., {Maguire}, K., {et~al.} 2022, \mnras, 513, 3035,
  \dodoi{10.1093/mnras/stac1045}

\bibitem[{{Maguire} {et~al.}(2012){Maguire}, {Pan}, {Le Guillou}, {Baumont},
  {Sullivan}, {Taubenberger}, {Valenti}, {Pastorello}, {Benetti}, {Smartt},
  {Smith}, {Young}, {Gal-Yam}, {Yaron}, {Baltay}, {Ellman}, {Hadjiyska},
  {McKinnon}, {Rabinowitz}, {Walker}, {Feindt}, {Kowalski}, \&
  {Nugent}}]{maguire12}
{Maguire}, K., {Pan}, Y., {Le Guillou}, L., {et~al.} 2012, The Astronomer's
  Telegram, 4624, 1

\bibitem[{{Malesani} {et~al.}(2018){Malesani}, {Rubin}, {Leloudas}, {Yaron}, \&
  {Knezevic}}]{malesani18}
{Malesani}, D., {Rubin}, A., {Leloudas}, G., {Yaron}, O., \& {Knezevic}, N.
  2018, Transient Name Server Classification Report, 2018-468, 1

\bibitem[{{Marino} {et~al.}(2013){Marino}, {Rosales-Ortega}, {S{\'a}nchez},
  {Gil de Paz}, {V{\'\i}lchez}, {Miralles-Caballero}, {Kehrig},
  {P{\'e}rez-Montero}, {Stanishev}, {Iglesias-P{\'a}ramo}, {D{\'\i}az},
  {Castillo-Morales}, {Kennicutt}, {L{\'o}pez-S{\'a}nchez}, {Galbany},
  {Garc{\'\i}a-Benito}, {Mast}, {Mendez-Abreu}, {Monreal-Ibero}, {Husemann},
  {Walcher}, {Garc{\'\i}a-Lorenzo}, {Masegosa}, {Del Olmo Orozco},
  {Mour{\~a}o}, {Ziegler}, {Moll{\'a}}, {Papaderos},
  {S{\'a}nchez-Bl{\'a}zquez}, {Gonz{\'a}lez Delgado}, {Falc{\'o}n-Barroso},
  {Roth}, {van de Ven}, \& {CALIFA Team}}]{marino13}
{Marino}, R.~A., {Rosales-Ortega}, F.~F., {S{\'a}nchez}, S.~F., {et~al.} 2013,
  \aap, 559, A114, \dodoi{10.1051/0004-6361/201321956}

\bibitem[{{Martin} {et~al.}(2005){Martin}, {Fanson}, {Schiminovich},
  {Morrissey}, {Friedman}, {Barlow}, {Conrow}, {Grange}, {Jelinsky},
  {Milliard}, {Siegmund}, {Bianchi}, {Byun}, {Donas}, {Forster}, {Heckman},
  {Lee}, {Madore}, {Malina}, {Neff}, {Rich}, {Small}, {Surber}, {Szalay},
  {Welsh}, \& {Wyder}}]{martin05}
{Martin}, D.~C., {Fanson}, J., {Schiminovich}, D., {et~al.} 2005, \apjl, 619,
  L1, \dodoi{10.1086/426387}

\bibitem[{{Masci} {et~al.}(2019){Masci}, {Laher}, {Rusholme}, {Shupe}, {Groom},
  {Surace}, {Jackson}, {Monkewitz}, {Beck}, {Flynn}, {Terek}, {Landry},
  {Hacopians}, {Desai}, {Howell}, {Brooke}, {Imel}, {Wachter}, {Ye}, {Lin},
  {Cenko}, {Cunningham}, {Rebbapragada}, {Bue}, {Miller}, {Mahabal}, {Bellm},
  {Patterson}, {Juri{\'c}}, {Golkhou}, {Ofek}, {Walters}, {Graham}, {Kasliwal},
  {Dekany}, {Kupfer}, {Burdge}, {Cannella}, {Barlow}, {Van Sistine}, {Giomi},
  {Fremling}, {Blagorodnova}, {Levitan}, {Riddle}, {Smith}, {Helou}, {Prince},
  \& {Kulkarni}}]{masci19}
{Masci}, F.~J., {Laher}, R.~R., {Rusholme}, B., {et~al.} 2019, \pasp, 131,
  018003, \dodoi{10.1088/1538-3873/aae8ac}

\bibitem[{{Matheson} {et~al.}(2008){Matheson}, {Kirshner}, {Challis}, {Jha},
  {Garnavich}, {Berlind}, {Calkins}, {Blondin}, {Balog}, {Bragg}, {Caldwell},
  {Dendy Concannon}, {Falco}, {Graves}, {Huchra}, {Kuraszkiewicz}, {Mader},
  {Mahdavi}, {Phelps}, {Rines}, {Song}, \& {Wilkes}}]{matheson08}
{Matheson}, T., {Kirshner}, R.~P., {Challis}, P., {et~al.} 2008, \aj, 135,
  1598, \dodoi{10.1088/0004-6256/135/4/1598}

\bibitem[{{Mattila} {et~al.}(2005){Mattila}, {Lundqvist}, {Sollerman}, {Kozma},
  {Baron}, {Fransson}, {Leibundgut}, \& {Nomoto}}]{mattila05}
{Mattila}, S., {Lundqvist}, P., {Sollerman}, J., {et~al.} 2005, \aap, 443, 649,
  \dodoi{10.1051/0004-6361:20052731}

\bibitem[{{Mazzali} {et~al.}(1998){Mazzali}, {Cappellaro}, {Danziger},
  {Turatto}, \& {Benetti}}]{mazzali98}
{Mazzali}, P.~A., {Cappellaro}, E., {Danziger}, I.~J., {Turatto}, M., \&
  {Benetti}, S. 1998, \apjl, 499, L49, \dodoi{10.1086/311345}

\bibitem[{{Mazzali} {et~al.}(1995){Mazzali}, {Danziger}, \&
  {Turatto}}]{mazzali95}
{Mazzali}, P.~A., {Danziger}, I.~J., \& {Turatto}, M. 1995, \aap, 297, 509

\bibitem[{{Meikle} {et~al.}(1996){Meikle}, {Cumming}, {Geballe}, {Lewis},
  {Walton}, {Balcells}, {Cimatti}, {Croom}, {Dhillon}, {Economou}, {Jenkins},
  {Knapen}, {Meadows}, {Morris}, {Perez-Fournon}, {Shanks}, {Smith}, {Tanvir},
  {Veilleux}, {Vilchez}, {Wall}, \& {Lucey}}]{meikle96}
{Meikle}, W.~P.~S., {Cumming}, R.~J., {Geballe}, T.~R., {et~al.} 1996, \mnras,
  281, 263, \dodoi{10.1093/mnras/281.1.263}

\bibitem[{{Milne} {et~al.}(2013){Milne}, {Brown}, {Roming}, {Bufano}, \&
  {Gehrels}}]{milne13}
{Milne}, P.~A., {Brown}, P.~J., {Roming}, P. W.~A., {Bufano}, F., \& {Gehrels},
  N. 2013, \apj, 779, 23, \dodoi{10.1088/0004-637X/779/1/23}

\bibitem[{{Morrell} {et~al.}(2024){Morrell}, {Phillips}, {Folatelli},
  {Stritzinger}, {Hamuy}, {Suntzeff}, \& {Hsiao}}]{morrell24}
{Morrell}, N., {Phillips}, M.~M., {Folatelli}, G., {et~al.} 2024, \apj, in
  print

\bibitem[{{NASA/IPAC Extragalactic Database (NED)}(2019)}]{ned19}
{NASA/IPAC Extragalactic Database (NED)}. 2019, NASA/IPAC Extragalactic
  Database (NED),  IPAC, \dodoi{10.26132/NED1}

\bibitem[{{Neill} {et~al.}(2009){Neill}, {Sullivan}, {Howell}, {Conley},
  {Seibert}, {Martin}, {Barlow}, {Foster}, {Friedman}, {Morrissey}, {Neff},
  {Schiminovich}, {Wyder}, {Bianchi}, {Donas}, {Heckman}, {Lee}, {Madore},
  {Milliard}, {Rich}, \& {Szalay}}]{neill09}
{Neill}, J.~D., {Sullivan}, M., {Howell}, D.~A., {et~al.} 2009, \apj, 707,
  1449, \dodoi{10.1088/0004-637X/707/2/1449}

\bibitem[{{Nicolas} {et~al.}(2021){Nicolas}, {Rigault}, {Copin}, {Graziani},
  {Aldering}, {Briday}, {Kim}, {Nordin}, {Perlmutter}, \& {Smith}}]{nicolas21}
{Nicolas}, N., {Rigault}, M., {Copin}, Y., {et~al.} 2021, \aap, 649, A74,
  \dodoi{10.1051/0004-6361/202038447}

\bibitem[{{Nomoto}(1982)}]{nomoto82}
{Nomoto}, K. 1982, \apj, 257, 780, \dodoi{10.1086/160031}

\bibitem[{{Nordin} {et~al.}(2019){Nordin}, {Brinnel}, {Giomi}, {Santen},
  {Gal-Yam}, {Yaron}, \& {Schulze}}]{nordin19}
{Nordin}, J., {Brinnel}, V., {Giomi}, M., {et~al.} 2019, Transient Name Server
  Discovery Report, 2019-930, 1

\bibitem[{{Nordin} {et~al.}(2018){Nordin}, {Aldering}, {Antilogus}, {Aragon},
  {Bailey}, {Baltay}, {Barbary}, {Bongard}, {Boone}, {Brinnel}, {Buton},
  {Childress}, {Chotard}, {Copin}, {Dixon}, {Fagrelius}, {Feindt}, {Fouchez},
  {Gangler}, {Hayden}, {Hillebrandt}, {Kim}, {Kowalski}, {Kuesters}, {Leget},
  {Lombardo}, {Lin}, {Pain}, {Pecontal}, {Pereira}, {Perlmutter}, {Rabinowitz},
  {Rigault}, {Runge}, {Rubin}, {Saunders}, {Smadja}, {Sofiatti}, {Suzuki},
  {Taubenberger}, {Tao}, \& {Thomas}}]{nordin18}
{Nordin}, J., {Aldering}, G., {Antilogus}, P., {et~al.} 2018, {VizieR Online
  Data Catalog: SNF20080514-002 and LSQ12fxd spectra (Nordin+, 2018)}, VizieR
  On-line Data Catalog: J/A+A/614/A71. Originally published in:
  2018A\&A...614A..71N, \dodoi{10.26093/cds/vizier.36140071}

\bibitem[{{Nugent} {et~al.}(1995){Nugent}, {Phillips}, {Baron}, {Branch}, \&
  {Hauschildt}}]{nugent95}
{Nugent}, P., {Phillips}, M., {Baron}, E., {Branch}, D., \& {Hauschildt}, P.
  1995, \apjl, 455, L147, \dodoi{10.1086/309846}

\bibitem[{{O'Brien} {et~al.}(2024){O'Brien}, {Kerzendorf}, {Fullard}, {Pakmor},
  {Buchner}, {Vogl}, {Chen}, {van der Smagt}, {Williamson}, \&
  {Singhal}}]{OBrien23}
{O'Brien}, J.~T., {Kerzendorf}, W.~E., {Fullard}, A., {et~al.} 2024, \apj, 964,
  137, \dodoi{10.3847/1538-4357/ad2358}

\bibitem[{{Pan} {et~al.}(2020){Pan}, {Foley}, {Jones}, {Filippenko}, \&
  {Kuin}}]{pan20}
{Pan}, Y.~C., {Foley}, R.~J., {Jones}, D.~O., {Filippenko}, A.~V., \& {Kuin},
  N.~P.~M. 2020, \mnras, 491, 5897, \dodoi{10.1093/mnras/stz3391}

\bibitem[{{Pan} {et~al.}(2015){Pan}, {Foley}, {Kromer}, {Fox}, {Zheng},
  {Challis}, {Clubb}, {Filippenko}, {Folatelli}, {Graham}, {Hillebrandt},
  {Kirshner}, {Lee}, {Pakmor}, {Patat}, {Phillips}, {Pignata}, {R{\"o}pke},
  {Seitenzahl}, {Silverman}, {Simon}, {Sternberg}, {Stritzinger},
  {Taubenberger}, {Vinko}, \& {Wheeler}}]{pan15}
{Pan}, Y.~C., {Foley}, R.~J., {Kromer}, M., {et~al.} 2015, \mnras, 452, 4307,
  \dodoi{10.1093/mnras/stv1605}

\bibitem[{{Parrent} {et~al.}(2012){Parrent}, {Howell}, {Friesen}, {Thomas},
  {Fesen}, {Milisavljevic}, {Bianco}, {Dilday}, {Nugent}, {Baron}, {Arcavi},
  {Ben-Ami}, {Bersier}, {Bildsten}, {Bloom}, {Cao}, {Cenko}, {Filippenko},
  {Gal-Yam}, {Kasliwal}, {Konidaris}, {Kulkarni}, {Law}, {Levitan}, {Maguire},
  {Mazzali}, {Ofek}, {Pan}, {Polishook}, {Poznanski}, {Quimby}, {Silverman},
  {Sternberg}, {Sullivan}, {Walker}, {Xu}, {Buton}, \& {Pereira}}]{parrent12}
{Parrent}, J.~T., {Howell}, D.~A., {Friesen}, B., {et~al.} 2012, \apjl, 752,
  L26, \dodoi{10.1088/2041-8205/752/2/L26}

\bibitem[{{Patat}(2017)}]{patat17}
{Patat}, F. 2017, in Handbook of Supernovae, ed. A.~W. {Alsabti} \&
  P.~{Murdin}, 1017, \dodoi{10.1007/978-3-319-21846-5_110}

\bibitem[{{Penney} \& {Hoeflich}(2014)}]{penney14}
{Penney}, R., \& {Hoeflich}, P. 2014, \apj, 795, 84,
  \dodoi{10.1088/0004-637X/795/1/84}

\bibitem[{{Pereira} {et~al.}(2013){Pereira}, {Thomas}, {Aldering}, {Antilogus},
  {Baltay}, {Benitez-Herrera}, {Bongard}, {Buton}, {Canto}, {Cellier-Holzem},
  {Chen}, {Childress}, {Chotard}, {Copin}, {Fakhouri}, {Fink}, {Fouchez},
  {Gangler}, {Guy}, {Hillebrandt}, {Hsiao}, {Kerschhaggl}, {Kowalski},
  {Kromer}, {Nordin}, {Nugent}, {Paech}, {Pain}, {P{\'e}contal}, {Perlmutter},
  {Rabinowitz}, {Rigault}, {Runge}, {Saunders}, {Smadja}, {Tao},
  {Taubenberger}, {Tilquin}, \& {Wu}}]{pereira13}
{Pereira}, R., {Thomas}, R.~C., {Aldering}, G., {et~al.} 2013, \aap, 554, A27,
  \dodoi{10.1051/0004-6361/201221008}

\bibitem[{{Phillips}(1993)}]{phillips93}
{Phillips}, M.~M. 1993, \apjl, 413, L105, \dodoi{10.1086/186970}

\bibitem[{{Phillips} {et~al.}(1992){Phillips}, {Wells}, {Suntzeff}, {Hamuy},
  {Leibundgut}, {Kirshner}, \& {Foltz}}]{phillips92}
{Phillips}, M.~M., {Wells}, L.~A., {Suntzeff}, N.~B., {et~al.} 1992, \aj, 103,
  1632, \dodoi{10.1086/116177}

\bibitem[{{Phillips} {et~al.}(1987){Phillips}, {Phillips}, {Heathcote},
  {Blanco}, {Geisler}, {Hamilton}, {Suntzeff}, {Jablonski}, {Steiner},
  {Cowley}, {Schmidtke}, {Wyckoff}, {Hutchings}, {Tonry}, {Strauss},
  {Thorstensen}, {Honey}, {Maza}, {Ruiz}, {Landolt}, {Uomoto}, {Rich},
  {Grindlay}, {Cohn}, {Smith}, {Lutz}, {Lavery}, \& {Saha}}]{phillips87}
{Phillips}, M.~M., {Phillips}, A.~C., {Heathcote}, S.~R., {et~al.} 1987, \pasp,
  99, 592, \dodoi{10.1086/132020}

\bibitem[{{Phillips} {et~al.}(2013){Phillips}, {Simon}, {Morrell}, {Burns},
  {Cox}, {Foley}, {Karakas}, {Patat}, {Sternberg}, {Williams}, {Gal-Yam},
  {Hsiao}, {Leonard}, {Persson}, {Stritzinger}, {Thompson}, {Campillay},
  {Contreras}, {Folatelli}, {Freedman}, {Hamuy}, {Roth}, {Shields}, {Suntzeff},
  {Chomiuk}, {Ivans}, {Madore}, {Penprase}, {Perley}, {Pignata}, {Preston}, \&
  {Soderberg}}]{phillips13}
{Phillips}, M.~M., {Simon}, J.~D., {Morrell}, N., {et~al.} 2013, \apj, 779, 38,
  \dodoi{10.1088/0004-637X/779/1/38}

\bibitem[{{Phillips} {et~al.}(2019){Phillips}, {Contreras}, {Hsiao}, {Morrell},
  {Burns}, {Stritzinger}, {Ashall}, {Freedman}, {Hoeflich}, {Persson}, {Piro},
  {Suntzeff}, {Uddin}, {Anais}, {Baron}, {Busta}, {Campillay}, {Castell{\'o}n},
  {Corco}, {Diamond}, {Gall}, {Gonzalez}, {Holmbo}, {Krisciunas}, {Roth},
  {Ser{\'o}n}, {Taddia}, {Torres}, {Anderson}, {Baltay}, {Folatelli},
  {Galbany}, {Goobar}, {Hadjiyska}, {Hamuy}, {Kasliwal}, {Lidman}, {Nugent},
  {Perlmutter}, {Rabinowitz}, {Ryder}, {Schmidt}, {Shappee}, \&
  {Walker}}]{phillips19}
{Phillips}, M.~M., {Contreras}, C., {Hsiao}, E.~Y., {et~al.} 2019, \pasp, 131,
  014001, \dodoi{10.1088/1538-3873/aae8bd}

\bibitem[{{Phillips} {et~al.}(2022){Phillips}, {Ashall}, {Burns}, {Contreras},
  {Galbany}, {Hoeflich}, {Hsiao}, {Morrell}, {Nugent}, {Uddin}, {Baron},
  {Freedman}, {Harris}, {Krisciunas}, {Kumar}, {Lu}, {Persson}, {Piro},
  {Polin}, {Shahbandeh}, {Stritzinger}, \& {Suntzeff}}]{phillips22}
{Phillips}, M.~M., {Ashall}, C., {Burns}, C.~R., {et~al.} 2022, \apj, 938, 47,
  \dodoi{10.3847/1538-4357/ac9305}

\bibitem[{{Piro} \& {Morozova}(2016)}]{piro16}
{Piro}, A.~L., \& {Morozova}, V.~S. 2016, \apj, 826, 96,
  \dodoi{10.3847/0004-637X/826/1/96}

\bibitem[{{Plewa} {et~al.}(2004){Plewa}, {Calder}, \& {Lamb}}]{plewa04}
{Plewa}, T., {Calder}, A.~C., \& {Lamb}, D.~Q. 2004, \apjl, 612, L37,
  \dodoi{10.1086/424036}

\bibitem[{{Polin} {et~al.}(2021){Polin}, {Nugent}, \& {Kasen}}]{polin21}
{Polin}, A., {Nugent}, P., \& {Kasen}, D. 2021, \apj, 906, 65,
  \dodoi{10.3847/1538-4357/abcccc}

\bibitem[{{Poznanski} {et~al.}(2012){Poznanski}, {Prochaska}, \&
  {Bloom}}]{poznanski12}
{Poznanski}, D., {Prochaska}, J.~X., \& {Bloom}, J.~S. 2012, \mnras, 426, 1465,
  \dodoi{10.1111/j.1365-2966.2012.21796.x}

\bibitem[{{Pravdo} {et~al.}(1999){Pravdo}, {Rabinowitz}, {Helin}, {Lawrence},
  {Bambery}, {Clark}, {Groom}, {Levin}, {Lorre}, {Shaklan}, {Kervin},
  {Africano}, {Sydney}, \& {Soohoo}}]{pravdo99}
{Pravdo}, S.~H., {Rabinowitz}, D.~L., {Helin}, E.~F., {et~al.} 1999, \aj, 117,
  1616, \dodoi{10.1086/300769}

\bibitem[{{Prieto} {et~al.}(2007){Prieto}, {Garnavich}, {Phillips}, {DePoy},
  {Parrent}, {Pooley}, {Dwarkadas}, {Baron}, {Bassett}, {Becker}, {Cinabro},
  {DeJongh}, {Dilday}, {Doi}, {Frieman}, {Hogan}, {Holtzman}, {Jha}, {Kessler},
  {Konishi}, {Lampeitl}, {Marriner}, {Marshall}, {Miknaitis}, {Nichol},
  {Riess}, {Richmond}, {Romani}, {Sako}, {Schneider}, {Smith}, {Takanashi},
  {Tokita}, {van der Heyden}, {Yasuda}, {Zheng}, {Wheeler}, {Barentine},
  {Dembicky}, {Eastman}, {Frank}, {Ketzeback}, {McMillan}, {Morrell},
  {Folatelli}, {Contreras}, {Burns}, {Freedman}, {Gonzalez}, {Hamuy},
  {Krzeminski}, {Madore}, {Murphy}, {Persson}, {Roth}, \&
  {Suntzeff}}]{prieto07}
{Prieto}, J.~L., {Garnavich}, P.~M., {Phillips}, M.~M., {et~al.} 2007, arXiv
  e-prints, arXiv:0706.4088, \dodoi{10.48550/arXiv.0706.4088}

\bibitem[{{Pskovskii}(1984)}]{pskovskii84}
{Pskovskii}, Y.~P. 1984, \sovast, 28, 658

\bibitem[{{Quimby} {et~al.}(2007){Quimby}, {H{\"o}flich}, \&
  {Wheeler}}]{quimby07}
{Quimby}, R., {H{\"o}flich}, P., \& {Wheeler}, J.~C. 2007, \apj, 666, 1083,
  \dodoi{10.1086/520527}

\bibitem[{{Rana} \& {Basu}(1992)}]{rana92}
{Rana}, N.~C., \& {Basu}, S. 1992, \aap, 265, 499

\bibitem[{{Rest} {et~al.}(2008){Rest}, {Matheson}, {Blondin}, {Bergmann},
  {Welch}, {Suntzeff}, {Smith}, {Olsen}, {Prieto}, {Garg}, {Challis}, {Stubbs},
  {Hicken}, {Modjaz}, {Wood-Vasey}, {Zenteno}, {Damke}, {Newman}, {Huber},
  {Cook}, {Nikolaev}, {Becker}, {Miceli}, {Covarrubias}, {Morelli}, {Pignata},
  {Clocchiatti}, {Minniti}, \& {Foley}}]{rest08}
{Rest}, A., {Matheson}, T., {Blondin}, S., {et~al.} 2008, \apj, 680, 1137,
  \dodoi{10.1086/587158}

\bibitem[{{Riess} {et~al.}(1996){Riess}, {Press}, \& {Kirshner}}]{riess96}
{Riess}, A.~G., {Press}, W.~H., \& {Kirshner}, R.~P. 1996, \apj, 473, 88,
  \dodoi{10.1086/178129}

\bibitem[{{Riess} {et~al.}(1999){Riess}, {Kirshner}, {Schmidt}, {Jha},
  {Challis}, {Garnavich}, {Esin}, {Carpenter}, {Grashius}, {Schild}, {Berlind},
  {Huchra}, {Prosser}, {Falco}, {Benson}, {Brice{\~n}o}, {Brown}, {Caldwell},
  {dell'Antonio}, {Filippenko}, {Goodman}, {Grogin}, {Groner}, {Hughes},
  {Green}, {Jansen}, {Kleyna}, {Luu}, {Macri}, {McLeod}, {McLeod}, {McNamara},
  {McLean}, {Milone}, {Mohr}, {Moraru}, {Peng}, {Peters}, {Prestwich},
  {Stanek}, {Szentgyorgyi}, \& {Zhao}}]{riess99}
{Riess}, A.~G., {Kirshner}, R.~P., {Schmidt}, B.~P., {et~al.} 1999, \aj, 117,
  707, \dodoi{10.1086/300738}

\bibitem[{{Roming} {et~al.}(2005){Roming}, {Kennedy}, {Mason}, {Nousek}, {Ahr},
  {Bingham}, {Broos}, {Carter}, {Hancock}, {Huckle}, {Hunsberger}, {Kawakami},
  {Killough}, {Koch}, {McLelland}, {Smith}, {Smith}, {Soto}, {Boyd},
  {Breeveld}, {Holland}, {Ivanushkina}, {Pryzby}, {Still}, \&
  {Stock}}]{roming05}
{Roming}, P. W.~A., {Kennedy}, T.~E., {Mason}, K.~O., {et~al.} 2005, \ssr, 120,
  95, \dodoi{10.1007/s11214-005-5095-4}

\bibitem[{{Roth} {et~al.}(2005){Roth}, {Kelz}, {Fechner}, {Hahn}, {Bauer},
  {Becker}, {B{\"o}hm}, {Christensen}, {Dionies}, {Paschke}, {Popow}, {Wolter},
  {Schmoll}, {Laux}, \& {Altmann}}]{roth05}
{Roth}, M.~M., {Kelz}, A., {Fechner}, T., {et~al.} 2005, \pasp, 117, 620,
  \dodoi{10.1086/429877}

\bibitem[{{Ruiz-Lapuente} {et~al.}(1992){Ruiz-Lapuente}, {Cappellaro},
  {Turatto}, {Gouiffes}, {Danziger}, {della Valle}, \&
  {Lucy}}]{ruiz-lapuente92}
{Ruiz-Lapuente}, P., {Cappellaro}, E., {Turatto}, M., {et~al.} 1992, \apjl,
  387, L33, \dodoi{10.1086/186299}

\bibitem[{{Salpeter}(1955)}]{salpeter55}
{Salpeter}, E.~E. 1955, \apj, 121, 161, \dodoi{10.1086/145971}

\bibitem[{{Sand} {et~al.}(2016){Sand}, {Hsiao}, {Banerjee}, {Marion},
  {Diamond}, {Joshi}, {Parrent}, {Phillips}, {Stritzinger}, \&
  {Venkataraman}}]{sand16}
{Sand}, D.~J., {Hsiao}, E.~Y., {Banerjee}, D.~P.~K., {et~al.} 2016, \apjl, 822,
  L16, \dodoi{10.3847/2041-8205/822/1/L16}

\bibitem[{{Sand} {et~al.}(2021){Sand}, {Sarbadhicary}, {Pellegrino}, {Misra},
  {Dastidar}, {Brown}, {Itagaki}, {Valenti}, {Swift}, {Andrews}, {Bostroem},
  {Burke}, {Chomiuk}, {Dong}, {Galbany}, {Graham}, {Hiramatsu}, {Howell},
  {Hsiao}, {Janzen}, {Jencson}, {Lundquist}, {McCully}, {Reichart}, {Smith},
  {Wang}, \& {Wyatt}}]{sand21}
{Sand}, D.~J., {Sarbadhicary}, S.~K., {Pellegrino}, C., {et~al.} 2021, \apj,
  922, 21, \dodoi{10.3847/1538-4357/ac20da}

\bibitem[{{Sasdelli} {et~al.}(2014){Sasdelli}, {Mazzali}, {Pian}, {Nomoto},
  {Hachinger}, {Cappellaro}, \& {Benetti}}]{sasdelli14}
{Sasdelli}, M., {Mazzali}, P.~A., {Pian}, E., {et~al.} 2014, \mnras, 445, 711,
  \dodoi{10.1093/mnras/stu1777}

\bibitem[{{Sauer} {et~al.}(2008){Sauer}, {Mazzali}, {Blondin}, {Stehle},
  {Benetti}, {Challis}, {Filippenko}, {Kirshner}, {Li}, \&
  {Matheson}}]{sauer08}
{Sauer}, D.~N., {Mazzali}, P.~A., {Blondin}, S., {et~al.} 2008, \mnras, 391,
  1605, \dodoi{10.1111/j.1365-2966.2008.14018.x}

\bibitem[{{Scalzo} {et~al.}(2012){Scalzo}, {Aldering}, {Antilogus}, {Aragon},
  {Bailey}, {Baltay}, {Bongard}, {Buton}, {Canto}, {Cellier-Holzem},
  {Childress}, {Chotard}, {Copin}, {Fakhouri}, {Gangler}, {Guy}, {Hsiao},
  {Kerschhaggl}, {Kowalski}, {Nugent}, {Paech}, {Pain}, {Pecontal}, {Pereira},
  {Perlmutter}, {Rabinowitz}, {Rigault}, {Runge}, {Smadja}, {Tao}, {Thomas},
  {Weaver}, {Wu}, \& {Nearby Supernova Factory}}]{scalzo12}
{Scalzo}, R., {Aldering}, G., {Antilogus}, P., {et~al.} 2012, \apj, 757, 12,
  \dodoi{10.1088/0004-637X/757/1/12}

\bibitem[{{Scalzo} {et~al.}(2014{\natexlab{a}}){Scalzo}, {Aldering},
  {Antilogus}, {Aragon}, {Bailey}, {Baltay}, {Bongard}, {Buton},
  {Cellier-Holzem}, {Childress}, {Chotard}, {Copin}, {Fakhouri}, {Gangler},
  {Guy}, {Kim}, {Kowalski}, {Kromer}, {Nordin}, {Nugent}, {Paech}, {Pain},
  {Pecontal}, {Pereira}, {Perlmutter}, {Rabinowitz}, {Rigault}, {Runge},
  {Saunders}, {Sim}, {Smadja}, {Tao}, {Taubenberger}, {Thomas}, {Weaver}, \&
  {Nearby Supernova Factory}}]{scalzo14a}
---. 2014{\natexlab{a}}, \mnras, 440, 1498, \dodoi{10.1093/mnras/stu350}

\bibitem[{{Scalzo} {et~al.}(2014{\natexlab{b}}){Scalzo}, {Childress}, {Tucker},
  {Yuan}, {Schmidt}, {Brown}, {Contreras}, {Morrell}, {Hsiao}, {Burns},
  {Phillips}, {Campillay}, {Gonzalez}, {Krisciunas}, {Stritzinger}, {Graham},
  {Parrent}, {Valenti}, {Lidman}, {Schaefer}, {Scott}, {Fraser}, {Gal-Yam},
  {Inserra}, {Maguire}, {Smartt}, {Sollerman}, {Sullivan}, {Taddia}, {Yaron},
  {Young}, {Taubenberger}, {Baltay}, {Ellman}, {Feindt}, {Hadjiyska},
  {McKinnon}, {Nugent}, {Rabinowitz}, \& {Walker}}]{scalzo14b}
{Scalzo}, R.~A., {Childress}, M., {Tucker}, B., {et~al.} 2014{\natexlab{b}},
  \mnras, 445, 30, \dodoi{10.1093/mnras/stu1723}

\bibitem[{{Schlafly} \& {Finkbeiner}(2011)}]{schlafly11}
{Schlafly}, E.~F., \& {Finkbeiner}, D.~P. 2011, \apj, 737, 103,
  \dodoi{10.1088/0004-637X/737/2/103}

\bibitem[{{Schlegel} {et~al.}(1998){Schlegel}, {Finkbeiner}, \&
  {Davis}}]{schlegel98}
{Schlegel}, D.~J., {Finkbeiner}, D.~P., \& {Davis}, M. 1998, \apj, 500, 525,
  \dodoi{10.1086/305772}

\bibitem[{{Schmidt} {et~al.}(1994){Schmidt}, {Kirshner}, {Leibundgut}, {Wells},
  {Porter}, {Ruiz-Lapuente}, {Challis}, \& {Filippenko}}]{schmidt94}
{Schmidt}, B.~P., {Kirshner}, R.~P., {Leibundgut}, B., {et~al.} 1994, \apjl,
  434, L19, \dodoi{10.1086/187562}

\bibitem[{{Seitenzahl} {et~al.}(2016){Seitenzahl}, {Kromer}, {Ohlmann},
  {Ciaraldi-Schoolmann}, {Marquardt}, {Fink}, {Hillebrandt}, {Pakmor},
  {R{\"o}pke}, {Ruiter}, {Sim}, \& {Taubenberger}}]{seitenzahl16}
{Seitenzahl}, I.~R., {Kromer}, M., {Ohlmann}, S.~T., {et~al.} 2016, \aap, 592,
  A57, \dodoi{10.1051/0004-6361/201527251}

\bibitem[{{Shappee} {et~al.}(2013){Shappee}, {Stanek}, {Pogge}, \&
  {Garnavich}}]{shappee13}
{Shappee}, B.~J., {Stanek}, K.~Z., {Pogge}, R.~W., \& {Garnavich}, P.~M. 2013,
  \apjl, 762, L5, \dodoi{10.1088/2041-8205/762/1/L5}

\bibitem[{{Shappee} {et~al.}(2014){Shappee}, {Prieto}, {Grupe}, {Kochanek},
  {Stanek}, {De Rosa}, {Mathur}, {Zu}, {Peterson}, {Pogge}, {Komossa}, {Im},
  {Jencson}, {Holoien}, {Basu}, {Beacom}, {Szczygie{\l}}, {Brimacombe},
  {Adams}, {Campillay}, {Choi}, {Contreras}, {Dietrich}, {Dubberley},
  {Elphick}, {Foale}, {Giustini}, {Gonzalez}, {Hawkins}, {Howell}, {Hsiao},
  {Koss}, {Leighly}, {Morrell}, {Mudd}, {Mullins}, {Nugent}, {Parrent},
  {Phillips}, {Pojmanski}, {Rosing}, {Ross}, {Sand}, {Terndrup}, {Valenti},
  {Walker}, \& {Yoon}}]{shappee14}
{Shappee}, B.~J., {Prieto}, J.~L., {Grupe}, D., {et~al.} 2014, \apj, 788, 48,
  \dodoi{10.1088/0004-637X/788/1/48}

\bibitem[{{Shappee} {et~al.}(2016){Shappee}, {Piro}, {Holoien}, {Prieto},
  {Contreras}, {Itagaki}, {Burns}, {Kochanek}, {Stanek}, {Alper}, {Basu},
  {Beacom}, {Bersier}, {Brimacombe}, {Conseil}, {Danilet}, {Dong}, {Falco},
  {Grupe}, {Hsiao}, {Kiyota}, {Morrell}, {Nicolas}, {Phillips}, {Pojmanski},
  {Simonian}, {Stritzinger}, {Szczygie{\l}}, {Taddia}, {Thompson},
  {Thorstensen}, {Wagner}, \& {Wo{\'z}niak}}]{shappee16}
{Shappee}, B.~J., {Piro}, A.~L., {Holoien}, T.~W.~S., {et~al.} 2016, \apj, 826,
  144, \dodoi{10.3847/0004-637X/826/2/144}

\bibitem[{{Shappee} {et~al.}(2019){Shappee}, {Holoien}, {Drout}, {Auchettl},
  {Stritzinger}, {Kochanek}, {Stanek}, {Shaya}, {Narayan}, {ASAS-SN}, {Brown},
  {Bose}, {Bersier}, {Brimacombe}, {Chen}, {Dong}, {Holmbo}, {Katz},
  {Mu{\~n}oz}, {Mutel}, {Post}, {Prieto}, {Shields}, {Tallon}, {Thompson},
  {Vallely}, {Villanueva}, {ATLAS}, {Denneau}, {Flewelling}, {Heinze}, {Smith},
  {Stalder}, {Tonry}, {Weiland}, {Kepler/K2}, {Barclay}, {Barentsen}, {Cody},
  {Dotson}, {Foerster}, {Garnavich}, {Gully-Santiago}, {Hedges}, {Howell},
  {Kasen}, {Margheim}, {Mushotzky}, {Rest}, {Tucker}, {Villar}, {Zenteno},
  {Kepler Spacecraft Team}, {Beerman}, {Bjella}, {Castillo}, {Coughlin},
  {Elsaesser}, {Flynn}, {Gangopadhyay}, {Griest}, {Hanley}, {Kampmeier},
  {Kloetzel}, {Kohnert}, {Labonde}, {Larsen}, {Larson}, {McCalmont-Everton},
  {McGinn}, {Migliorini}, {Moffatt}, {Muszynski}, {Nystrom}, {Osborne},
  {Packard}, {Peterson}, {Redick}, {Reedy}, {Ross}, {Spencer}, {Steward}, {Van
  Cleve}, {Cardoso}, {Weschler}, {Wheaton}, {Pan-STARRS}, {Bulger}, {Chambers},
  {Flewelling}, {Huber}, {Lowe}, {Magnier}, {Schultz}, {Waters}, {Willman},
  {PTSS/TNTS}, {Baron}, {Chen}, {Derkacy}, {Huang}, {Li}, {Li}, {Li}, {Mo},
  {Rui}, {Sai}, {Wang}, {Wang}, {Wang}, {Xiang}, {Zhang}, {Zhang}, {Zhang},
  {Zhang}, {Zhang}, {Zhao}, {Brown}, {Hermes}, {Nordin}, {Points}, {S{\'o}dor},
  {Strampelli}, \& {Zenteno}}]{shappee19}
{Shappee}, B.~J., {Holoien}, T.~W.~S., {Drout}, M.~R., {et~al.} 2019, \apj,
  870, 13, \dodoi{10.3847/1538-4357/aaec79}

\bibitem[{{Sharma} {et~al.}(2023){Sharma}, {Sollerman}, {Fremling}, {Kulkarni},
  {De}, {Irani}, {Schulze}, {Strotjohann}, {Gal-Yam}, {Maguire}, {Perley},
  {Bellm}, {Kool}, {Brink}, {Bruch}, {Deckers}, {Dekany}, {Dugas},
  {Filippenko}, {Goldwasser}, {Graham}, {Graham}, {Groom}, {Hankins},
  {Jencson}, {Johansson}, {Karambelkar}, {Kasliwal}, {Masci}, {Medford},
  {Neill}, {Nir}, {Riddle}, {Rigault}, {Schweyer}, {Terwel}, {Yan}, {Yang}, \&
  {Yao}}]{sharma23}
{Sharma}, Y., {Sollerman}, J., {Fremling}, C., {et~al.} 2023, \apj, 948, 52,
  \dodoi{10.3847/1538-4357/acbc16}

\bibitem[{{Shen} {et~al.}(2013){Shen}, {Guillochon}, \& {Foley}}]{shen13}
{Shen}, K.~J., {Guillochon}, J., \& {Foley}, R.~J. 2013, \apjl, 770, L35,
  \dodoi{10.1088/2041-8205/770/2/L35}

\bibitem[{{Silverman} {et~al.}(2013{\natexlab{a}}){Silverman}, {Ganeshalingam},
  \& {Filippenko}}]{silverman13a}
{Silverman}, J.~M., {Ganeshalingam}, M., \& {Filippenko}, A.~V.
  2013{\natexlab{a}}, \mnras, 430, 1030, \dodoi{10.1093/mnras/sts674}

\bibitem[{{Silverman} {et~al.}(2012){Silverman}, {Foley}, {Filippenko},
  {Ganeshalingam}, {Barth}, {Chornock}, {Griffith}, {Kong}, {Lee}, {Leonard},
  {Matheson}, {Miller}, {Steele}, {Barris}, {Bloom}, {Cobb}, {Coil},
  {Desroches}, {Gates}, {Ho}, {Jha}, {Kandrashoff}, {Li}, {Mandel}, {Modjaz},
  {Moore}, {Mostardi}, {Papenkova}, {Park}, {Perley}, {Poznanski}, {Reuter},
  {Scala}, {Serduke}, {Shields}, {Swift}, {Tonry}, {Van Dyk}, {Wang}, \&
  {Wong}}]{silverman12}
{Silverman}, J.~M., {Foley}, R.~J., {Filippenko}, A.~V., {et~al.} 2012, \mnras,
  425, 1789, \dodoi{10.1111/j.1365-2966.2012.21270.x}

\bibitem[{{Silverman} {et~al.}(2013{\natexlab{b}}){Silverman}, {Nugent},
  {Gal-Yam}, {Sullivan}, {Howell}, {Filippenko}, {Arcavi}, {Ben-Ami}, {Bloom},
  {Cenko}, {Cao}, {Chornock}, {Clubb}, {Coil}, {Foley}, {Graham}, {Griffith},
  {Horesh}, {Kasliwal}, {Kulkarni}, {Leonard}, {Li}, {Matheson}, {Miller},
  {Modjaz}, {Ofek}, {Pan}, {Perley}, {Poznanski}, {Quimby}, {Steele},
  {Sternberg}, {Xu}, \& {Yaron}}]{silverman13b}
{Silverman}, J.~M., {Nugent}, P.~E., {Gal-Yam}, A., {et~al.}
  2013{\natexlab{b}}, \apjs, 207, 3, \dodoi{10.1088/0067-0049/207/1/3}

\bibitem[{{Silverman} {et~al.}(2013{\natexlab{c}}){Silverman}, {Nugent},
  {Gal-Yam}, {Sullivan}, {Howell}, {Filippenko}, {Pan}, {Cenko}, \&
  {Hook}}]{silverman13c}
---. 2013{\natexlab{c}}, \apj, 772, 125, \dodoi{10.1088/0004-637X/772/2/125}

\bibitem[{{Sivaraman} {et~al.}(1991){Sivaraman}, {Prabhu}, {Anupama},
  {Schmeer}, {Smith}, {Royer}, {Lubcke}, {Villi}, \& {Cortini}}]{sivaraman91}
{Sivaraman}, K.~R., {Prabhu}, T.~P., {Anupama}, G.~C., {et~al.} 1991, \iaucirc,
  5255, 1

\bibitem[{{Smartt} {et~al.}(2015){Smartt}, {Valenti}, {Fraser}, {Inserra},
  {Young}, {Sullivan}, {Pastorello}, {Benetti}, {Gal-Yam}, {Knapic},
  {Molinaro}, {Smareglia}, {Smith}, {Taubenberger}, {Yaron}, {Anderson},
  {Ashall}, {Balland}, {Baltay}, {Barbarino}, {Bauer}, {Baumont}, {Bersier},
  {Blagorodnova}, {Bongard}, {Botticella}, {Bufano}, {Bulla}, {Cappellaro},
  {Campbell}, {Cellier-Holzem}, {Chen}, {Childress}, {Clocchiatti},
  {Contreras}, {Dall'Ora}, {Danziger}, {de Jaeger}, {De Cia}, {Della Valle},
  {Dennefeld}, {Elias-Rosa}, {Elman}, {Feindt}, {Fleury}, {Gall},
  {Gonzalez-Gaitan}, {Galbany}, {Morales Garoffolo}, {Greggio}, {Guillou},
  {Hachinger}, {Hadjiyska}, {Hage}, {Hillebrandt}, {Hodgkin}, {Hsiao}, {James},
  {Jerkstrand}, {Kangas}, {Kankare}, {Kotak}, {Kromer}, {Kuncarayakti},
  {Leloudas}, {Lundqvist}, {Lyman}, {Hook}, {Maguire}, {Manulis}, {Margheim},
  {Mattila}, {Maund}, {Mazzali}, {McCrum}, {McKinnon}, {Moreno-Raya},
  {Nicholl}, {Nugent}, {Pain}, {Pignata}, {Phillips}, {Polshaw}, {Pumo},
  {Rabinowitz}, {Reilly}, {Romero-Ca{\~n}izales}, {Scalzo}, {Schmidt},
  {Schulze}, {Sim}, {Sollerman}, {Taddia}, {Tartaglia}, {Terreran},
  {Tomasella}, {Turatto}, {Walker}, {Walton}, {Wyrzykowski}, {Yuan}, \&
  {Zampieri}}]{smartt15}
{Smartt}, S.~J., {Valenti}, S., {Fraser}, M., {et~al.} 2015, \aap, 579, A40,
  \dodoi{10.1051/0004-6361/201425237}

\bibitem[{{Smitka} {et~al.}(2015){Smitka}, {Brown}, {Suntzeff}, {Zhang},
  {Zhai}, {Wang}, {Mo}, \& {Zhang}}]{smitka15}
{Smitka}, M.~T., {Brown}, P.~J., {Suntzeff}, N.~B., {et~al.} 2015, \apj, 813,
  30, \dodoi{10.1088/0004-637X/813/1/30}

\bibitem[{{Sparks} {et~al.}(1999){Sparks}, {Macchetto}, {Panagia}, {Boffi},
  {Branch}, {Hazen}, \& {Della Valle}}]{sparks99}
{Sparks}, W.~B., {Macchetto}, F., {Panagia}, N., {et~al.} 1999, \apj, 523, 585,
  \dodoi{10.1086/307766}

\bibitem[{{Spyromilio} {et~al.}(1992){Spyromilio}, {Meikle}, {Allen}, \&
  {Graham}}]{spyromilio92}
{Spyromilio}, J., {Meikle}, W.~P.~S., {Allen}, D.~A., \& {Graham}, J.~R. 1992,
  \mnras, 258, 53P, \dodoi{10.1093/mnras/258.1.53P}

\bibitem[{{Stahl} {et~al.}(2019){Stahl}, {Zheng}, {de Jaeger}, {Filippenko},
  {Bigley}, {Blanchard}, {Blanchard}, {Brink}, {Cargill}, {Casper}, {Channa},
  {Choi}, {Choksi}, {Chu}, {Clubb}, {Cohen}, {Ellison}, {Falcon}, {Fazeli},
  {Fuller}, {Ganeshalingam}, {Gates}, {Gould}, {Halevi}, {Hayakawa},
  {Hestenes}, {Jeffers}, {Joubert}, {Kandrashoff}, {Kim}, {Kim}, {Kislak},
  {Kleiser}, {Kong}, {de Kouchkovsky}, {Krishnan}, {Kumar}, {Leja}, {Leonard},
  {Li}, {Li}, {Lu}, {Mason}, {Molloy}, {Pina}, {Rex}, {Ross}, {Stegman},
  {Tang}, {Thrasher}, {Wang}, {Wilkins}, {Yuk}, {Yunus}, \& {Zhang}}]{stahl19}
{Stahl}, B.~E., {Zheng}, W., {de Jaeger}, T., {et~al.} 2019, \mnras, 490, 3882,
  \dodoi{10.1093/mnras/stz2742}

\bibitem[{{Stahl} {et~al.}(2020){Stahl}, {Zheng}, {de Jaeger}, {Brink},
  {Filippenko}, {Silverman}, {Cenko}, {Clubb}, {Graham}, {Halevi}, {Kelly},
  {Kleiser}, {Shivvers}, {Yuk}, {Cobb}, {Fox}, {Kandrashoff}, {Kong},
  {Mauerhan}, {Wang}, \& {Wang}}]{stahl20}
---. 2020, \mnras, 492, 4325, \dodoi{10.1093/mnras/staa102}

\bibitem[{{Stanek}(2017)}]{stanek17}
{Stanek}, K.~Z. 2017, Transient Name Server Discovery Report, 2017-449, 1

\bibitem[{{Stritzinger} {et~al.}(2018){Stritzinger}, {Shappee}, {Piro},
  {Ashall}, {Baron}, {Hoeflich}, {Holmbo}, {Holoien}, {Phillips}, {Burns},
  {Contreras}, {Morrell}, \& {Tucker}}]{stritzinger18}
{Stritzinger}, M.~D., {Shappee}, B.~J., {Piro}, A.~L., {et~al.} 2018, \apjl,
  864, L35, \dodoi{10.3847/2041-8213/aadd46}

\bibitem[{{Sullivan} {et~al.}(2006){Sullivan}, {Le Borgne}, {Pritchet},
  {Hodsman}, {Neill}, {Howell}, {Carlberg}, {Astier}, {Aubourg}, {Balam},
  {Basa}, {Conley}, {Fabbro}, {Fouchez}, {Guy}, {Hook}, {Pain},
  {Palanque-Delabrouille}, {Perrett}, {Regnault}, {Rich}, {Taillet}, {Baumont},
  {Bronder}, {Ellis}, {Filiol}, {Lusset}, {Perlmutter}, {Ripoche}, \&
  {Tao}}]{sullivan06}
{Sullivan}, M., {Le Borgne}, D., {Pritchet}, C.~J., {et~al.} 2006, \apj, 648,
  868, \dodoi{10.1086/506137}

\bibitem[{{Taddia} {et~al.}(2012){Taddia}, {Stritzinger}, {Phillips}, {Burns},
  {Heinrich-Josties}, {Morrell}, {Sollerman}, {Valenti}, {Anderson}, {Boldt},
  {Campillay}, {Castellon}, {Contreras}, {Folatelli}, {Freedman}, {Hamuy},
  {Krzeminski}, {Leloudas}, {Maeda}, {Persson}, {Roth}, \&
  {Suntzeff}}]{taddia12}
{Taddia}, F., {Stritzinger}, M.~D., {Phillips}, M.~M., {et~al.} 2012, \aap,
  545, L7, \dodoi{10.1051/0004-6361/201220105}

\bibitem[{{Taubenberger}(2017)}]{taubenberger17}
{Taubenberger}, S. 2017, in Handbook of Supernovae, ed. A.~W. {Alsabti} \&
  P.~{Murdin}, 317, \dodoi{10.1007/978-3-319-21846-5_37}

\bibitem[{{Taubenberger} {et~al.}(2013){Taubenberger}, {Kromer}, {Hachinger},
  {Mazzali}, {Benetti}, {Nugent}, {Scalzo}, {Pakmor}, {Stanishev},
  {Spyromilio}, {Bufano}, {Sim}, {Leibundgut}, \&
  {Hillebrandt}}]{taubenberger13}
{Taubenberger}, S., {Kromer}, M., {Hachinger}, S., {et~al.} 2013, \mnras, 432,
  3117, \dodoi{10.1093/mnras/stt668}

\bibitem[{{Thormann} {et~al.}(2009){Thormann}, {Sugerman}, \&
  {Lonsdale}}]{thormann09}
{Thormann}, A., {Sugerman}, B., \& {Lonsdale}, S. 2009, in American
  Astronomical Society Meeting Abstracts, Vol. 213, American Astronomical
  Society Meeting Abstracts \#213, 412.08

\bibitem[{{Tody}(1986)}]{tody86}
{Tody}, D. 1986, in Society of Photo-Optical Instrumentation Engineers (SPIE)
  Conference Series, Vol. 627, Instrumentation in astronomy VI, ed. D.~L.
  {Crawford}, 733, \dodoi{10.1117/12.968154}

\bibitem[{{Tomasella} {et~al.}(2017){Tomasella}, {Benetti}, {Cappellaro},
  {Pastorello}, {Ochner}, {Turatto}, \& {Terreran}}]{tomasella17}
{Tomasella}, L., {Benetti}, S., {Cappellaro}, E., {et~al.} 2017, The
  Astronomer's Telegram, 10306, 1

\bibitem[{{Tonry}(2011)}]{tonry11}
{Tonry}, J.~L. 2011, \pasp, 123, 58, \dodoi{10.1086/657997}

\bibitem[{{Townsley} {et~al.}(2019){Townsley}, {Miles}, {Shen}, \&
  {Kasen}}]{townsley19}
{Townsley}, D.~M., {Miles}, B.~J., {Shen}, K.~J., \& {Kasen}, D. 2019, \apjl,
  878, L38, \dodoi{10.3847/2041-8213/ab27cd}

\bibitem[{{Tucker} {et~al.}(2022{\natexlab{a}}){Tucker}, {Ashall}, {Shappee},
  {Kochanek}, {Stanek}, \& {Garnavich}}]{tucker22a}
{Tucker}, M.~A., {Ashall}, C., {Shappee}, B.~J., {et~al.} 2022{\natexlab{a}},
  \apjl, 926, L25, \dodoi{10.3847/2041-8213/ac4fbd}

\bibitem[{{Tucker} {et~al.}(2020){Tucker}, {Shappee}, {Vallely}, {Stanek},
  {Prieto}, {Botyanszki}, {Kochanek}, {Anderson}, {Brown}, {Galbany},
  {Holoien}, {Hsiao}, {Kumar}, {Kuncarayakti}, {Morrell}, {Phillips},
  {Stritzinger}, \& {Thompson}}]{tucker20}
{Tucker}, M.~A., {Shappee}, B.~J., {Vallely}, P.~J., {et~al.} 2020, \mnras,
  493, 1044, \dodoi{10.1093/mnras/stz3390}

\bibitem[{{Tucker} {et~al.}(2022{\natexlab{b}}){Tucker}, {Shappee}, {Huber},
  {Payne}, {Do}, {Hinkle}, {de Jaeger}, {Ashall}, {Desai}, {Hoogendam},
  {Aldering}, {Auchettl}, {Baranec}, {Bulger}, {Chambers}, {Chun}, {Hodapp},
  {Lowe}, {McKay}, {Rampy}, {Rubin}, \& {Tonry}}]{tucker22b}
{Tucker}, M.~A., {Shappee}, B.~J., {Huber}, M.~E., {et~al.} 2022{\natexlab{b}},
  \pasp, 134, 124502, \dodoi{10.1088/1538-3873/aca719}

\bibitem[{{Tully} {et~al.}(2016){Tully}, {Courtois}, \& {Sorce}}]{tully16}
{Tully}, R.~B., {Courtois}, H.~M., \& {Sorce}, J.~G. 2016, \aj, 152, 50,
  \dodoi{10.3847/0004-6256/152/2/50}

\bibitem[{{Uddin} {et~al.}(2020){Uddin}, {Burns}, {Phillips}, {Suntzeff},
  {Contreras}, {Hsiao}, {Morrell}, {Galbany}, {Stritzinger}, {Hoeflich},
  {Ashall}, {Piro}, {Freedman}, {Persson}, {Krisciunas}, \& {Brown}}]{uddin20}
{Uddin}, S.~A., {Burns}, C.~R., {Phillips}, M.~M., {et~al.} 2020, \apj, 901,
  143, \dodoi{10.3847/1538-4357/abafb7}

\bibitem[{{Uddin} {et~al.}(2023){Uddin}, {Burns}, {Phillips}, {Suntzeff},
  {Freedman}, {Brown}, {Morrell}, {Hamuy}, {Krisciunas}, {Wang}, {Hsiao},
  {Goobar}, {Perlmutter}, {Lu}, {Stritzinger}, {Anderson}, {Ashall},
  {Hoeflich}, {Shappee}, {Persson}, {Piro}, {Baron}, {Contreras}, {Galbany},
  {Kumar}, {Shahbandeh}, {Davis}, {Anais}, {Busta}, {Campillay},
  {Castell{\'o}n}, {Corco}, {Diamond}, {Gall}, {Gonzalez}, {Holmbo}, {Roth},
  {Ser{\'o}n}, {Taddia}, {Torres}, {Baltay}, {Folatelli}, {Hadjiyska},
  {Kasliwal}, {Nugent}, {Rabinowitz}, \& {Ryder}}]{uddin23}
---. 2023, arXiv e-prints, arXiv:2308.01875, \dodoi{10.48550/arXiv.2308.01875}

\bibitem[{{Wang} {et~al.}(2007){Wang}, {Baade}, \& {Patat}}]{wang07}
{Wang}, L., {Baade}, D., \& {Patat}, F. 2007, Science, 315, 212,
  \dodoi{10.1126/science.1121656}

\bibitem[{{Wang} \& {Wheeler}(2008)}]{wang08}
{Wang}, L., \& {Wheeler}, J.~C. 2008, \araa, 46, 433,
  \dodoi{10.1146/annurev.astro.46.060407.145139}

\bibitem[{{Wang} {et~al.}(2024){Wang}, {Hu}, {Wang}, {Yang}, {Gomez}, {Chen},
  {Hu}, {Chen}, {Mo}, {Wang}, {Baade}, {Hoeflich}, {Wheeler}, {Pignata},
  {Burke}, {Hiramatsu}, {Howell}, {McCully}, {Pellegrino}, {Galbany}, {Hsiao},
  {Sand}, {Zhang}, {Uddin}, {Anderson}, {Ashall}, {Cheng}, {Gromadzki},
  {Inserra}, {Lin}, {Morrell}, {Morales-Garoffolo}, {M{\"u}ller-Bravo},
  {Nicholl}, {Gonzalez}, {Phillips}, {Pineda-Garc{\'\i}a}, {Sai}, {Smith},
  {Shahbandeh}, {Srivastav}, {Stritzinger}, {Yang}, {Young}, {Yu}, \&
  {Zhang}}]{wang24}
{Wang}, L., {Hu}, M., {Wang}, L., {et~al.} 2024, Nature Astronomy,
  \dodoi{10.1038/s41550-024-02197-9}

\bibitem[{{Wang} {et~al.}(2009){Wang}, {Filippenko}, {Ganeshalingam}, {Li},
  {Silverman}, {Wang}, {Chornock}, {Foley}, {Gates}, {Macomber}, {Serduke},
  {Steele}, \& {Wong}}]{wang09}
{Wang}, X., {Filippenko}, A.~V., {Ganeshalingam}, M., {et~al.} 2009, \apjl,
  699, L139, \dodoi{10.1088/0004-637X/699/2/L139}

\bibitem[{{Wheeler} {et~al.}(1991){Wheeler}, {Smith}, \& {Gilmore}}]{wheeler91}
{Wheeler}, C., {Smith}, V., \& {Gilmore}, A.~C. 1991, \iaucirc, 5256, 1

\bibitem[{{Wheeler} {et~al.}(1998){Wheeler}, {H{\"o}flich}, {Harkness}, \&
  {Spyromilio}}]{wheeler98}
{Wheeler}, J.~C., {H{\"o}flich}, P., {Harkness}, R.~P., \& {Spyromilio}, J.
  1998, \apj, 496, 908, \dodoi{10.1086/305427}

\bibitem[{{Wilk} {et~al.}(2020){Wilk}, {Hillier}, \& {Dessart}}]{wilk20}
{Wilk}, K.~D., {Hillier}, D.~J., \& {Dessart}, L. 2020, \mnras, 494, 2221,
  \dodoi{10.1093/mnras/staa640}

\bibitem[{{Wood-Vasey} {et~al.}(2002){Wood-Vasey}, {Aldering}, \&
  {Nugent}}]{wood-vasey02}
{Wood-Vasey}, W.~M., {Aldering}, G., \& {Nugent}, P. 2002, \iaucirc, 8019, 2

\bibitem[{{Wood-Vasey} {et~al.}(2004){Wood-Vasey}, {Wang}, \&
  {Aldering}}]{wood-vasey04}
{Wood-Vasey}, W.~M., {Wang}, L., \& {Aldering}, G. 2004, \apj, 616, 339,
  \dodoi{10.1086/424826}

\bibitem[{{Xu} {et~al.}(2017){Xu}, {Li}, {Li}, {Li}, {Wang}, {Tan}, {Zhao},
  {Wang}, \& {Zhang}}]{xu17}
{Xu}, Z., {Li}, W., {Li}, B., {et~al.} 2017, Transient Name Server Discovery
  Report, 2017-1155, 1

\bibitem[{{Yamaoka} {et~al.}(1992){Yamaoka}, {Nomoto}, {Shigeyama}, \&
  {Thielemann}}]{yamaoka92}
{Yamaoka}, H., {Nomoto}, K., {Shigeyama}, T., \& {Thielemann}, F.-K. 1992,
  \apjl, 393, L55, \dodoi{10.1086/186450}

\bibitem[{{Yang} {et~al.}(2022){Yang}, {Wang}, {Suntzeff}, {Hu}, {Aldoroty},
  {Brown}, {Krisciunas}, {Arcavi}, {Burke}, {Galbany}, {Hiramatsu},
  {Hosseinzadeh}, {Howell}, {McCully}, {Pellegrino}, \& {Valenti}}]{yang22}
{Yang}, J., {Wang}, L., {Suntzeff}, N., {et~al.} 2022, \apj, 938, 83,
  \dodoi{10.3847/1538-4357/ac8c97}

\bibitem[{{Yang} {et~al.}(2023){Yang}, {Baade}, {Hoeflich}, {Wang}, {Cikota},
  {Chen}, {Burke}, {Hiramatsu}, {Pellegrino}, {Howell}, {McCully}, {Valenti},
  {Schulze}, {Gal-Yam}, {Wang}, {Filippenko}, {Maeda}, {Bulla}, {Yao}, {Maund},
  {Patat}, {Spyromilio}, {Wheeler}, {Rau}, {Hu}, {Li}, {Andrews}, {Galbany},
  {Sand}, {Shahbandeh}, {Hsiao}, \& {Wang}}]{yang23}
{Yang}, Y., {Baade}, D., {Hoeflich}, P., {et~al.} 2023, \mnras, 519, 1618,
  \dodoi{10.1093/mnras/stac3477}

\bibitem[{{Yaron} \& {Gal-Yam}(2012)}]{yaron12}
{Yaron}, O., \& {Gal-Yam}, A. 2012, \pasp, 124, 668, \dodoi{10.1086/666656}

\bibitem[{{Yin} {et~al.}(2018){Yin}, {Jones}, {Pan}, {Foley}, \&
  {Siellez}}]{yin18}
{Yin}, Y., {Jones}, D.~O., {Pan}, Y.~C., {Foley}, R.~J., \& {Siellez}, K. 2018,
  Transient Name Server Classification Report, 2018-1012, 1

\bibitem[{{Zhang} {et~al.}(2017{\natexlab{a}}){Zhang}, {Lu}, {Wang}, {Xiang},
  {Rui}, {Lin}, \& {Xu}}]{zhang17}
{Zhang}, J., {Lu}, K., {Wang}, X., {et~al.} 2017{\natexlab{a}}, The
  Astronomer's Telegram, 10707, 1

\bibitem[{{Zhang} {et~al.}(2016{\natexlab{a}}){Zhang}, {Wang}, {Sasdelli},
  {Zhang}, {Liu}, {Mazzali}, {Meng}, {Maeda}, {Chen}, {Huang}, {Zhao}, {Zhang},
  {Zhai}, {Pian}, {Wang}, {Chang}, {Yi}, {Wang}, {Wang}, {Xin}, {Wang}, {Lun},
  {Zheng}, {Zhang}, {Fan}, \& {Bai}}]{zhang16a}
{Zhang}, J.-J., {Wang}, X.-F., {Sasdelli}, M., {et~al.} 2016{\natexlab{a}},
  \apj, 817, 114, \dodoi{10.3847/0004-637X/817/2/114}

\bibitem[{{Zhang} {et~al.}(2017{\natexlab{b}}){Zhang}, {Wang}, {Wang}, {Li},
  {Tan}, {Li}, {Xu}, {Zhao}, {Wang}, {Xiang}, {Rui}, \& {Yang}}]{zhang17b}
{Zhang}, J.~J., {Wang}, J.~G., {Wang}, F.~X., {et~al.} 2017{\natexlab{b}}, The
  Astronomer's Telegram, 10895, 1

\bibitem[{{Zhang} {et~al.}(2016{\natexlab{b}}){Zhang}, {Wang}, {Zhang},
  {Zhang}, {Ganeshalingam}, {Li}, {Filippenko}, {Zhao}, {Zheng}, {Bai}, {Chen},
  {Chen}, {Huang}, {Mo}, {Rui}, {Song}, {Sai}, {Li}, {Wang}, \&
  {Wu}}]{zhang16b}
{Zhang}, K., {Wang}, X., {Zhang}, J., {et~al.} 2016{\natexlab{b}}, \apj, 820,
  67, \dodoi{10.3847/0004-637X/820/1/67}

\end{thebibliography}
\bibliographystyle{aasjournal}


%\bibliography{sample631}{}
%\bibliographystyle{aasjournal}



\end{document}